\pdfoutput=1
\documentclass[11pt,twoside,a4paper,cmspaper,final,collab]{cms-tdr}
\def\svnVersion{a0885fa}\def\svnDate{2021/11/11}\def\cmsCernNoTag{CERN-EP-2021-136}\def\cmsCernDate{\today}\def\cmsMessage{Published in the Journal of High Energy Physics as \href{http://dx.doi.org/10.1007/JHEP11(2021)153}{\doi{10.1007/JHEP11(2021)153}.}}
\begin{document}\cmsNoteHeader{EXO-20-004}

\newcommand{\Zvv}{\ensuremath{\PZ\to\PGn\PGn}\xspace}
\newcommand{\Zll}{\ensuremath{\PZ\to\Pell\Pell}\xspace}
\newcommand{\Vqq}{\ensuremath{\PV\to\PQq\PQq}\xspace}
\newcommand{\Wlv}{\ensuremath{\PW\to \Pell\PGn}\xspace}
\newcommand{\Zlljets}{\ensuremath{\PZ(\Pell\Pell)+\text{jets}}\xspace}
\newcommand{\Zjets}{\ensuremath{\PZ+\text{jets}}\xspace}
\newcommand{\Wjets}{\ensuremath{\PW+\text{jets}}\xspace}
\newcommand{\Wlvjets}{\ensuremath{\PW(\Pell\PGn)+\text{jets}}\xspace}
\newcommand{\phojets}{\ensuremath{\PGg+\text{jets}}\xspace}
\newcommand{\Vjets}{\ensuremath{\PV+\text{jets}}\xspace}
\newcommand{\Vgamma}{\ensuremath{\PV\PGg}\xspace}
\newcommand{\dmsimp}{\textsc{DMsimp}\xspace}
\newcommand{\dphitkpf}{\ensuremath{\Delta\phi(\mathrm{PF},\text{charged})}\xspace}
\newcommand{\ptvecjet}{\ensuremath{\ptvec^{\kern1pt\text{jet}}}\xspace}
\newcommand{\msd}{\ensuremath{m_\mathrm{SD}}\xspace}

\newcommand{\dpfcalo}{\ensuremath{\Delta\ptmiss(\text{PF--calorimeter})}\xspace}
\newcommand{\recoilvec}{\ensuremath{\vec{U}}\xspace}

\newcommand{\lambdalq}{\ensuremath{\lambda_\mathrm{LQ}}\xspace}
\newcommand{\lambdafp}{\ensuremath{\lambda_\mathrm{FP}}\xspace}
\newcommand{\gq}{\ensuremath{g_\PQq}\xspace}
\newcommand{\gchi}{\ensuremath{g_\chi}\xspace}

\newcommand{\dphijm}{\ensuremath{\Delta\phi(\ptvecmiss, \ptvec^\text{j})}\xspace}
\newcommand{\mmed}{\ensuremath{m_\text{med}}\xspace}
\newcommand{\mlq}{\ensuremath{m_\text{LQ}}\xspace}
\newcommand{\mdm}{\ensuremath{m_\text{DM}}\xspace}
\newcommand{\mphi}{\ensuremath{m_\Phi}\xspace}
\renewcommand{\MD}{\ensuremath{M_\text{D}}\xspace}

\newcommand{\delphes}{{\textsc{Delphes}}\xspace}
\newcommand{\madanalysis}{{\textsc{MadAnalysis}}\xspace}

\cmsNoteHeader{EXO-20-004}
\title{Search for new particles in events with energetic jets and large missing transverse momentum in proton-proton collisions at \texorpdfstring{$\sqrt{s} = 13\TeV$}{sqrt(s) = 13 TeV}}

\date{\today}

\abstract{
 A search is presented for new particles produced at the LHC in proton-proton collisions at $\sqrt{s}=13\TeV$, using events with energetic jets and large missing transverse momentum. The analysis is based on a data sample corresponding to an integrated luminosity of 101\fbinv, collected in 2017--2018 with the CMS detector. Machine learning techniques are used to define separate categories for events with narrow jets from initial-state radiation and  events with large-radius jets consistent with a hadronic decay of a \PW or  \PZ boson. A statistical combination is made with an earlier search based on a data sample of 36\fbinv, collected in 2016. No significant excess of events is observed with respect to the standard model background expectation determined from control samples in data. The results are interpreted in terms of limits on the branching fraction of an invisible decay of the Higgs boson, as well as constraints on simplified models of dark matter, on first-generation scalar leptoquarks decaying to quarks and neutrinos, and on models with large extra dimensions. Several of the new limits, specifically for spin-1 dark matter mediators, pseudoscalar mediators, colored mediators, and leptoquarks, are the most restrictive to date.}

\hypersetup{
pdfauthor={CMS Collaboration},
pdftitle={Search for new particles in events with energetic jets and large missing transverse momentum in proton-proton collisions at sqrt(s) = 13 TeV},
pdfsubject={CMS},
pdfkeywords={CMS, beyond standard model, dark matter, jets}}

\maketitle
\newboolean{vjets_mc_is_nlo}
\setboolean{vjets_mc_is_nlo}{true}

\section{Introduction}
The standard model (SM) of particle physics has been widely recognized as a very successful, yet incomplete theory. Many important features of the universe, such as gravity and the existence of dark matter (DM), are not described in the SM. It is therefore paramount to search for evidence of physics beyond the SM (BSM). Attempts at finding BSM physics often center around the production of new, hypothetical particles, which subsequently decay to the observable SM particles. In this search, we aim at scenarios that are hidden from such searches, because the decay products of BSM particles are not necessarily detectable.

\begin{figure*}[hbtp]
    \centering
        \includegraphics[width=0.245\textwidth]{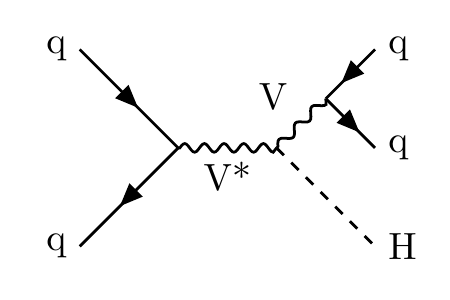}
        \includegraphics[width=0.245\textwidth]{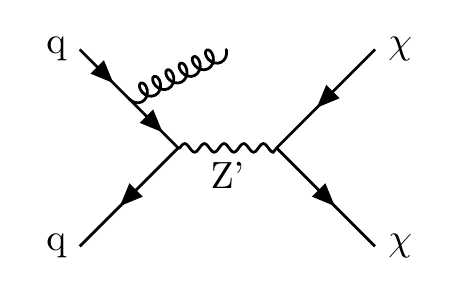}
        \includegraphics[width=0.245\textwidth]{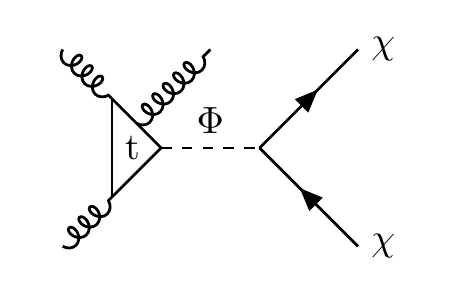}
        \includegraphics[width=0.245\textwidth]{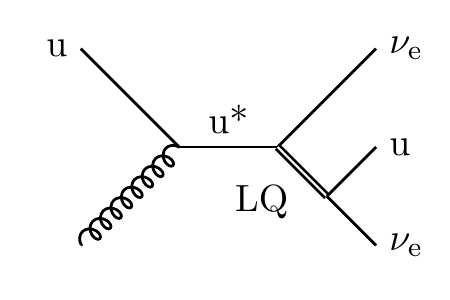}
        \caption{
            Representative Feynman diagrams for a number of signal models:  Higgs production in association with an SM vector boson (left), colorless spin-1 and spin-0 mediators (middle left and right, respectively), single leptoquark production (right). In all cases, subdominant production modes not pictured here are taken into account, as described in the text.}
        \label{fig:feyn}

\end{figure*}

Scenarios with new particles that are not directly observable in collider detectors are motivated by many BSM theories. One of the strongest motivations stems from the idea of particle DM. Over the last decades, cosmological evidence for the existence of DM has been steadily accumulating~\cite{FNAL_review}, yet with few hints as to its nature or detailed properties. One theoretically attractive model of DM is that of a thermally produced weakly interacting massive particle (WIMP). If such a particle has just the right mass and couplings, the abundance of DM in the universe, as well as many of the observed phenomena commonly ascribed to DM, can be explained. In this search, multiple scenarios of DM production are considered. A Higgs portal scenario~\cite{Higgs-portal1,Higgs-portal2,Higgs-portal3} is tested, in which DM particles are produced in decays of the Higgs boson~\cite{ATLAS-Higgs,CMS-Higgs1,CMS-Higgs2}. Many of the properties of the new boson have already been measured with impressive precision, but a decay branching fraction $\mathcal{B}$ to nondetectable particles of up to about 20\% is allowed by the current constraints~\cite{ATLAS-Hinv,CMS-Hinv}. Beyond the Higgs portal scenario, simplified models of DM production~\cite{Abercrombie:2015wmb} via new bosonic mediators with spin 0 or 1 are explored. Colorless mediators coupled to a pair of quarks and to a pair of DM particles are considered, as well as colored mediators, which decay into a single quark together with a single DM candidate. The latter scenario is referred to as a ``fermion portal''~\cite{Bai:2013iqa,DiFranzo:2013vra}. In addition to a search for DM, a scenario with large extra dimensions proposed by Arkani-Hamed, Dimopoulos, and Dvali (ADD)~\cite{ArkaniHamed:1998rs,ArkaniHamed:1998nn} is tested. In this model, the existence of additional spatial dimensions beyond the known three could explain the large difference in strength between the gravitational and electroweak (EW) interactions. In this scenario, gravitons can be produced in proton-proton ($\Pp\Pp$) collisions via their enhanced couplings to quarks or gluons and avoid detection by escaping in the additional dimensions. Representative Feynman diagrams for a subset of these signal models are shown in the first three panels of Fig.~\ref{fig:feyn}.

In these models the final-state particles are not detectable, but one needs a visible detector signature to be able to identify and record such events. We use energetic hadronic jets accompanying the invisible particles to select signal candidates. The experimental signature therefore comprises one or more energetic jets and large missing transverse momentum (\ptmiss). While the \ptmiss is the intrinsic result of BSM or SM particles escaping a detector without leaving any trace, hadronic jets derive from either initial-state gluon radiation or hadronic decays of energetic heavy SM vector bosons (\PV) produced in association with BSM particles. Production in association with a \PV boson is particularly important for the Higgs portal scenario, where the Higgs boson couples directly to the vector boson. For energetic \PV bosons, the hadronic decay products are Lorentz boosted in the laboratory frame and are reconstructed as a single large-radius jet with a characteristic substructure. Machine learning algorithms based on artificial neural networks are used in order to identify such signatures and efficiently suppress the overwhelming background coming from quantum chromodynamics (QCD) production of jets~\cite{Sirunyan:2020lcu}. Separate signal categories are defined for events with and without an identified \PV candidate. Several control samples in data are used to constrain background contributions to the signal regions.

The chosen experimental signature can also be used to probe other BSM scenarios with new particles decaying into final states with visible and invisible particles. One such scenario probed by the present search is the production of leptoquarks (LQs). The LQs are hypothetical scalar or vector particles that carry both baryon and lepton numbers~\cite{LQ1,LQ2,LQ3}. Here, a scenario with a single scalar LQ type is considered. This first-generation LQ decays into an up quark and an electron neutrino (\PGne), and can be either produced in pairs~\cite{Diaz:2017lit} via a coupling to gluons, or singly~\cite{Hewett:1987yg,Eboli:1987vb} in association with a \PGne, through its coupling to the up quark and \PGne. Both processes result in a jets + \ptmiss signature. A representative Feynman diagram for single LQ production is shown in the last panel of Fig.~\ref{fig:feyn}.

Searches for new phenomena in events with jets and \ptmiss at $\sqrt{s} = 13\TeV$ have been previously published by the CMS~\cite{Sirunyan:2017jix} and ATLAS~\cite{Aaboud:2018xdl,Aad:2021egl} Collaborations. The search is carried out with the CMS detector at the CERN LHC, in $\Pp\Pp$ collisions at $\sqrt{s} = 13\TeV$, using a data set collected in 2017--2018, corresponding to an integrated luminosity of 101\fbinv.
Compared to Refs.~\cite{Sirunyan:2017jix}, we have tripled the amount of analyzed data and enhanced the analysis sensitivity by means of improved identification of hadronically decaying \PV bosons. While such decays were previously selected using $N$-subjettiness~\cite{Thaler:2010tr}, we now use a criteria based on a deep neural network. We have further extended the sensitivity by combining the new results with those from Ref.~\cite{Sirunyan:2017jix}, which are based on a data set of 36\fbinv, yielding a total data set of 137\fbinv, equivalent in size to that of Ref.~\cite{Aad:2021egl}.

This paper is organized as follows. After discussing the CMS detector in Section 2 and the simulated samples in Section 3, we describe the event selection in Section 4, followed by the background estimation in Section 5. Section 6 contains the results of the analysis and their interpretation in the context of the above scenarios. We summarize the paper in Section 7. Tabulated results, as well as extensive material for use in reinterpretation, are provided in HEPData~\cite{hepdata}. To further aid reinterpretation, an implementation of the analysis selection is provided in the \textsc{MadAnalysis} framework~\cite{Conte:2014zja,Dumont:2014tja,DVN/IRF7ZL_2021}. Information related to the validation of this implementation is provided as supplementary material.

\section{The CMS detector and event reconstruction}
The central feature of the CMS apparatus is a superconducting solenoid of 6\unit{m} internal diameter, providing a magnetic field of 3.8\unit{T}. Within the solenoid volume are a silicon pixel and strip tracker, a lead tungstate crystal electromagnetic calorimeter (ECAL), and a brass and scintillator hadron calorimeter (HCAL), each composed of a barrel and two endcap sections. Forward calorimeters extend the pseudorapidity ($\eta$) coverage provided by the barrel and endcap detectors. Muons are detected in gas-ionization detectors embedded in the steel flux-return yoke outside the solenoid.

The silicon tracker measures charged particles within the pseudorapidity range $\abs{\eta} < 2.5$. During the LHC running period when the data used in this paper were recorded, the silicon tracker consisted of 1856 silicon pixel and 15\,148 silicon strip detector modules.

In the region $\abs{\eta} < 1.74$, the HCAL cells have widths of 0.087 in pseudorapidity and 0.087 in azimuth ($\phi$). In the $\eta$-$\phi$ plane, and for $\abs{\eta} < 1.48$, the HCAL cells map on to $5{\times}5$ arrays of ECAL crystals to form calorimeter towers projecting radially outwards from close to the nominal interaction point. For $\abs{\eta} > 1.74$, the coverage of the towers increases progressively to a maximum of 0.174 in $\Delta \eta$ and $\Delta \phi$. The hadron forward (HF) calorimeter uses steel as an absorber and quartz fibers as the sensitive material. The two halves of the HF are located 11.2\unit{m} from the interaction region, one on each end, and together they provide coverage in the range $3.0 < \abs{\eta} < 5.2$.

Events of interest are selected using a two-tiered trigger system. The first level (L1), composed of custom hardware processors, uses information from the calorimeters and muon detectors to select events at a rate of around 100\unit{kHz} within a fixed latency of about 4\mus~\cite{Sirunyan:2020zal}. The second level, known as the high-level trigger (HLT), consists of a farm of processors running a version of the full event reconstruction software optimized for fast processing, and reduces the event rate to around 1\unit{kHz} before data storage~\cite{Khachatryan:2016bia}.

A more detailed description of the CMS detector, together with a definition of the coordinate system used and the relevant kinematic variables, can be found in Ref.~\cite{Chatrchyan:2008zzk}.

The candidate vertex with the largest value of summed physics-object transverse momenta $\pt^2$ is taken to be the primary vertex (PV) of the $\Pp\Pp$ interaction. The physics objects are the jets, clustered using the jet finding algorithm~\cite{Cacciari:2008gp,Cacciari:2011ma} with the tracks assigned to candidate vertices as inputs, and the associated missing transverse momentum, taken as the negative vector sum of the \pt of those jets.

A particle-flow (PF) algorithm~\cite{CMS-PRF-14-001} aims to reconstruct and identify each individual particle in an event, with an optimized combination of information from the various elements of the CMS detector.  In this process, the identification of the PF candidate type (photon, electron, muon, and charged and neutral hadrons) plays an important role in the determination of the particle direction and energy. The energy of photons is obtained from the ECAL measurement. The energy of electrons is determined from a combination of the electron momentum at the PV as determined by the tracker, the energy of the corresponding ECAL cluster, and the energy sum of all bremsstrahlung photons spatially compatible with originating from the electron track. The energy of muons is obtained from the curvature of the corresponding track. The energy of charged hadrons is determined from a combination of their momentum measured in the tracker and the matching ECAL and HCAL energy deposits, corrected for the response function of the calorimeters to hadronic showers. Finally, the energy of neutral hadrons is obtained from the corresponding corrected ECAL and HCAL energies.

For each event, hadronic jets are clustered from the PF candidates using the infrared- and collinear-safe anti-\kt algorithm~\cite{Cacciari:2008gp, Cacciari:2011ma} with a distance parameter of 0.4 or 0.8. Depending on the respective distance parameter, these jets are referred to as ``AK4'' or ``AK8'' jets. Jet momentum is determined as the vectorial sum of all particle momenta in the jet, and is found from simulation to be, on average, within 5 to 10\% of the true momentum over the entire \pt spectrum and detector acceptance~\cite{Khachatryan:2016kdb}. Additional $\Pp\Pp$ interactions within the same or nearby bunch crossings (pileup) can contribute additional tracks and calorimetric energy depositions to the jet momentum. To mitigate this effect, charged particles identified as not originating from the PV are discarded and an offset correction is applied to correct for the remaining neutral pileup contributions~\cite{Khachatryan:2016kdb}. Jet energy corrections are derived from simulation to bring the measured response of jets to that of particle-level jets on average. In situ measurements of the momentum balance in the dijet, $\PGg + \text{jet}$, $\PZ + \text{jet}$, and multijet events are used to account for any residual differences in the jet energy scale (JES) and jet energy resolution (JER) in data and simulation~\cite{Khachatryan:2016kdb}.  The jet energy resolution amounts typically to 15--20\% at 30\GeV, 10\% at 100\GeV, and 5\% at 1\TeV~\cite{Khachatryan:2016kdb}. Additional selection criteria~\cite{CMS-PAS-JME-16-003} are applied to each jet to remove jets potentially dominated by anomalous contributions from various subdetector components or reconstruction failures.

The missing transverse momentum vector \ptvecmiss is computed as the negative vector sum of the transverse momenta of all the PF candidates in an event, and its magnitude is denoted as \ptmiss. The \ptvecmiss is modified to account for corrections to the energy scale and resolution of the reconstructed jets in the event~\cite{Sirunyan:2019kia}.  Anomalous high-\ptmiss events can be due to a variety of reconstruction failures, detector malfunctions, or noncollision backgrounds. Such events are rejected by dedicated filters that are designed to eliminate more than 85--90\% of the spurious high-\ptmiss events with a signal efficiency exceeding 99.9\%~\cite{Sirunyan:2019kia}.

Large-radius AK8 jets are used for the identification of hadronic decays of \PW and \PZ bosons. The pileup-per-particle identification (PUPPI) algorithm~\cite{Bertolini:2014bba} is used to mitigate the effect of pileup at the reconstructed-particle level, making use of local shape information, event pileup properties, and tracking information. Charged particles identified as not originating from the PV are discarded. For each neutral particle, a local shape variable is computed using the surrounding charged particles within the tracker acceptance ($\abs{\eta} < 2.5$) compatible with the PV, and using both charged and neutral particles in the region outside of the tracker coverage. The momenta of the neutral particles are then rescaled according to their probability to originate from the PV deduced from the local shape variable, avoiding the need for jet-based pileup corrections~\cite{CMS-PAS-JME-16-003}. The modified mass drop tagger algorithm~\cite{Dasgupta:2013ihk,Butterworth:2008iy}, also known as the soft-drop (SD) algorithm, with the angular exponent $\beta = 0$, soft cutoff threshold $z_{\text{cut}} < 0.1$, and characteristic radius $R_{0} = 0.8$~\cite{Larkoski:2014wba}, is applied to remove soft, wide-angle radiation from the jet.

\section{Simulated samples}
Monte Carlo (MC) simulated event samples are used to model signal and background contributions to all the analysis regions. In all cases, parton showering, hadronization, and underlying event properties are modeled using \PYTHIA~\cite{Sjostrand:2014zea} version 8.202 or later with the underlying event tune CP5~\cite{Sirunyan:2019dfx}. Simulation of interactions between particles and the CMS detector is based on \GEANTfour~\cite{AGOSTINELLI2003250}. The same reconstruction algorithms used for data are applied to simulated samples. The NNPDF3.1 next-to-next-to-leading order (NNLO) set of parton distribution functions (PDFs)~\cite{Ball:2017nwa} is used for the generation of all samples.

For the \Vjets processes, predictions with up to two partons in the final state are obtained at next-to-leading order (NLO) in QCD using \MGvATNLO version 2.4.2~\cite{Alwall:2014hca} with the FxFx matching scheme~\cite{Frederix:2012ps} between the jets from the matrix element calculations and the parton shower. The \phojets samples are simulated at NLO in QCD with up to one additional parton using \MGvATNLO version 2.6.5. This version is also used for all other \MGvATNLO samples, unless indicated otherwise. Samples of events with top quark pairs are generated at NLO in QCD with up to two additional partons in the matrix element calculations using \MGvATNLO and the FxFx jet matching scheme. Their cross sections are normalized to the inclusive cross section of the top quark pair production at NNLO in QCD~\cite{Czakon:2013goa}. Events with single top quarks are simulated using \POWHEG 2.0~\cite{Alioli:2009je,Re:2010bp} and normalized to the inclusive cross section calculated at NNLO in QCD~\cite{Kidonakis:2010ux} for single top quarks produced in association with a \PW boson, and NLO in QCD~\cite{Aliev:2010zk,Kant:2014oha} for production in association with a quark. Production of diboson events ($\PW\PW$, $\PW\PZ$, and $\PZ\PZ$) is simulated at leading order (LO) in QCD using \PYTHIA, and normalized to the cross sections at NNLO precision for $\PW\PW$ production~\cite{Gehrmann:2014fva} and at NLO precision for the others~\cite{Campbell:1999ah}. The production of $\PW\gamma$ and $\PZ\gamma$ events is simulated using \MGvATNLO at NLO in QCD. Samples of QCD multijet events are generated at LO using \PYTHIA.

For the Higgs portal signal model, \POWHEG is used to generate separate signal samples for the different production modes of the Higgs boson: via gluon fusion~\cite{Bagnaschi:2011tu}, in association with a SM vector boson (VH)~\cite{Luisoni:2013kna}, and via vector boson fusion (VBF)~\cite{Nason:2009ai}. The samples are generated by enforcing decays of the SM Higgs boson to neutrinos, and are normalized to the SM cross sections evaluated at next-to-NNLO in QCD and NLO in EW corrections for the gluon fusion production, and at NNLO in QCD and NLO in EW for the VBF and VH modes~\cite{deFlorian:2016spz}. Events for the simplified model scenarios of DM production are generated using \MGvATNLO and the \dmsimp model implementation~\cite{Mattelaer:2015haa,Backovic:2015soa,Neubert:2015fka}. For the case of spin-1 mediators, events with a pair of DM particles and either one or two additional partons are generated at NLO in QCD, and the FxFx jet matching is used. The couplings between the mediator and quarks, as well as between the mediator and the DM particles, are set to $\gq=0.25$ and $\gchi=1.0$, respectively, as recommended by the LHC Dark Matter Working Group~\cite{Albert:2017onk}. For DM production via spin-0 mediators, which is loop-induced, signal samples are generated at LO with one additional parton in the matrix element calculations, and the respective couplings are set to $\gq=\gchi=1.0$~\cite{Albert:2017onk}. Separate samples are generated for different coupling types (vector, axial vector, scalar, and pseudoscalar), as well as for different mass hypotheses for the mediator and DM particles. Signal events for the fermion portal scenario are generated using \MGvATNLO and the \textsc{S3D\_uR} implementation of Ref.~\cite{Arina:2020udz}. In this case, the mediator is assumed to couple to right-handed up quarks and a Dirac fermion DM candidate with a coupling of $\lambdafp=1$. The single and pair production of scalar LQs are simulated at LO in QCD using \MGvATNLO version 2.6.0 with an implementation provided by the authors of Ref.~\cite{Diaz:2017lit}. Decays of each LQ to an up quark and an electron neutrino are enforced, and separate samples are generated for the LQ mass values between $0.5$ and $2.5\TeV$, as well as for the LQ-\PQu-\PGne coupling values \lambdalq ranging from 0.01 to 1.5, depending on the LQ mass. Finally, events with graviton production in the ADD scenario are generated at LO using \PYTHIA~\cite{Ask:2009pv}. In this case, samples of signal events are generated for the number of extra dimensions $d$ between 2 and 7, and the values of the fundamental Planck scale $\MD$ between 5 and 15\TeV.

\section{Event selection}
The key feature of the analysis is the extensive use of control data samples for the purpose of precise prediction of the background contributions in the signal regions (SRs), which contain events with high-\pt jets and large \ptmiss. The leading SM background contributions originate from \Zvv and \Wlv production ($\ell=\Pe,\PGm,\PGt$), the properties of which are constrained using control regions (CRs) with charged leptons that are enriched in \Zll and \Wlv events, respectively. Additionally, CRs enriched in \phojets events are defined. The \Vjets events in these CRs share many kinematic properties of the processes in the SRs and are used to constrain the latter. The CR and SR definitions share as many of the selection criteria as possible, in order to ensure that minimal selection biases are introduced. For each SR, five CRs are defined: dielectron and dimuon CRs enriched in $\Zll$ events, single-electron and single-muon CRs enriched in $\Wlv$ events, and a fifth CR enriched in \phojets events.

The SR events are selected using a trigger with a \ptmiss requirement of at least $120\GeV$. The trigger requirement for the SRs is based on an online calculation of \ptmiss based on all PF candidates reconstructed at the HLT, except for muons. Events with high-\pt muons are therefore also assigned large online \ptmiss, and the same trigger is used to collect data populating the single-muon and dimuon CRs. The control samples with electrons are selected based on two different single-electron triggers requiring of $\pt>35$ (32) \GeV for 2017 (2018) and $\pt>115\GeV$, and on a single-photon trigger with a requirement of $\pt>200\GeV$. The single-electron triggers differ in their usage of isolation requirements: while the lower threshold trigger requires electrons to be well isolated, the higher-threshold trigger does not, which gives an improved efficiency at high \pt. Similarly, the single-photon trigger avoids the reliance on the online track reconstruction and increases the overall efficiency for electrons with $\pt>200\GeV$. The photon trigger is also used to select events for the photon control samples. During the 2017 data taking, a gradual shift in the timing of the inputs of the ECAL L1 trigger in the region at $\abs{\eta} > 2.0$ caused a specific trigger inefficiency. For events containing an electron or a photon (a jet) with $\pt\gtrsim50$ (100)\GeV in this region, the efficiency loss is up to $\approx$10--20\%, depending on \pt, $\eta$, and time. Correction factors are computed from data and applied to the acceptance evaluated by simulation for the 2017 samples.

At the analysis level, a requirement of $\ptmiss>250\GeV$ is applied to the SR events in order to ensure a \ptmiss trigger efficiency of at least $95\%$. Events are separated into three mutually exclusive categories based on the properties of the highest \pt (``leading'') jet in the event: low-purity mono-\PV, high-purity mono-\PV, and monojet. For the mono-\PV categories, the leading AK8 jet is required to have $\pt>250\GeV$ and $\abs{\eta}<2.4$. In order to preferentially select events where an AK8 jet originates from a hadronic decay of a \PW or \PZ boson, the jet is further required to be \PV tagged with the \textsc{DeepAK8} algorithm~\cite{Sirunyan:2020lcu} and to have an SD-corrected mass of $65 < m_\text{SD} < 120\GeV$. The \textsc{DeepAK8} algorithm employs a deep neural network to differentiate between jets from vector boson, top quark, and Higgs boson decays, as well as jets originating from QCD radiation. The inputs to the neural network are features of up to 100 jet constituent PF candidates of a given jet and features related to up to seven secondary vertices reconstructed in a given collision event. For each jet, the output of the neural network is one numerical score for each of the jet classes, representing the likelihood that the jet originates from that class. In this analysis, separation between vector boson and QCD jets is sought, and a binary score is constructed by taking the ratio of the vector boson score to the sum of vector boson and QCD scores. The assignment to low- and high-purity mono-V categories is then based on the binary score of the leading jet. The high-purity category selects genuine V jets (QCD jets) with an efficiency of 30 (0.7)\% at a jet \pt of 250\GeV, rising to 40 (0.7)\% at 800\GeV. For jets failing the high-purity selection, the low-purity selection has an efficiency of 40 (7)\% at 250\GeV, falling to 30 (5)\% at 800\GeV. Compared to the $N$-subjettiness-based selection employed in the previous analysis~\cite{Sirunyan:2017jix}, the \textsc{DeepAK8} tagger reduces the rate of QCD jets incorrectly identified as vector boson jets by a factor of five to ten depending on jet \pt without reducing the efficiency for genuine V jets. Events that do not pass the mono-\PV selection are considered for the monojet category. In this case, the leading AK4 jet in the event is required to have $\pt>100\GeV$, $\abs{\eta} < 2.4$, and to pass quality criteria based on the composition of the jet in terms of different types of PF candidates, such as a minimum charged-hadron energy fraction of 10\% and a maximum neutral-hadron energy fraction of 80\%~\cite{CMS-PAS-JME-16-003}.

In all categories, further requirements are imposed in order to suppress reducible background processes. Events are rejected if they contain a well-reconstructed and isolated electron (photon) with $\pt>10$ (15)\GeV and $\abs{\eta}<2.5$, or a muon with $\pt>10\GeV$ and $\abs{\eta}<2.4$~\cite{Sirunyan_2021,Sirunyan:2018fpa}. Hadronically decaying \PGt leptons are identified using the ``hadrons-plus-strips'' algorithm and a multivariate classifier at a working point corresponding to an efficiency of 70\% for genuine \PGt decays and 0.5--3\% for jets from QCD production, depending on jet \pt~\cite{Sirunyan:2018pgf}. Events with a hadronically decaying \PGt lepton candidate with $\pt>18\GeV$ and $\abs{\eta}<2.3$ are removed. These requirements efficiently reject events with leptonic decays of the \PV  bosons and top quarks, as well as backgrounds with photons. Contributions from top quark processes are further suppressed by rejecting events with AK4 jets that have $\pt>20\GeV$, $\abs{\eta}<2.4$, and are identified to have originated from the hadronization of a bottom quark (``\PQb-tagged jets'') using the \textsc{DeepCSV} algorithm with a ``medium" working point, corresponding to correctly identifying a \PQb jet with a probability of $80\%$ and misidentifying a light-flavor quark or gluon jet with a probability of 10\%~\cite{Sirunyan:2017ezt}. Finally, topological requirements are applied in order to reject contributions from QCD multijet events. These events do not have \ptmiss from genuine sources and require a \ptmiss  mismeasurement in order to pass the SR selections, which can happen in two main ways. In the first case, the energy of a jet in the event could be misreconstructed either as a result of an interaction between the jet with poorly instrumented or inactive parts of the detector, or because of failures in the readout of otherwise functioning detector modules. In these cases, artificial \ptmiss is generated with a characteristically small azimuthal angle difference between the misreconstructed jet \ptvec and the \ptvecmiss vectors. Such events are rejected by requiring $\Delta\phi(\ptvecjet,\ptvecmiss) > 0.5$. In the second case, large \ptmiss is generated due to failures of the PF reconstruction, which are suppressed by considering an alternative calculation of \ptmiss based on calorimeter energy clusters and muon candidates, rather than the full set of all PF candidates. While the calorimeter-based \ptmiss has significantly worse resolution than PF \ptmiss, it is much simpler and more robust. To reduce the multijet background caused by PF reconstruction failures, events are required to have $\dpfcalo = \abs{\ptmiss(\text{PF})/\ptmiss(\text{calorimeter}) - 1}<0.5$. A similar criterion is constructed using an alternative \ptmiss calculation based exclusively on charged-particle candidates. Since charged particles are only reconstructed within the coverage of the pixel tracking detector, this \ptmiss variant is robust against noise and PU contributions in the forward calorimeters. Events in the SR are required to have a maximum angular separation in the transverse plane between the regular and charged-particle candidate \ptmiss vectors of $\dphitkpf<2$. Finally, a section of the HCAL was not functioning during a part of the 2018 data taking period corresponding to 65\% of the total integrated luminosity recorded in that year, leading to irrecoverable mismeasurement in a localized region of the detector ($-1.57<\phi<-0.87$, $-3.0<\eta<-1.3$). To avoid contamination from such mismeasurement, events where any jet with $\pt>30\GeV$ is found in the corresponding $\eta$-$\phi$ region are rejected in the analysis of the 2018 data set. Events where the mismeasurement is so severe that a jet is fully lost in this region are found to contribute at low values of $\ptmiss<470\GeV$ and to have a characteristic signature in $\phi(\ptvecmiss)$. Such events are rejected by requiring that $\phi(\ptvecmiss)\notin[-1.62,-0.62]$ if $\ptmiss<470\GeV$. The value of 470\GeV is the boundary of the optimal signal region binning just above this contamination region.

In each of the CRs, the same selection criteria are applied as for the corresponding SR (monojet, or low- or high-purity mono-\PV), with two exceptions: the charged-lepton and photon rejection criteria are inverted to allow the exact number of desired leptons or photons for each CR, and the \ptvecmiss vector used in the SR definition is replaced by the hadronic recoil vector \recoilvec. The hadronic recoil is defined as the vectorial sum of the \ptvecmiss vector and the transverse momentum vectors of the selected charged lepton(s) or the photon in each event. The hadronic recoil therefore acts as a proxy of the momentum of the \PV boson or a photon in each CR, convolved with the \ptmiss resolution, which is equivalent to the role of \ptmiss in the SRs. In order to enhance the purity of the CRs, specific additional selection criteria are applied. For the charged-lepton CRs, at least one of the leptons is required to pass a more strict set of quality criteria and have $\pt>40$ (20)\GeV electrons (muons), while the photon in the photon CR is required to have $\pt>230\GeV$ in order to ensure high trigger efficiency. Additionally, events in the single-lepton CRs are required to have a transverse mass $\mT=\sqrt{\smash[b]{2\ptmiss\pt^\ell(1-\cos[\Delta\phi(\ptvecmiss,\ptvec^\ell)])}}<160\GeV$, and events in the single-electron CR are required to have $\ptmiss>50\GeV$ in order to reject contributions from QCD multijet events. Finally, in order to enrich the dilepton CRs with \PZ events, the two leptons are required to have opposite signs and to have an invariant mass in the range $60<m_{\ell\ell}<120\GeV$, consistent with the mass of the \PZ boson~\cite{PDG2020}.

The event selection criteria for the signal regions of the different analysis categories, and the topological selection differences between regions in the same category are respectively summarized in Tables~\ref{tab:supp_selection} and~\ref{tab:supp_regions} in Appendix~\ref{app:appendix_selection_tables}.

\section{Background estimation}
{\tolerance=800
Background estimation and signal extraction are performed simultaneously, using a joint maximum likelihood (ML) fit across all SRs and the corresponding single-lepton, dilepton, and photon CRs. For each analysis category, a likelihood function is constructed to model the expected background contributions in each recoil variable bin of the SR and CRs, as well as the expected signal yield in each bin of the SR. The best fit background model, as well as the best fit signal strength, are obtained by maximizing the joint likelihood function of all categories.
\par}
\subsection{Likelihood function}
The likelihood function is defined in the same way as described in Ref.~\cite{Sirunyan:2017hci} and previously used in Ref.~\cite{Sirunyan:2017jix}. Separate approaches are adopted to estimate the dominant (\Zjets, \Wjets, \phojets) and subdominant (\ttbar, diboson, and QCD multijet) backgrounds.

The predictions for the dominant backgrounds are based on the yield of \Zvv events in each bin of the SR. The per-bin yields for this process are defined as free parameters of the likelihood function. The yields for the \Wjets contribution to the SR, as well as the yields of the \phojets process in the photon CR and the \Zll process in the dilepton CRs, are defined relative to the \Zvv yields by introducing a set of per-bin transfer factors. The yields of \Wlv events in the single-lepton CRs are similarly related via transfer factors to the \Wlv event yields in the SRs. This choice of transfer factors takes into account the correlations between the \Vjets background contributions in all regions. In all cases, the central values of the transfer factors are obtained from the ratios of the simulated recoil spectra of the respective processes in the SRs to those in CRs. For the minor backgrounds, such as \ttbar and QCD multijet production, the nominal expected yield per region is obtained directly from simulation (top quark and diboson backgrounds, as well as QCD multijet production in the single-lepton CRs) or by dedicated estimates based on control samples in data (QCD multijet production in the SRs and photon CRs). Contributions from triboson processes are negligible.

Systematic uncertainties are incorporated in the likelihood function as nuisance parameters, as described in more detail below. In the case of the \Vjets processes, the nuisance parameters affect the values of the transfer factors in each recoil variable bin and thus control the ratios of the contributions from different processes, as well as the ratios of the yields in the SRs to those in various CRs. For the subdominant background processes, the yields in each bin are directly parameterized in terms of the nuisance parameters. The final free parameter of the likelihood function is the signal strength modifier $\mu$, which---for a given signal hypothesis---controls the signal normalization relative to the theoretical cross section.

The likelihood method relies on the accurate predictions of the ratios between the dominant backgrounds in the SRs and CRs, as well as on the absolute normalization and shape of the recoil distributions for the subdominant backgrounds. To achieve the most accurate possible predictions for these quantities, weights are applied to each simulated event to take into account both experimental and theoretical effects not present in the MC simulated samples. The experimental corrections are related to the trigger efficiencies, identification and reconstruction efficiencies of charged leptons, photons and \PQb-tagged jets, and the pileup distribution in simulation. Theoretical corrections are applied to the \Vjets processes in order to model the effects of NLO terms in the perturbative EW corrections~\cite{Lindert:2017olm}. The corrections are parameterized as functions of the generator-level boson \pt and are evaluated separately for the \Wlvjets, \Zlljets, and \phojets processes. For the diboson processes ($\PW\PW$, $\PW\PZ$, and $\PZ\PZ$), EW and QCD NLO corrections are applied differentially in the boson \pt. The EW corrections are obtained from Ref.~\cite{Baglio:2013toa}, while the QCD corrections are derived from simulated samples generated with \MGvATNLO and \POWHEG. The EW NLO corrections for the $\PW\gamma$ and $\PZ\gamma$ processes are similarly obtained from Refs.~\cite{Denner:2014bna, Denner:2015fca}.

\begin{figure*}[hbtp]
    \centering
        \includegraphics[width=0.4\textwidth]{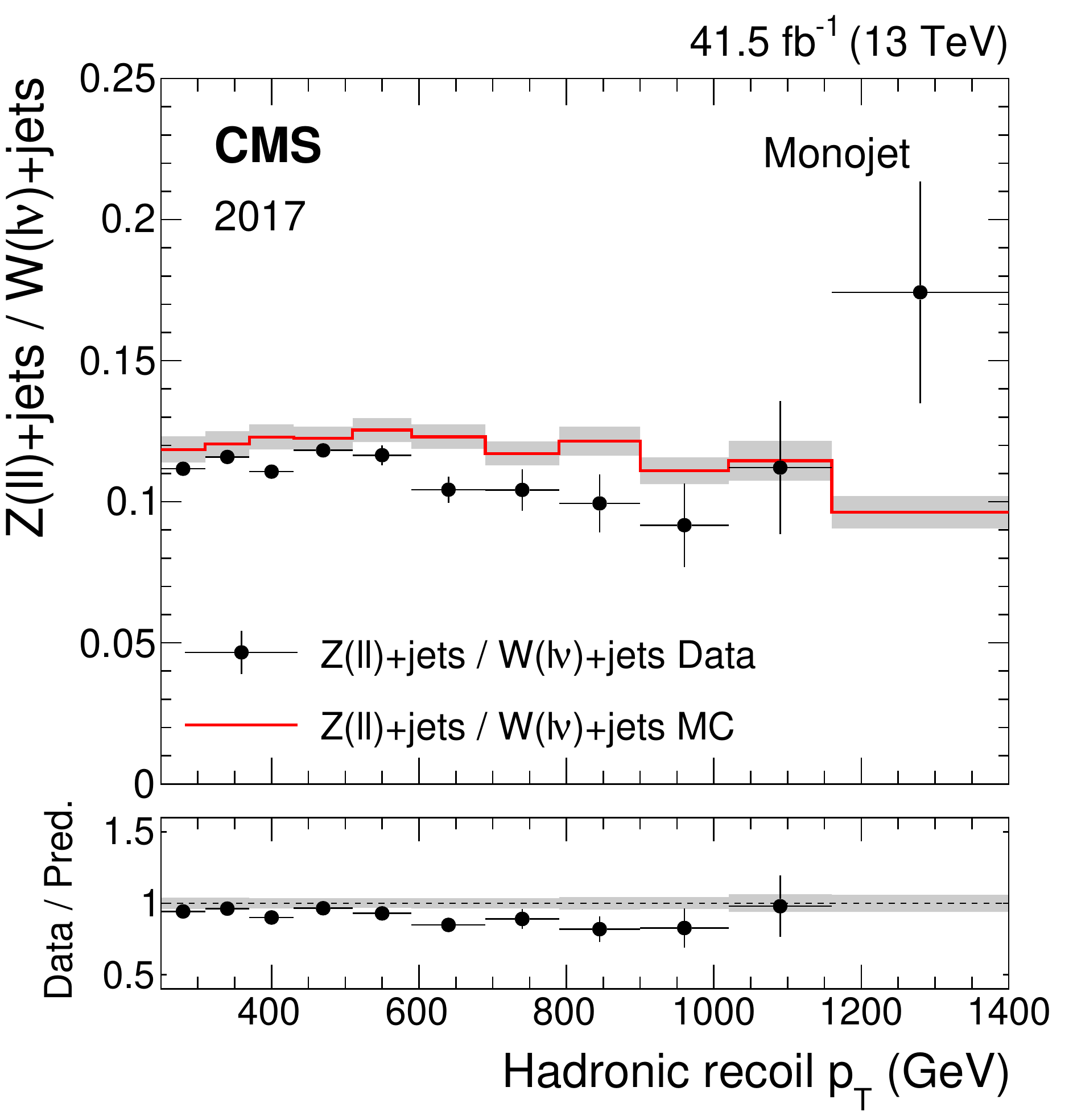}
        \includegraphics[width=0.4\textwidth]{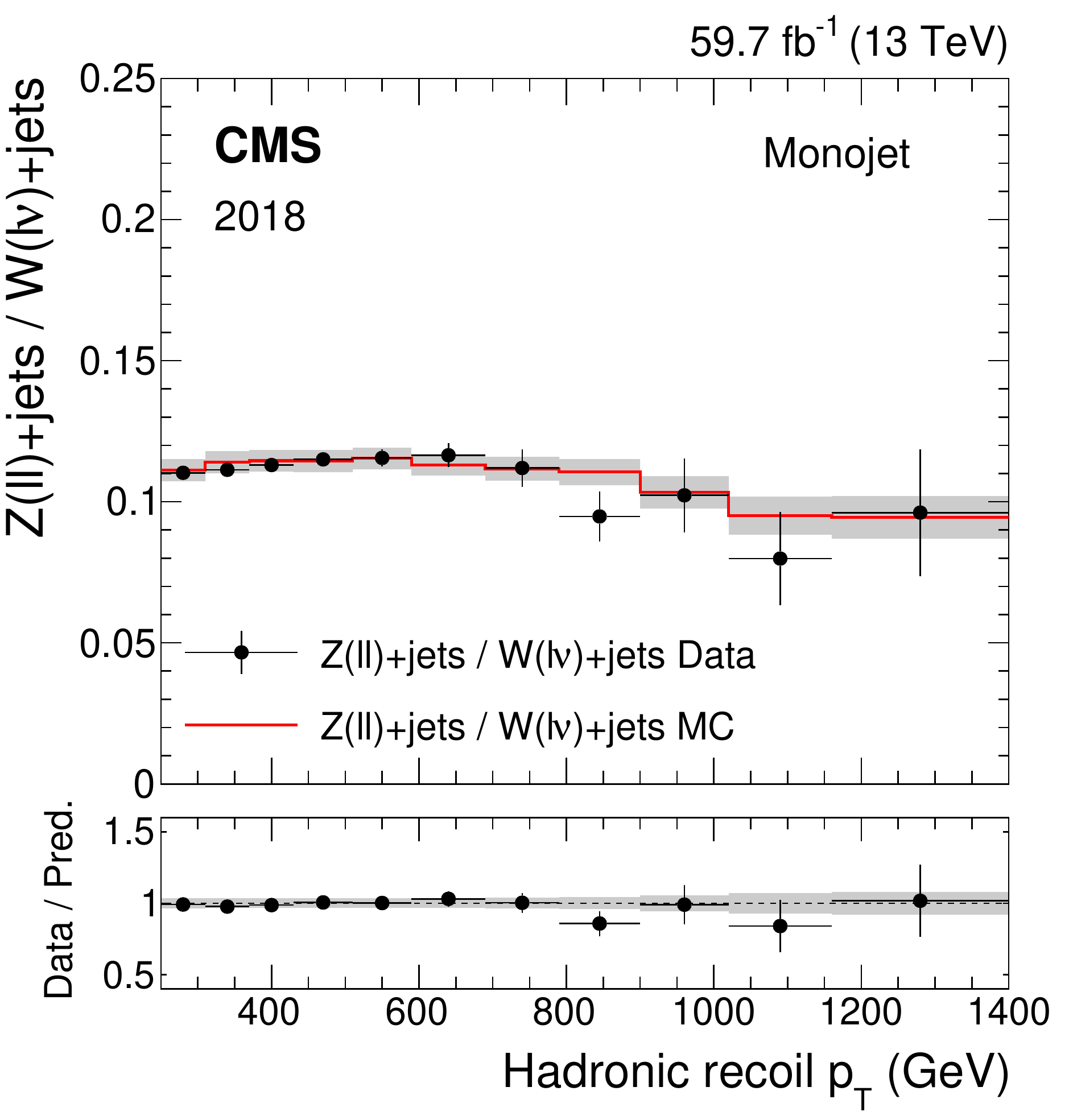}
        \\
        \includegraphics[width=0.4\textwidth]{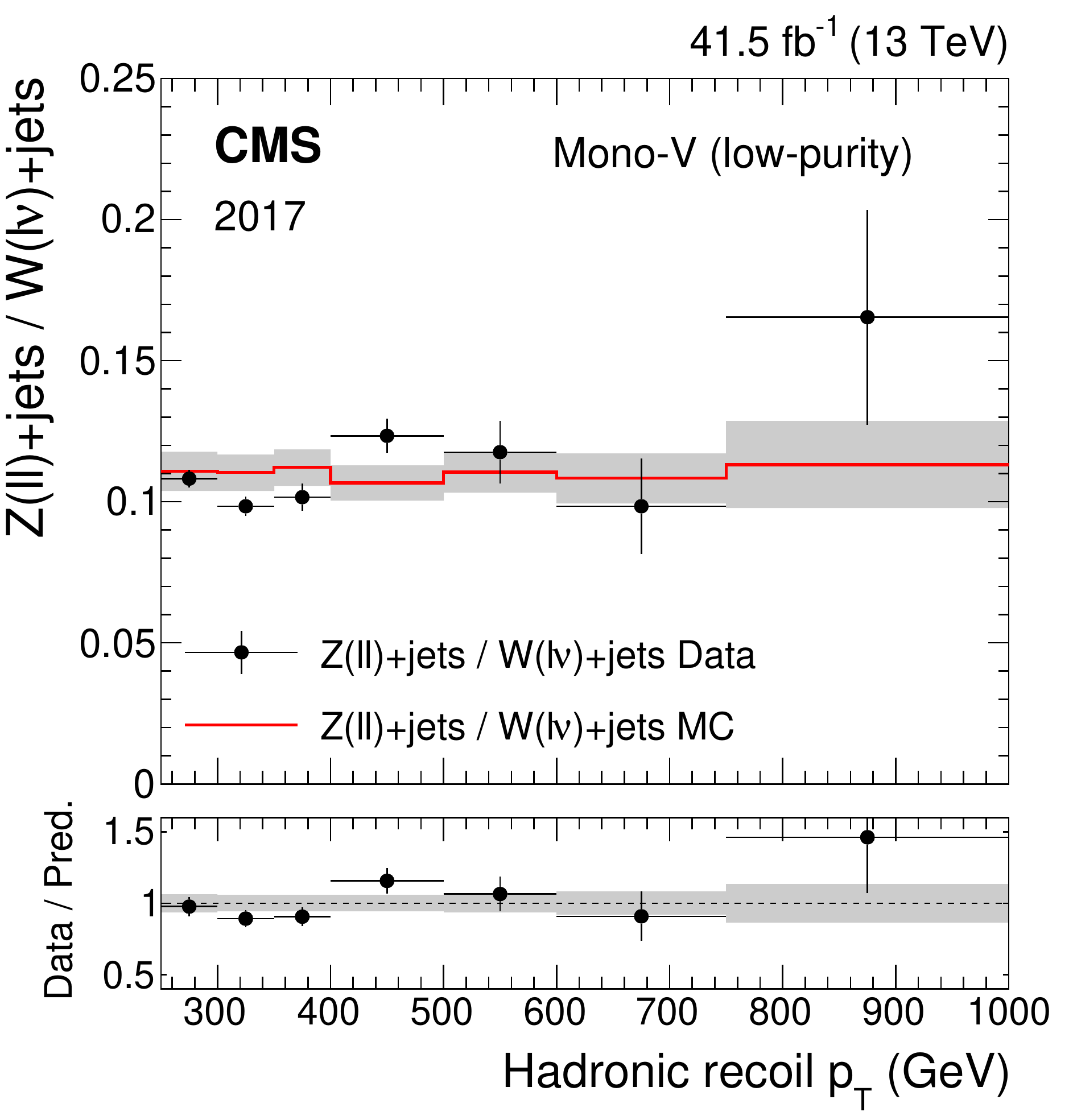}
        \includegraphics[width=0.4\textwidth]{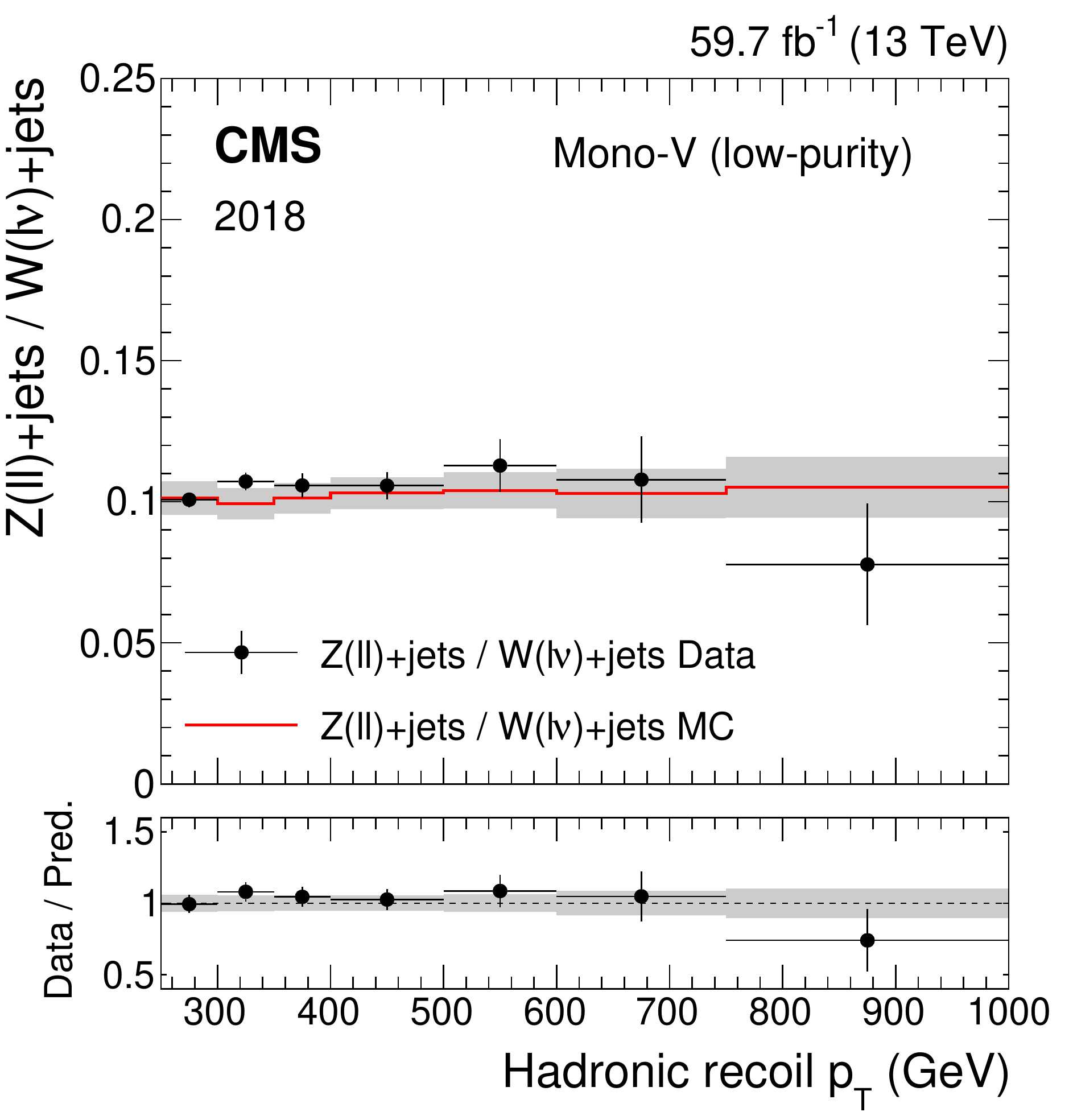}
        \\
        \includegraphics[width=0.4\textwidth]{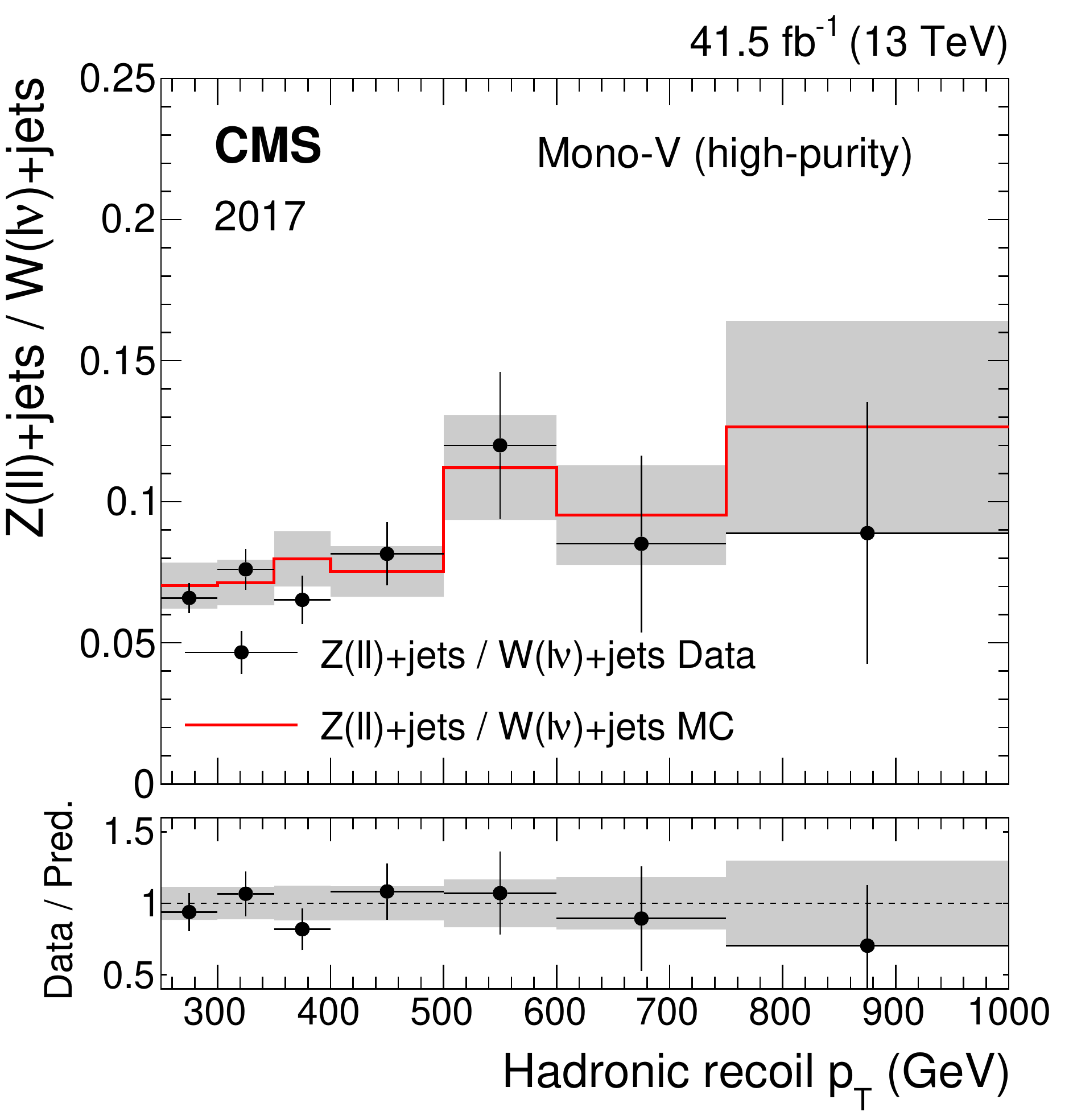}
        \includegraphics[width=0.4\textwidth]{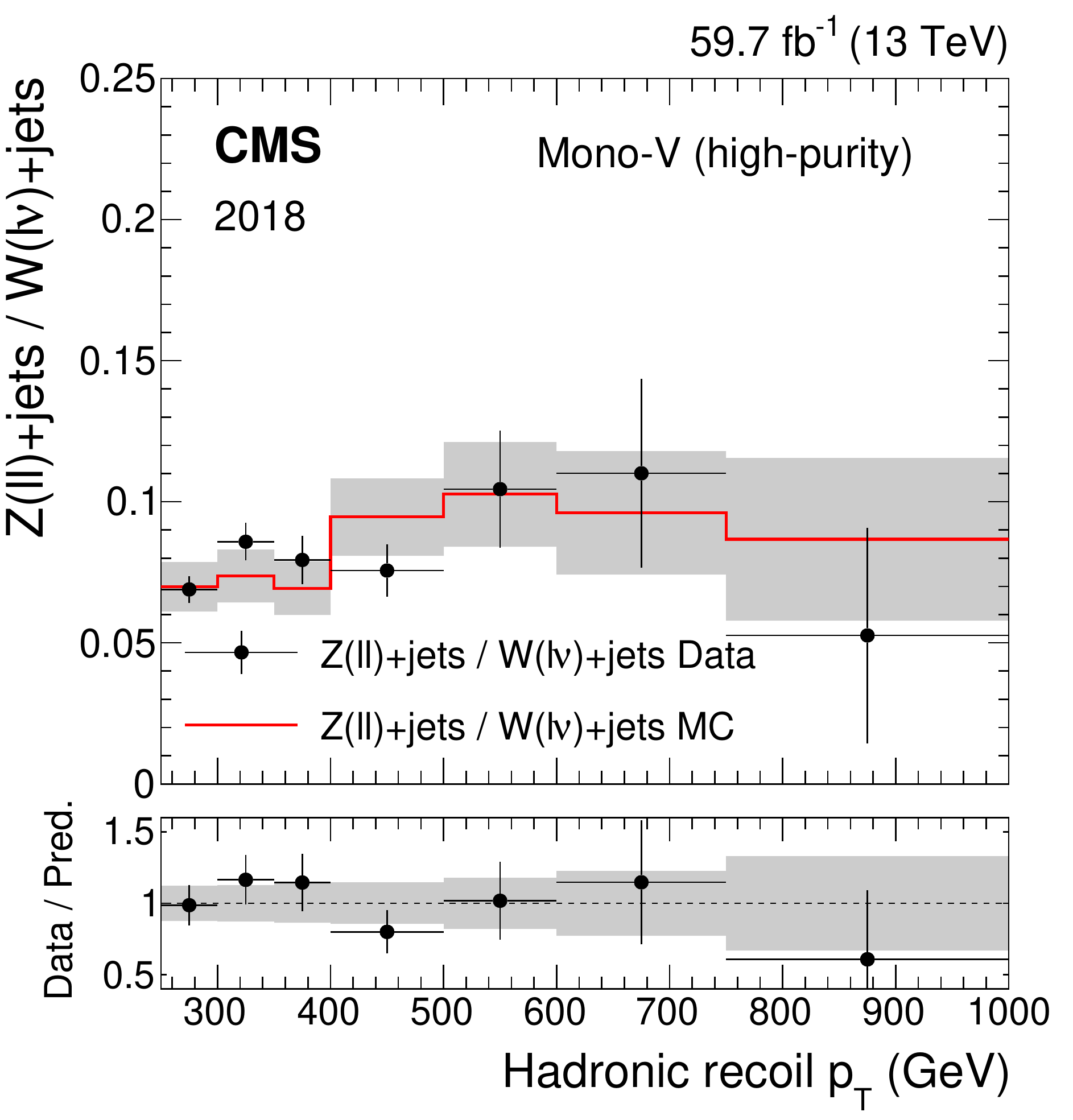}
        \\
        \caption{
            Ratio of the dilepton to single-lepton control region yields predicted using simulation (red solid line), and observed in data (black points). The gray band represents the total uncertainty in the ratio. In the lower panels, the ratio of data over prediction is shown. From upper to lower, the rows show the monojet, low-purity, and high-purity mono-V categories, while the left (right) column represents the 2017 (2018) data set.
        }
        \label{fig:cr_ratios_z_over_w}

\end{figure*}

\begin{figure*}[hbtp]
    \centering
        \includegraphics[width=0.4\textwidth]{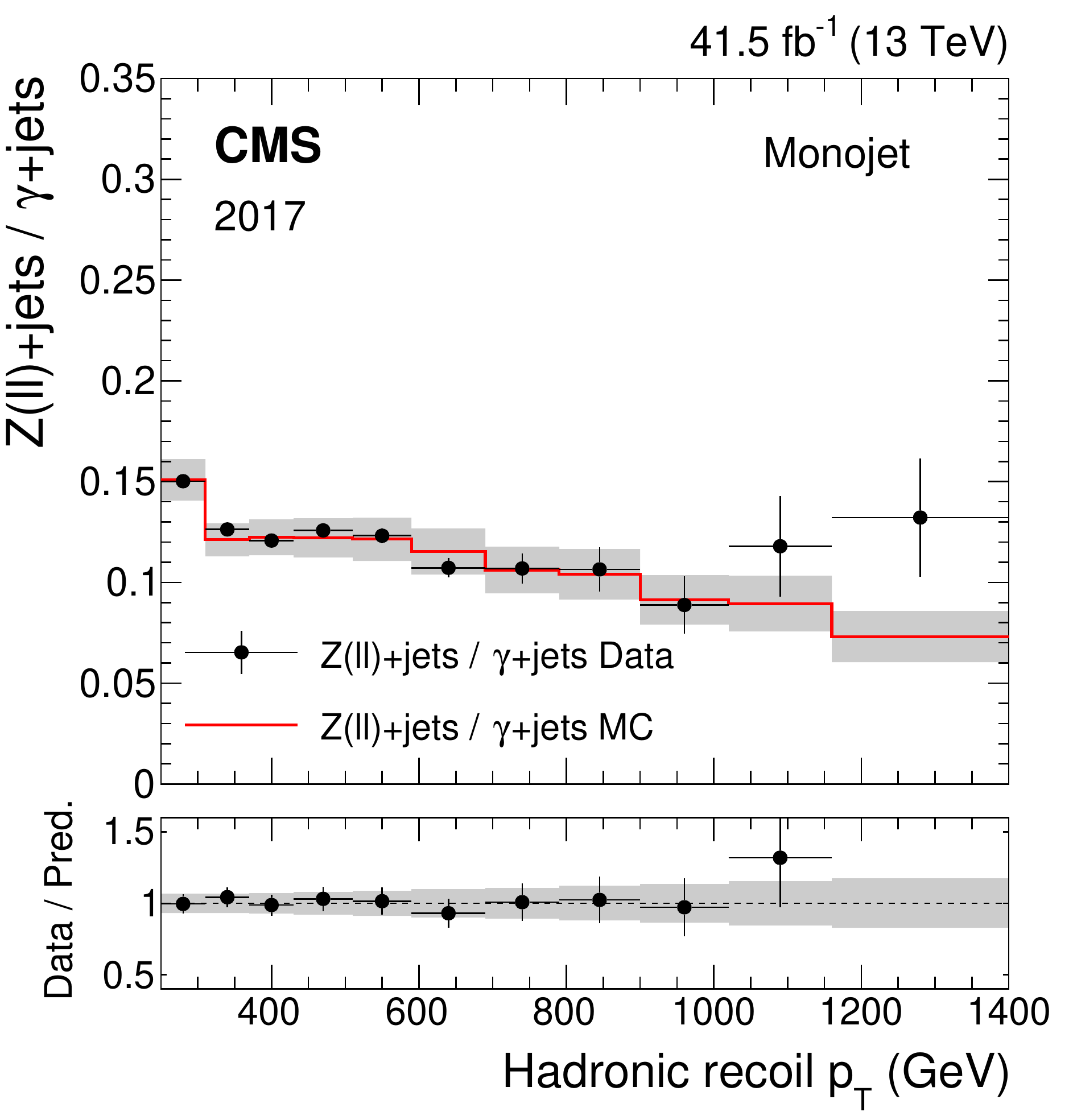}
        \includegraphics[width=0.4\textwidth]{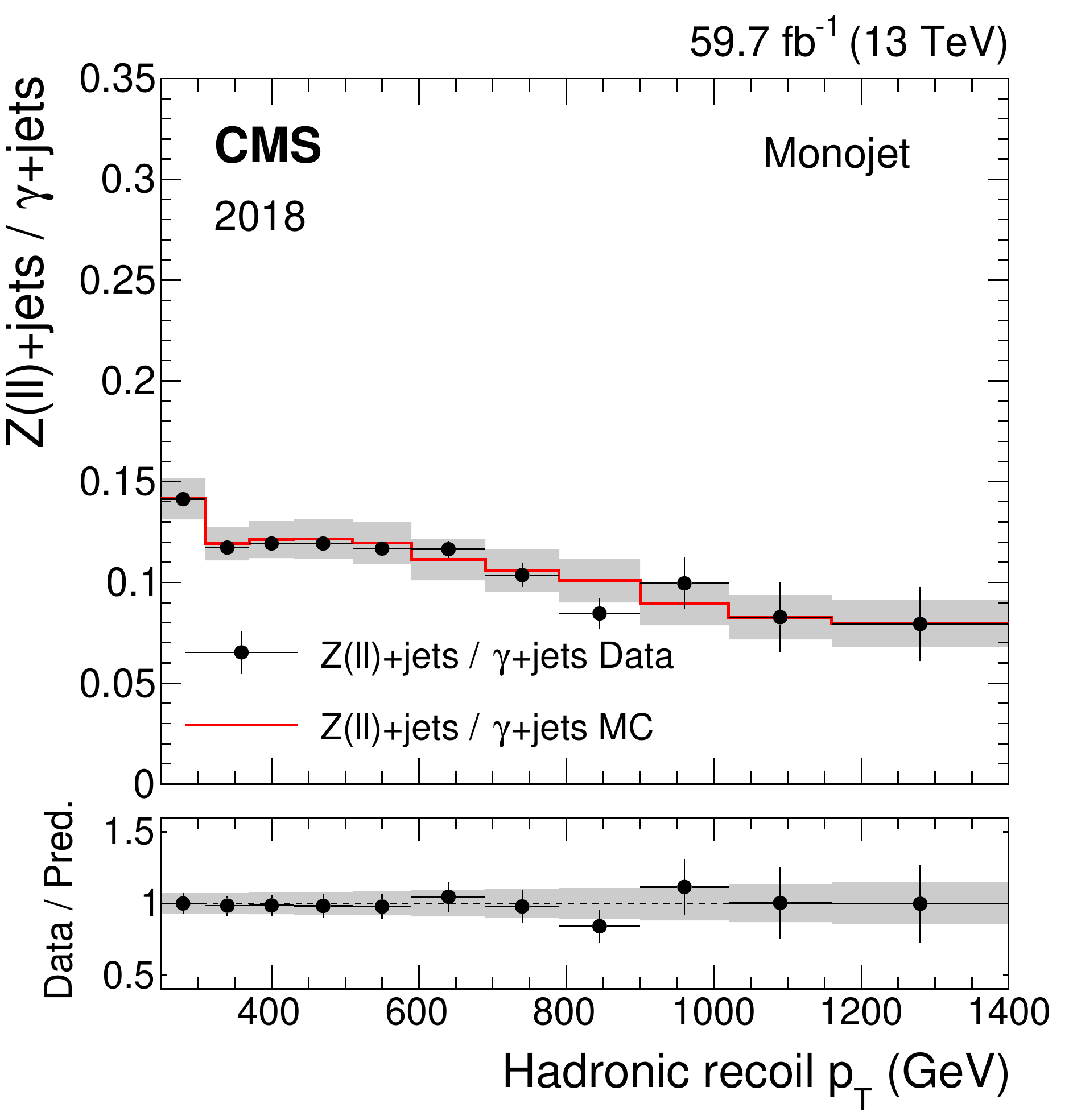}
        \\
        \includegraphics[width=0.4\textwidth]{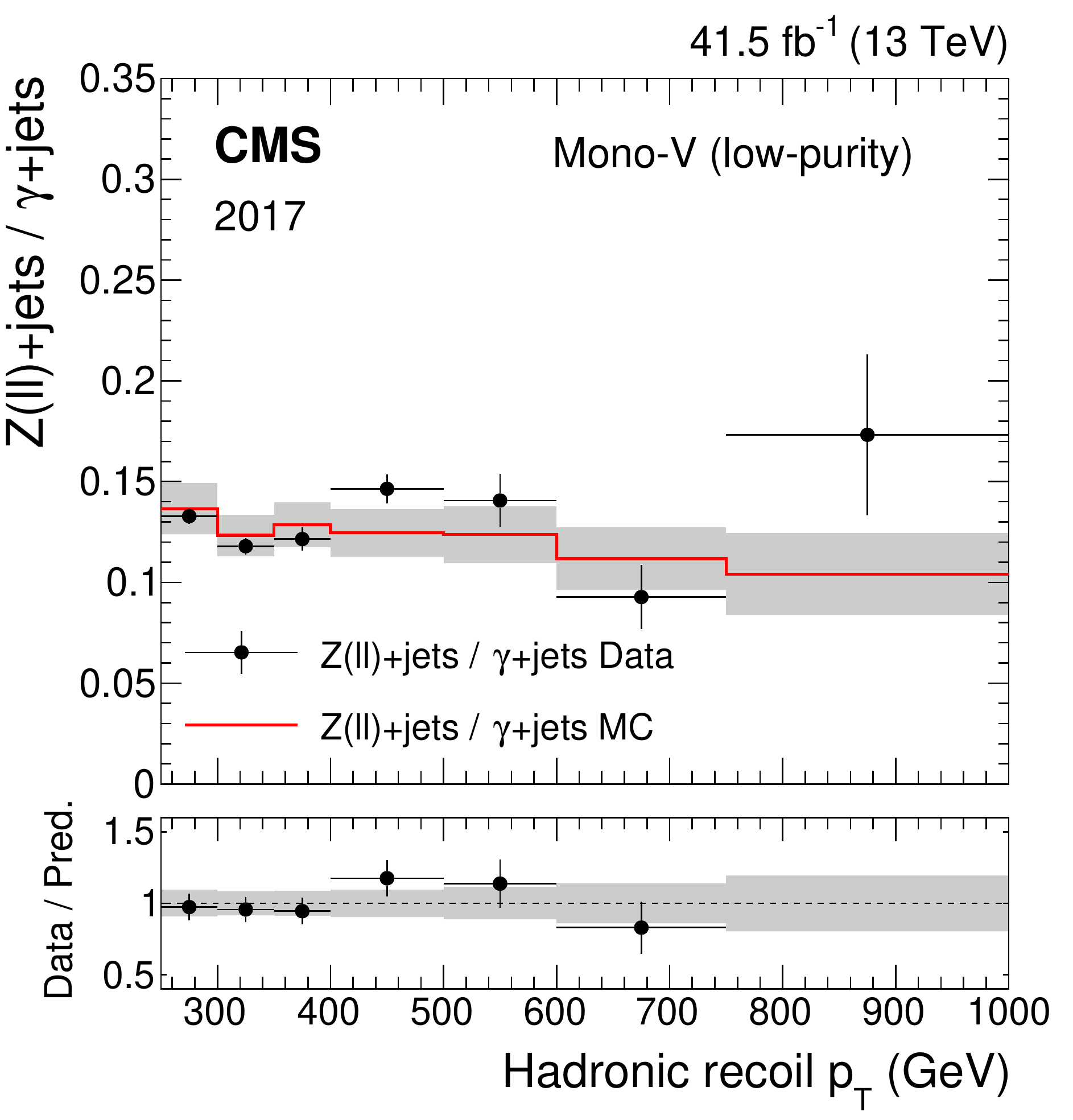}
        \includegraphics[width=0.4\textwidth]{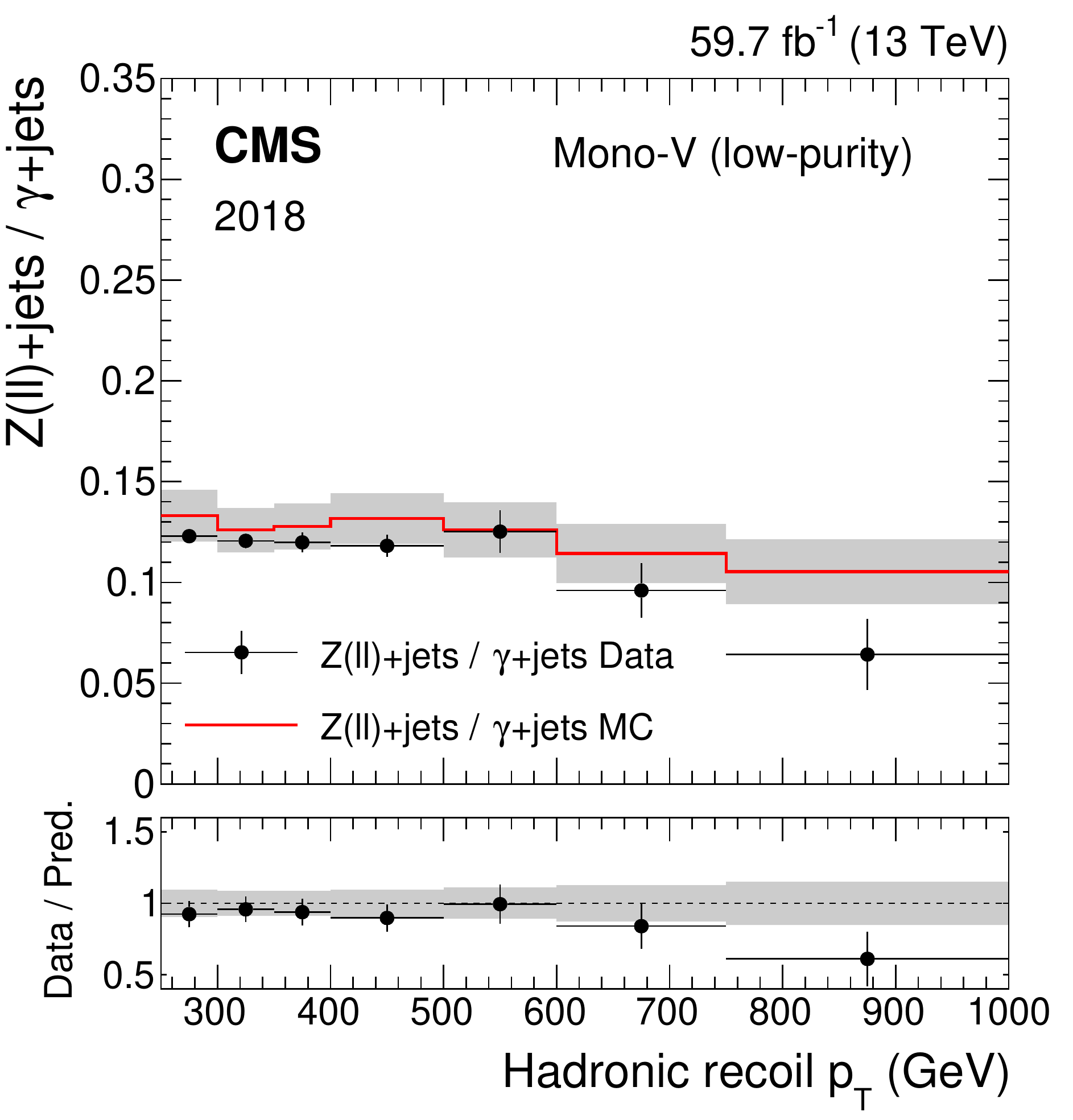}
        \\
        \includegraphics[width=0.4\textwidth]{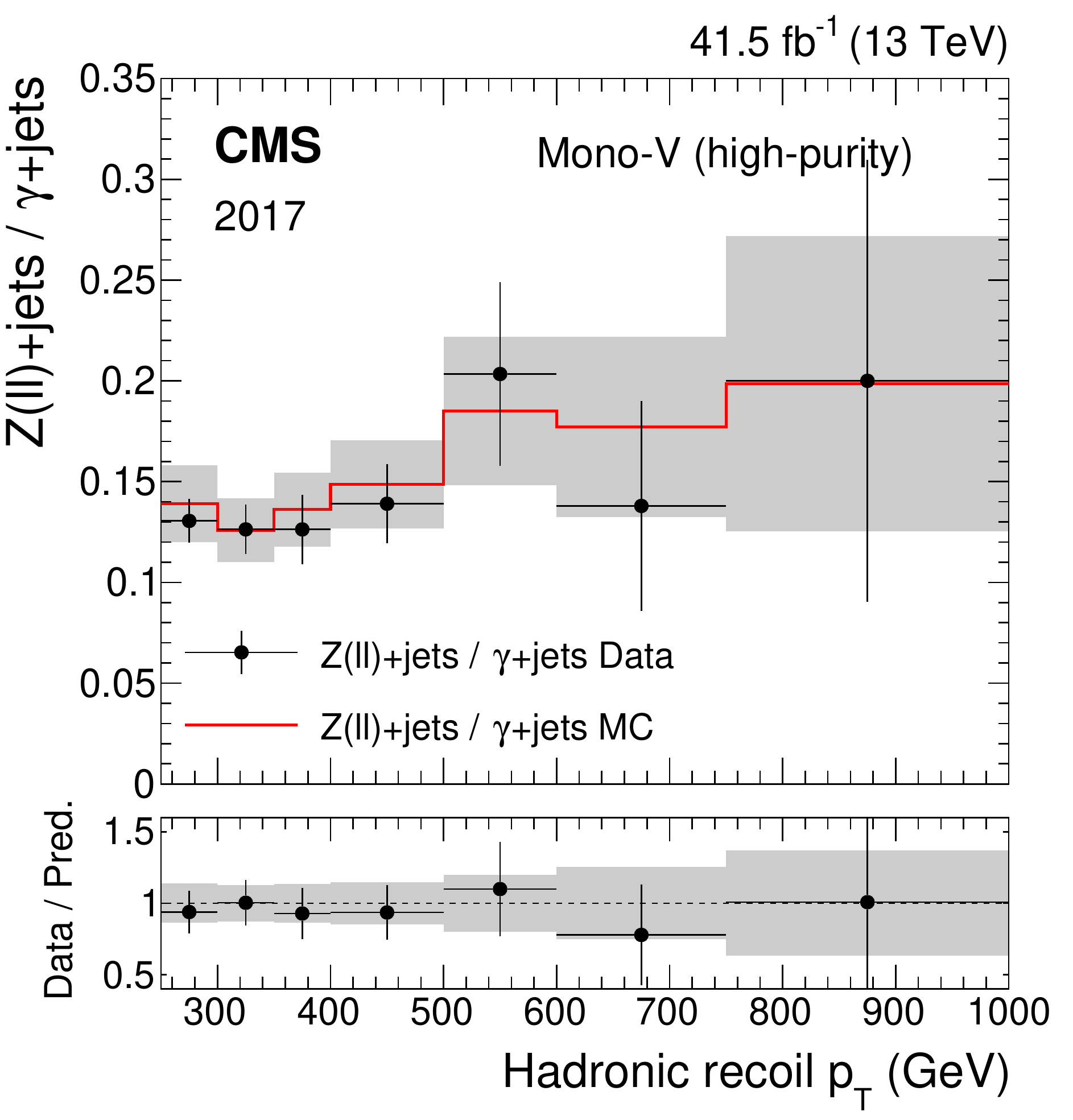}
        \includegraphics[width=0.4\textwidth]{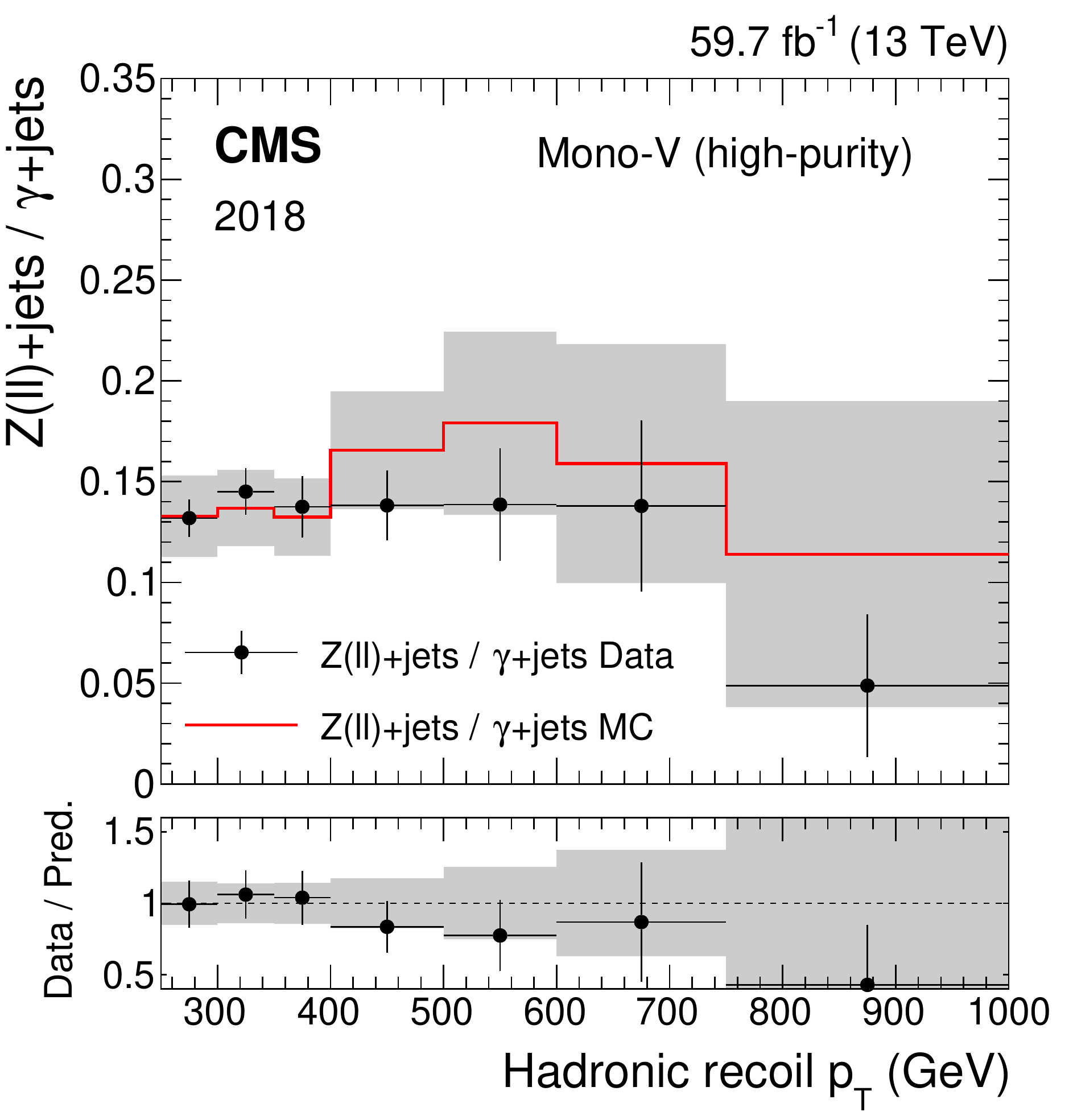}
        \\
        \caption{
            Ratio of the dilepton to photon control region yields predicted using simulation (red solid line), and observed in data (black points). The gray band represents the total uncertainty in the ratio. In the lower panels, the ratio of data over prediction is shown. From upper to lower, the rows show the monojet, low-purity, and high-purity mono-V categories, while the left (right) column represents the 2017 (2018) data set.
        }
        \label{fig:cr_ratios_z_over_g}

\end{figure*}

The validity of the predictions is checked by considering the differential ratio of yields in the CRs. The yield ratio serves as a proxy for the ratios of the different \Vjets processes, which the fit relies on. The yield ratios between the dilepton and single-lepton CRs, and between the dilepton and photon CRs are shown in Figs.~\ref{fig:cr_ratios_z_over_w} and~\ref{fig:cr_ratios_z_over_g}, respectively. Good agreement is observed between prediction and data. In the monojet categories, it is found that the rate of $\Wlv$ events is initially underpredicted relative to $\Zll$ and $\gamma$ events. This underprediction is corrected in the ML fit, mostly via an adjustment of the nuisance parameters related to the experimental efficiencies for leptons and photons, as well as those related to the noncanceling components of the QCD higher-order corrections.

\subsection{Estimation of the QCD multijet background}
The contributions from QCD multijet events in each SR and the corresponding photon CR are estimated from data. Multijet events do not carry large intrinsic \ptmiss, and therefore could only contribute to the SR if one of the hadronic jets in an event is significantly misreconstructed or partially lost, leading to the \ptvecmiss vector and the transverse momentum vector of the jet being aligned. The contribution from such events is estimated from a CR that is enriched in multijet events by inverting the requirement on \dphijm relative to the SR. The recoil spectrum of multijet events in the SR is obtained by multiplying the spectrum in data in this CR by a transfer factor obtained from simulation. The nonmultijet background components, as predicted from simulation, are subtracted from data before applying the transfer factor. The performance of the method is tested by splitting the low-\dphijm CR into parts across different boundaries in \dphijm (\eg, for a boundary of 0.25, the regions would be $\dphijm<0.25$ and $0.25<\dphijm<0.5$) and verifying that an estimate based on the low-\dphijm part of the region ($\dphijm<0.25$ in the above example) can correctly predict the QCD multijet background contribution in the high-\dphijm part ($0.25<\dphijm<0.5$). The method is found to predict correctly the QCD background contribution to approximately 25\% for various choices of \dphijm boundaries, a value which is assigned as a normalization uncertainty in the QCD multijet background estimate in the SR. Uncertainties related to the finite size of multijet samples, as well as to the choice of the transfer factor binning, are taken into account and may affect the normalization and shape of the background estimate by between 10 and 50\% depending on \ptmiss.

In the photon CR, multijet events can contribute if a jet is misreconstructed as an isolated photon. The fraction of photons resulting from jet misreconstruction is estimated from the distribution of the lateral shower width of the photons. The distribution of this variable shows a characteristic peak for genuine photons, while being significantly more flat for the contribution from jets misreconstructed as photons. A template fit is performed to the distribution in data in order to extract the relative contributions of the two components. Templates for genuine photons are obtained from simulation, while templates for misreconstructed jets are taken from a CR in data with an inverted photon isolation requirement that is enriched in QCD multijet events. The fraction of photons originating from jet misreconstruction is found to range between 3.5\% at $\pt=200\GeV$ and 1\% at $800\GeV$. A prediction for the recoil distribution in QCD multijet events in the photon CR is obtained by weighting the photon candidate spectrum in data by the misreconstructed jet fraction evaluated at the respective \pt of the photon candidates. A 25\% uncertainty is assigned to the normalization of the QCD multijet background to account for mismodeling of the shower width in simulation. The uncertainty is estimated by repeating the measurement while varying the binning of the shower width distribution used for fitting, which serves to modulate the effect of the mismodeling. The statistical uncertainty in the determination of the differential recoil shape is taken into account and ranges from less than 1\% at low recoil values up to 10 (20)\% at a recoil value of 1.4\TeV in the 2017 (2018) data set.

\subsection{Systematic uncertainties}
The inputs to the ML fit are subject to various experimental and theoretical uncertainties. The overall experimental uncertainty is dominated by the uncertainties in the efficiency of identifying and reconstructing lepton and photon candidates, as well as the uncertainty in the trigger efficiency. The uncertainties in the efficiencies of reconstructing and identifying electron candidates are 1.0 and 2.5\%, respectively. For muons, the corresponding uncertainties are 1\%, with an additional 1\% uncertainty in the efficiency of the isolation criteria. Finally, for photons, the uncertainty in the reconstruction efficiency is negligible, and the uncertainty in the identification efficiency ranges between 4\% at $\pt=200\GeV$ and 12\% at 1\TeV. The uncertainties in the identification efficiency of lepton candidates are further propagated to the estimate of the contribution from background processes in the SRs, where events with identified leptons are rejected. These uncertainties predominantly affect the \Wlv process, and their magnitude is taken to be 1--2\% of the total \Wlv yield for the identification of \PGt leptons, 1.5\% for electrons, and less than 0.5\% for muons. The uncertainty in the photon energy calibration modeling is 1\% of the photon momentum, leading to an effect on the background yield in the photon control region of up to 3\% at low recoil values. The uncertainty in the \PQb tagging efficiency leads to an uncertainty of 6\% in the normalization of background processes with top quarks, and 2\% in the normalization of the diboson and QCD multijet processes. The uncertainties in the trigger efficiency are 2\% for both the electron or photon triggers, and 1\% per identified muon for the \ptmiss trigger for recoil values of less than 400\GeV, and negligible above this threshold. The muon multiplicity dependence of the \ptmiss trigger uncertainty reflects the differences in the reconstruction of muons at the trigger and offline levels, which affect the calculation of the hadronic recoil value. Uncertainties of 75\% are assigned to the normalization of the QCD multijet background contributions in the single-lepton regions, which are estimated from LO simulation. Finally, additional uncertainties of 20\% each are assigned to the rate of the Drell--Yan events entering the single-lepton CRs and of the \phojets events entering the single-electron CRs.

The theoretical uncertainties in the transfer factors related to higher-order effects in the QCD and EW perturbative expansions are calculated according to the prescription given in Ref.~\cite{Lindert:2017olm} and implemented, as described in Ref.~\cite{Sirunyan:2017jix}. The uncertainty related to the modeling of PDFs is estimated using the replicas provided in the PDF4LHC15 PDF set~\cite{Butterworth:2015oua,Dulat:2015mca,Harland-Lang:2014zoa,Ball:2014uwa}. Additionally, uncertainties of 10\% each are assigned to the cross sections of the diboson and top quark processes, and a further 10\% normalization uncertainty is assigned to account for the differences in the \pt spectrum of simulated and observed top quark events~\cite{Czakon:2015owf}. For the diboson and $\Vgamma$ processes, additional uncertainties related to unknown mixed QCD-EW NLO corrections are estimated based on the product of the individual EW and QCD correction terms. These uncertainties range between 1 and 10\%, depending on the process and boson \pt.

The likelihood functions obtained for the monojet and mono-\PV categories, as well as for the two data taking years, are combined in order to maximize the statistical power of the analysis. The results based on the data set analyzed here, which corresponds to an integrated luminosity of 101\fbinv, are further combined with the results of an earlier analysis~\cite{Sirunyan:2017jix} based on a data set collected at the same center-of-mass energy in 2016 and corresponding to an integrated luminosity of 36\fbinv. The combination is performed by defining a combined likelihood describing all the analysis regions in all data sets. For this purpose, the effects of all theoretical uncertainties are assumed to be correlated. Most experimental uncertainties are dominated by the inherent precision of auxiliary measurements specific to each data set and are thus assumed to be uncorrelated between different data taking years. The experimental uncertainties related to the JES and JER, as well as those related to the determination of the integrated luminosity are partially correlated between the data taking years, which is taken into account by splitting the total uncertainty into its correlated and uncorrelated components. In order to harmonize the theoretical signal treatment between the data sets, the signal templates from Ref.~\cite{Sirunyan:2017jix} are replaced by the templates derived from simulated samples with generator configurations identical to those used in the analysis of the more recent data sets. Use of the more accurate generator worsens the excluded cross sections based on the 2016 data set alone by up to 13\%, depending on the signal hypothesis. The effect is reduced to a few percent level in the fully combined final result.

\section{Results and interpretation}

\begin{figure*}[hbtp!]
    \centering
        \includegraphics[width=0.40\textwidth]{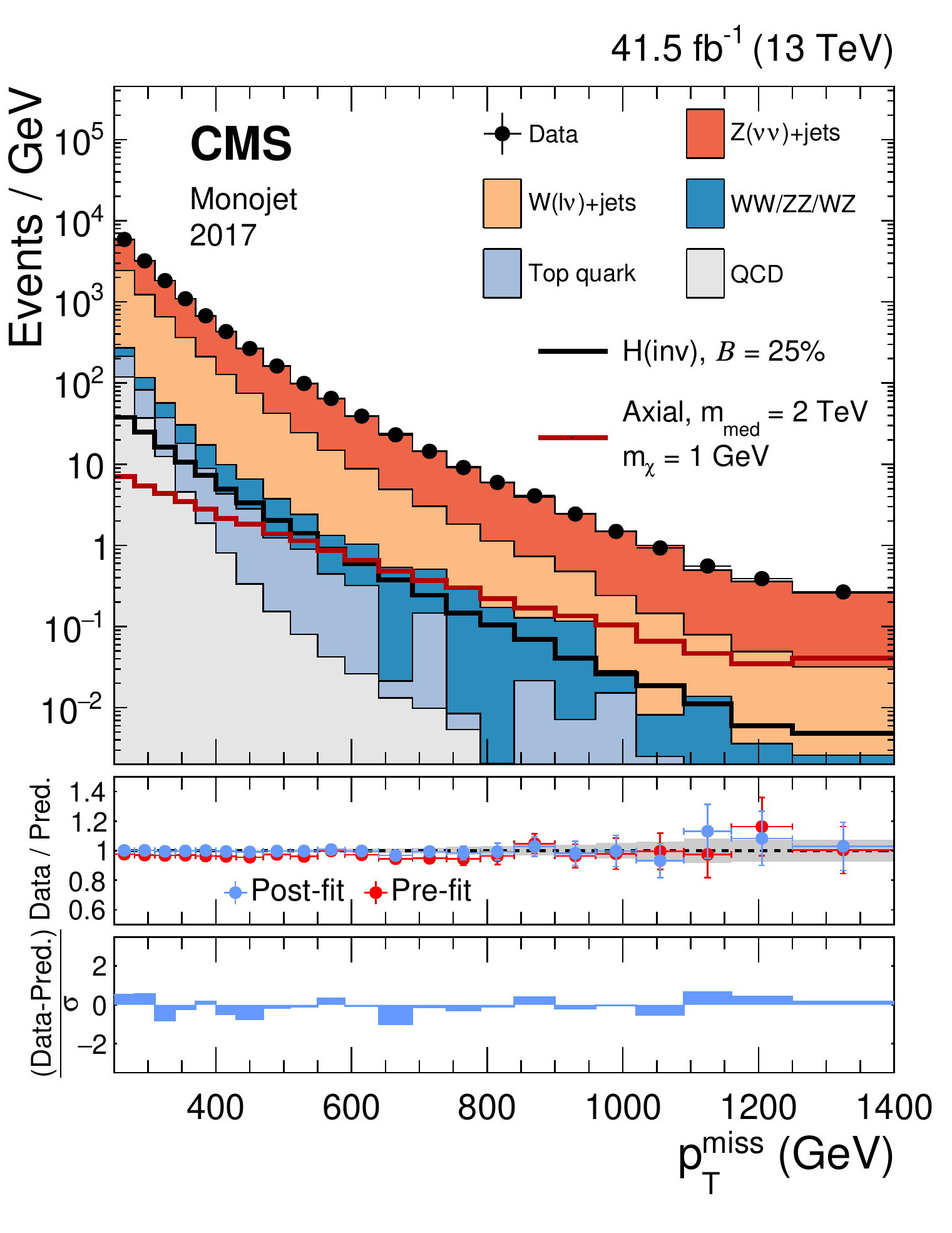}
        \includegraphics[width=0.40\textwidth]{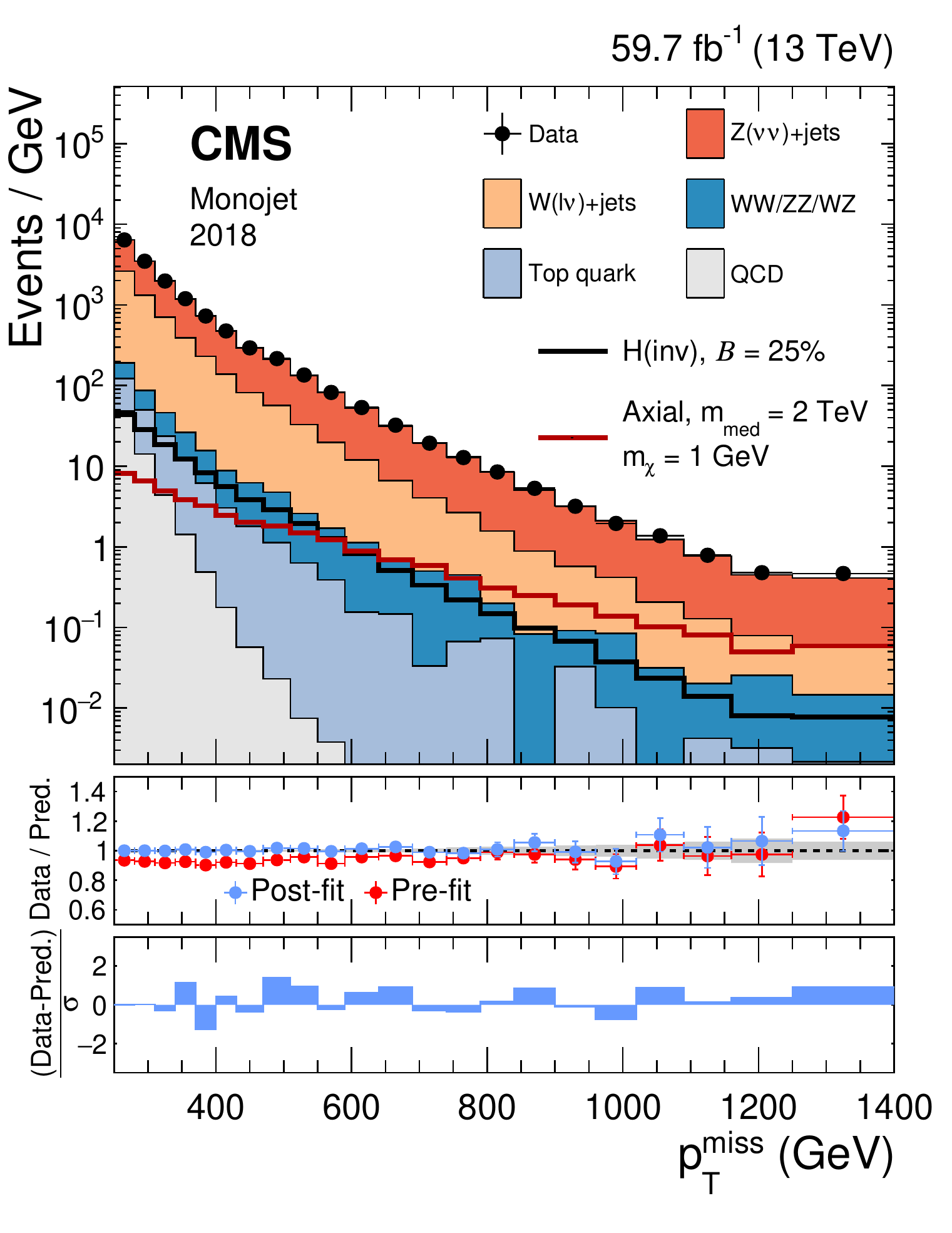}
        \caption{
            Comparison between data and the background prediction in the monojet signal region before and after the simultaneous fit.
            The fit includes all control regions and the signal region in all categories and both data taking years, and the background-only fit model is used.
            The resulting distributions are shown separately for 2017 (left) and 2018 (right).
            Templates for two signal hypotheses are shown overlaid as black and dark red solid lines.
            The last bin includes the overflow.
            In the middle panels, ratios of data to the pre-fit background
            prediction (red solid points) and post-fit background
            prediction (blue solid points) are shown.
            The gray band in the middle panels indicates the post-fit uncertainty
            after combining all the systematic uncertainties. Finally, the distribution of the pulls, defined as the
            difference between data and the post-fit background prediction divided by the quadratic sum of the
            post-fit uncertainty in the prediction and statistical uncertainty in data, is shown in the lower panels.
        }
        \label{fig:postfit_sr_j}

\end{figure*}

\begin{figure*}[hbtp!]
    \centering
      \includegraphics[width=0.40\textwidth]{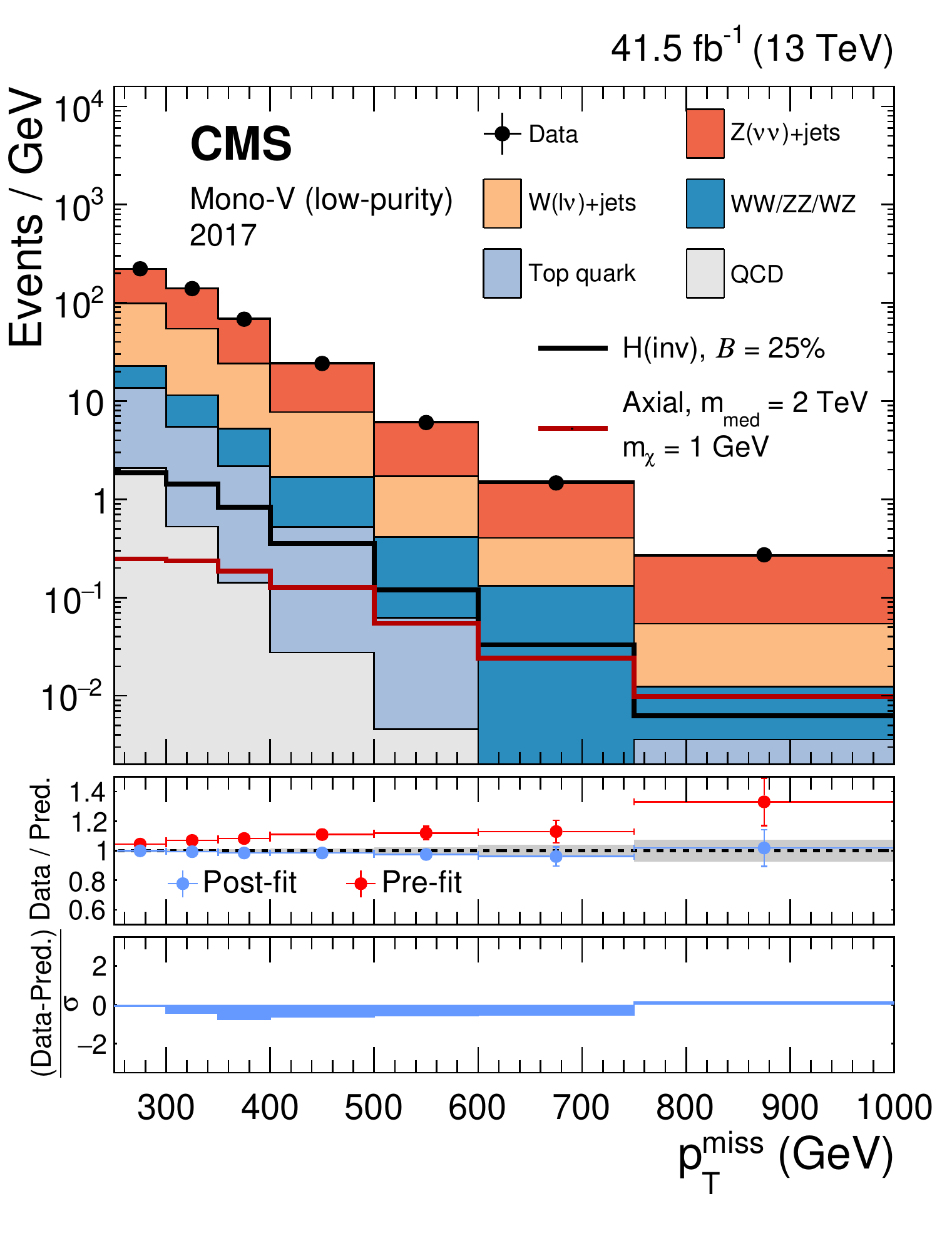}
        \includegraphics[width=0.40\textwidth]{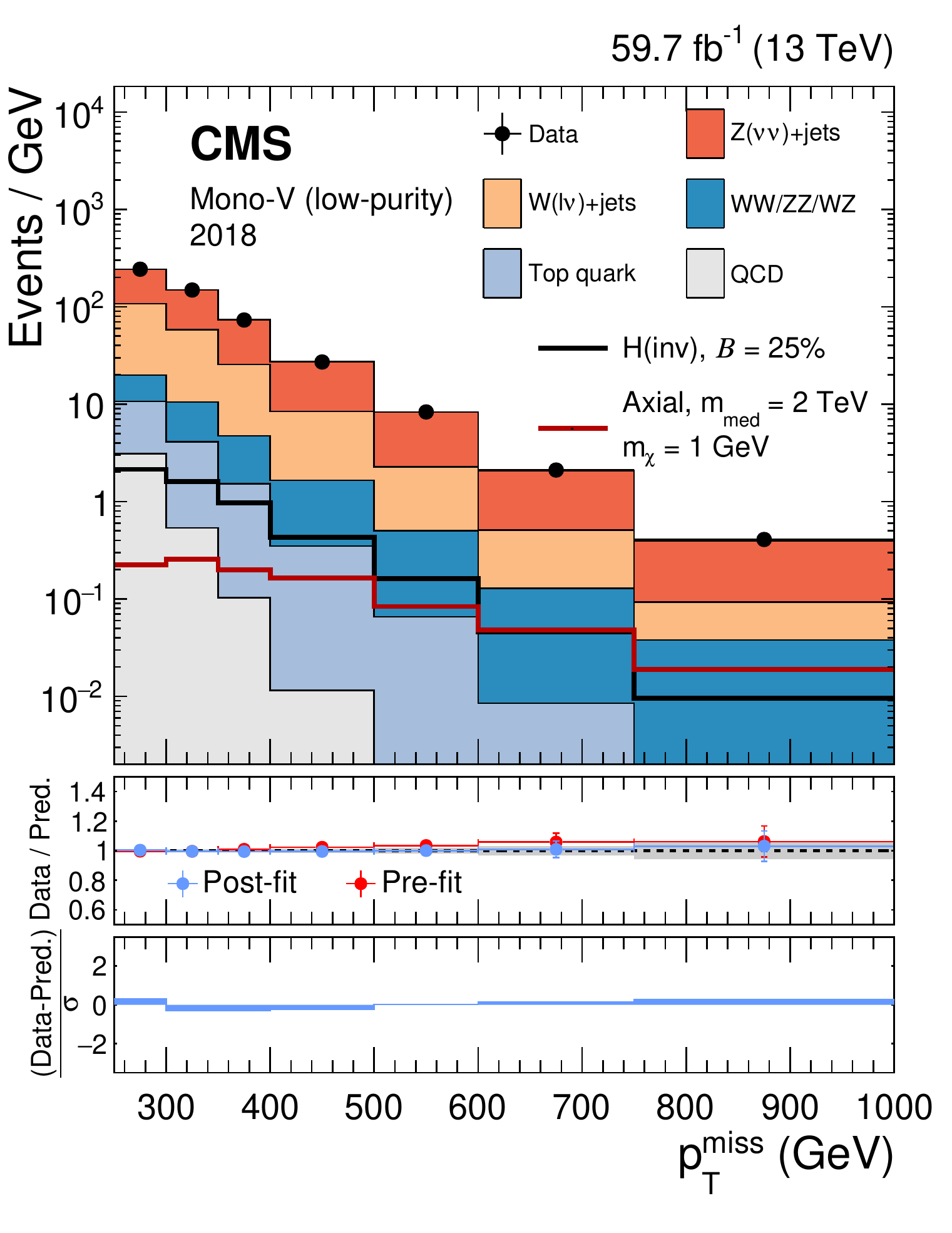}
        \includegraphics[width=0.40\textwidth]{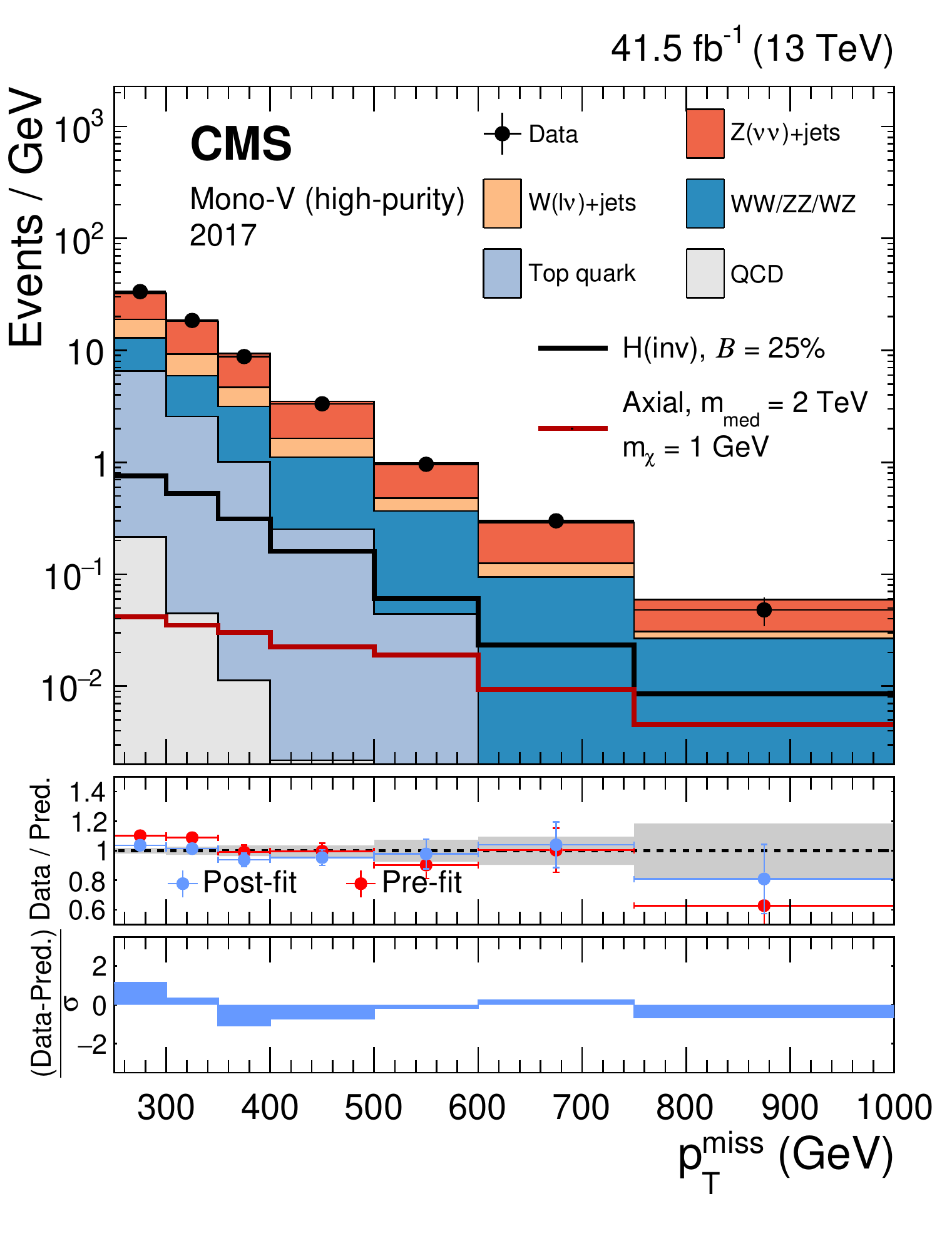}
        \includegraphics[width=0.40\textwidth]{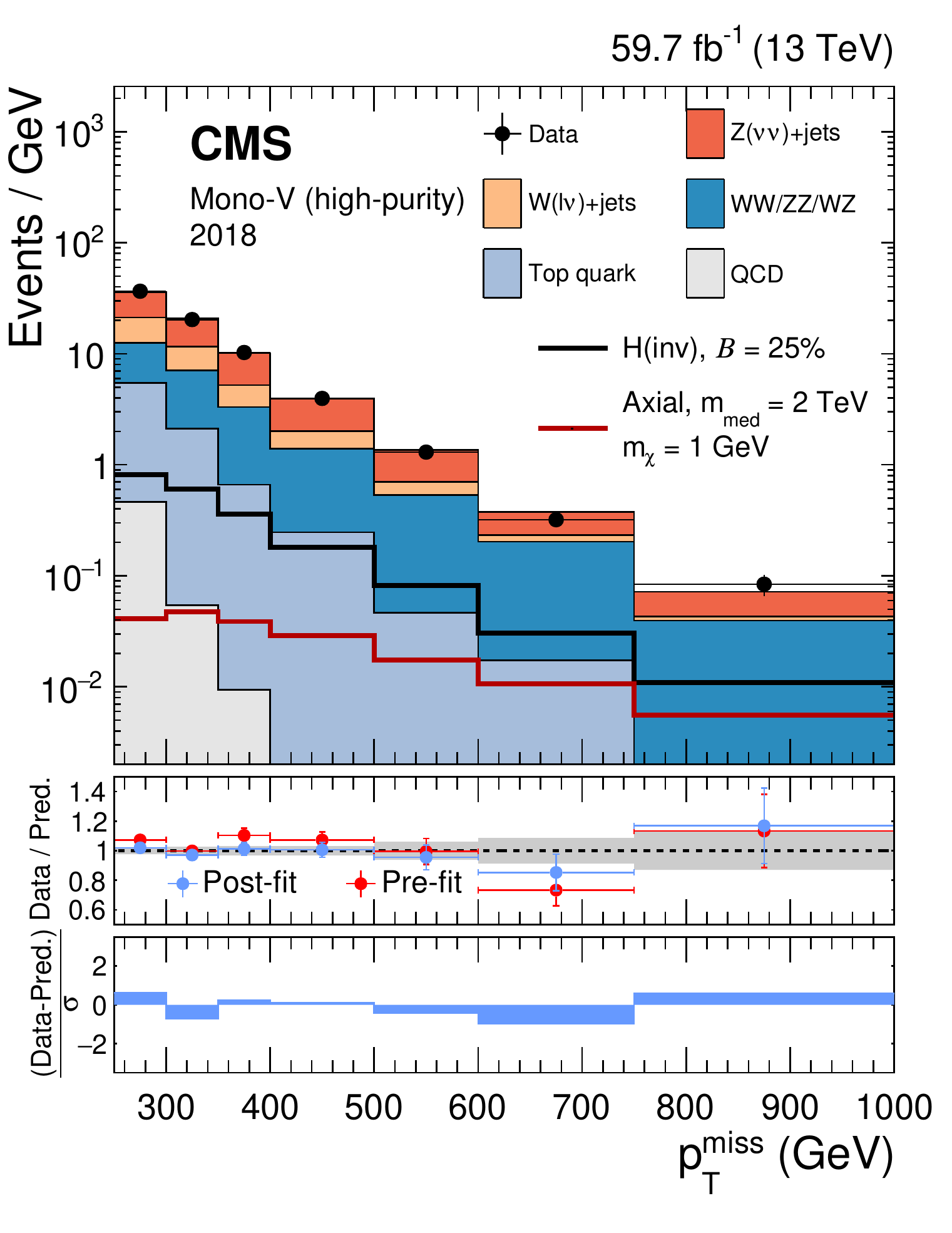}
        \caption{
            Comparison between data and the background prediction in the mono-V signal regions before and after the simultaneous fit.
            The fit includes all control regions and the signal region in all categories and both data taking years, and the background-only fit model is used.
            The resulting distributions are shown separately for 2017 (left column) and 2018 (right column), as well as for the low- and high-purity categories (upper and lower rows, respectively). Templates for two signal hypothesis are shown overlaid as black and dark red solid lines.
            The last bin includes the overflow.
            In the middle panels, ratios of data to the pre-fit background
            prediction (red solid points) and post-fit background
            prediction (blue solid points) are shown.
            The gray band in the middle panels indicates the post-fit uncertainty
            after combining all the systematic uncertainties. Finally, the distribution of the pulls, defined as the
            difference between data and the post-fit background prediction divided by the quadratic sum of the
            post-fit uncertainty in the prediction and statistical uncertainty in data, is shown in the lower panels.
        }
        \label{fig:postfit_sr_v}

\end{figure*}

The ML fit is performed by combining the analysis categories as well as the 2017 and 2018 data sets. The \ptmiss distributions in the SRs before (``pre-fit'') and after (``post-fit'') the fit are shown in Fig.~\ref{fig:postfit_sr_j} for the monojet category and in Fig.~\ref{fig:postfit_sr_v} for the low-purity and high-purity mono-V categories. In all cases, good agreement is observed between the background-only post-fit result and the data. The corresponding distributions for the CRs are shown in Figs.~\ref{fig:postfit_zll_photon_monojet}--\ref{fig:postfit_cr_wln_monovtight} in Appendix~\ref{app:appendix_recoil_cr}.

In the following, signal strength exclusion limits are presented for different signal hypotheses. Unless explicitly stated, all data sets and categories are included. The exclusion limits are calculated using the asymptotic approximation of the \CLs\ method~\cite{CLS1,CLS2,Cowan:2010js}. In this method, a signal-plus-background fit is performed for each signal hypothesis in addition to the background-only fit. In the signal fits, the nuisance parameters are profiled, and the resulting best fit nuisance parameters vary for the different signal hypotheses. Consequently, different nonzero best fit values for the signal strength can be obtained for different signals even if the background-only fit succeeds in modeling the data. In the exclusion limits, this feature is represented by differences between the observed and expected limits.


\subsection{Higgs portal interpretation}

\begin{figure*}[hbt]
    \centering
        \includegraphics[width=0.6\textwidth]{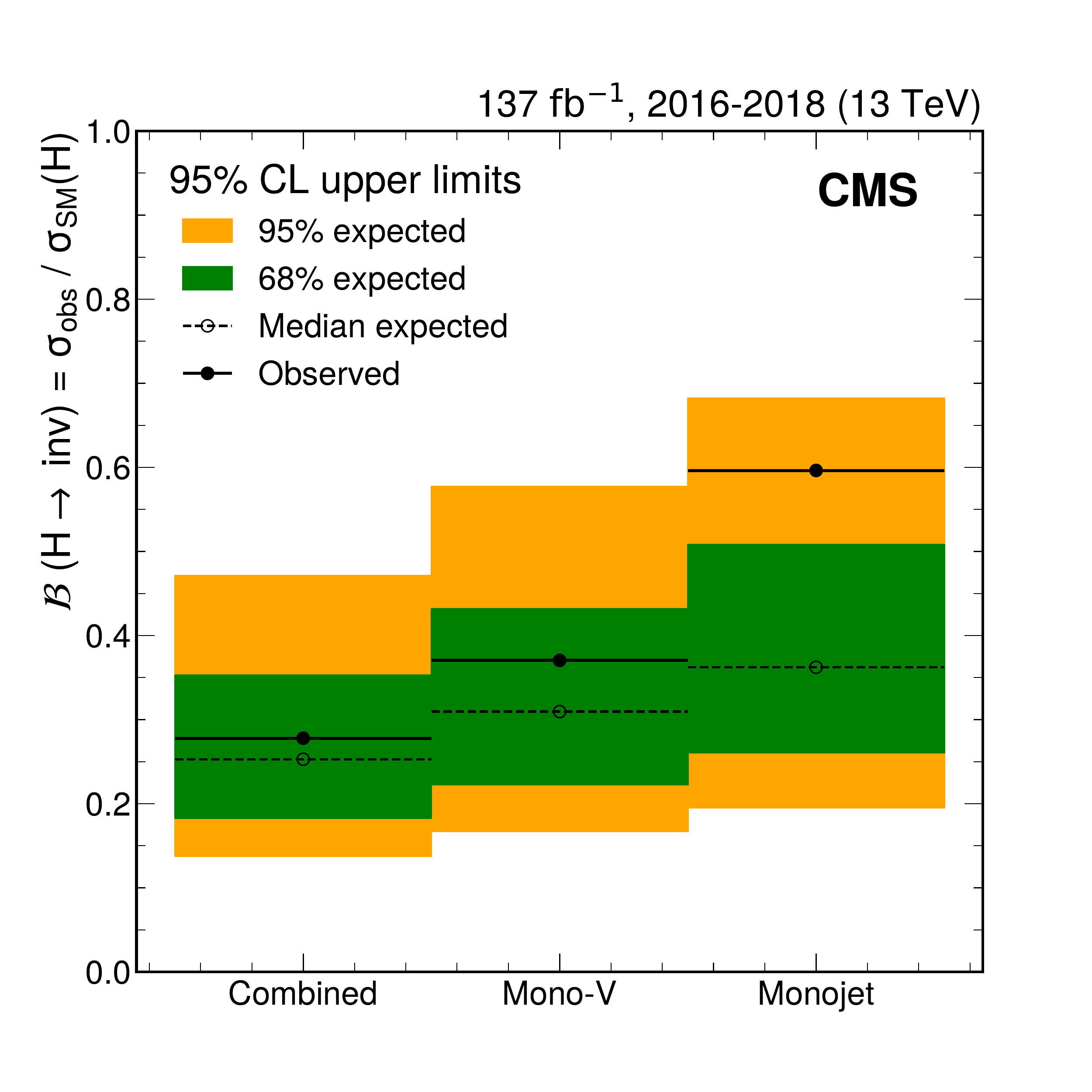}
        \caption{
                Upper limits at 95\% \CL on the branching fraction $\mathcal{B}$ of the Higgs boson to invisible final states.
                The results are shown separately for the monojet and mono-V categories, as well as for their combination.
                The final combined limit is $27.8\%$ ($25.3\%$ expected).
                }
        \label{fig:results_hinv}

\end{figure*}

The results are interpreted in terms of the exclusion limits at 95\% confidence level (\CL) on the branching fraction of an otherwise SM-like Higgs boson to particles without detectable detector interactions (invisible decays). The limits are derived assuming the SM production cross section for the Higgs boson~\cite{deFlorian:2016spz}. In the monojet category, values of $\mathcal{B}(\PH \to \text{inv.})$ larger than $59.6\%$ are excluded ($36.2\%$ expected). In the combination of the mono-V categories, branching fractions of more than $37.0\%$ are excluded ($31.0\%$ expected). Finally, the combination of all categories yields an exclusion limit of $\mathcal{B}(\PH \to \text{inv.}) < 27.8\%$ ($25.3\%$ expected).
These limits are summarized in Fig.~\ref{fig:results_hinv}. The result from the combination of the mono-V and monojet channels exhibits a closer agreement between the expected and observed exclusions than either of the two channels individually. This is a result of correlations in the background model between the categories. A year-by-year breakdown of the sensitivity is shown in Fig.~\ref{fig:supp_hinv} in Appendix~\ref{app:appendix_hinv_split}.
Compared to the previous result in the same channel from Ref.~\cite{Sirunyan:2017jix}, which is included here, the exclusion limit is improved by a factor of 1.9 (1.6 expected), and represents the most stringent limit from the combined gluon-fusion and $\PV(\PQq\PQq)\PH$ channels to date. The current best limit is 19\% from Ref.~\cite{CMS-Hinv}, in which multiple analyses based on data sets of up to 36\fbinv are combined, including Ref.~\cite{Sirunyan:2017jix}.

\subsection{Interpretation in a DM simplified model with a colorless mediator}

\begin{figure*}[hbt!]
    \centering
        \includegraphics[width=0.6\textwidth]{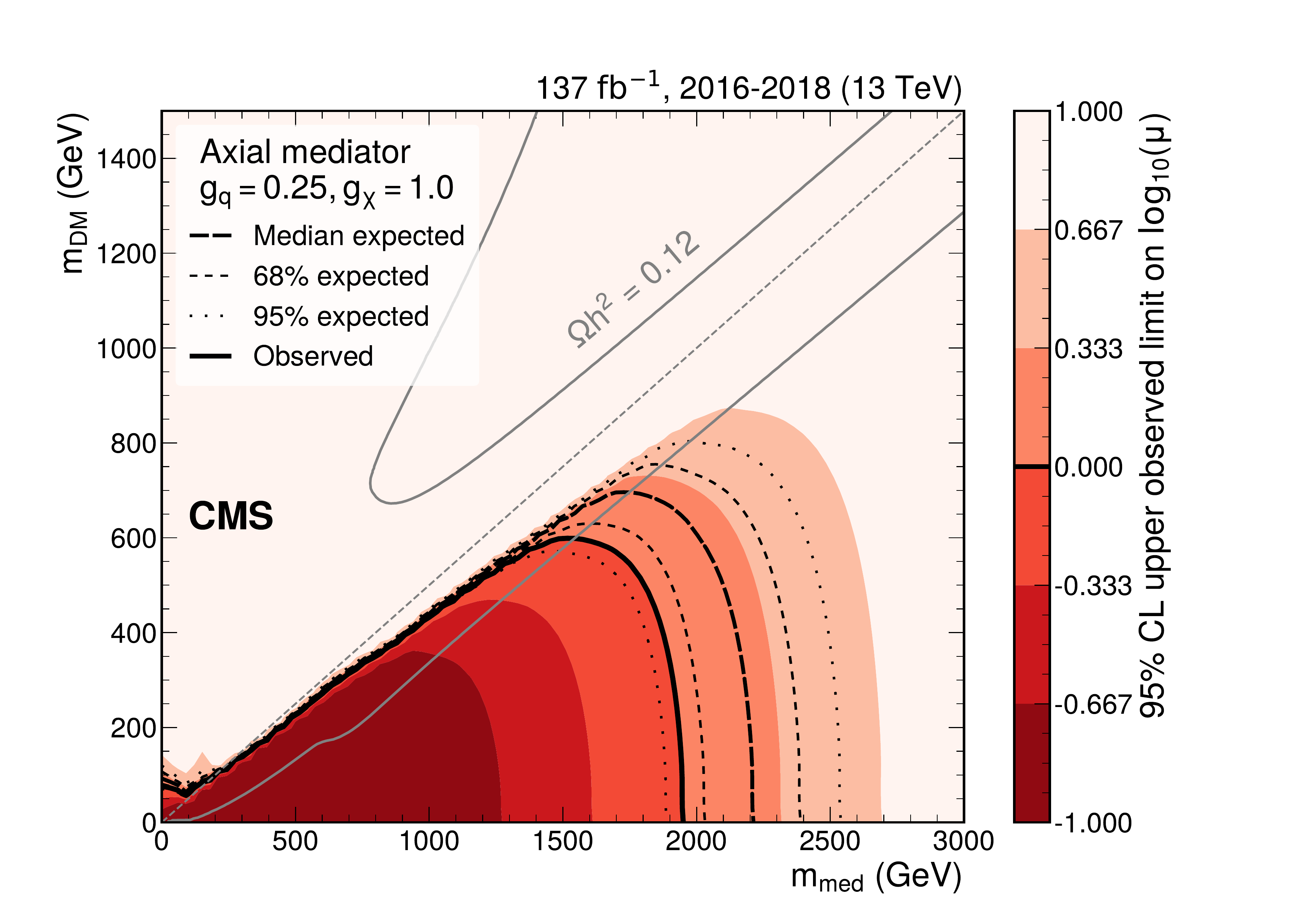}
        \includegraphics[width=0.6\textwidth]{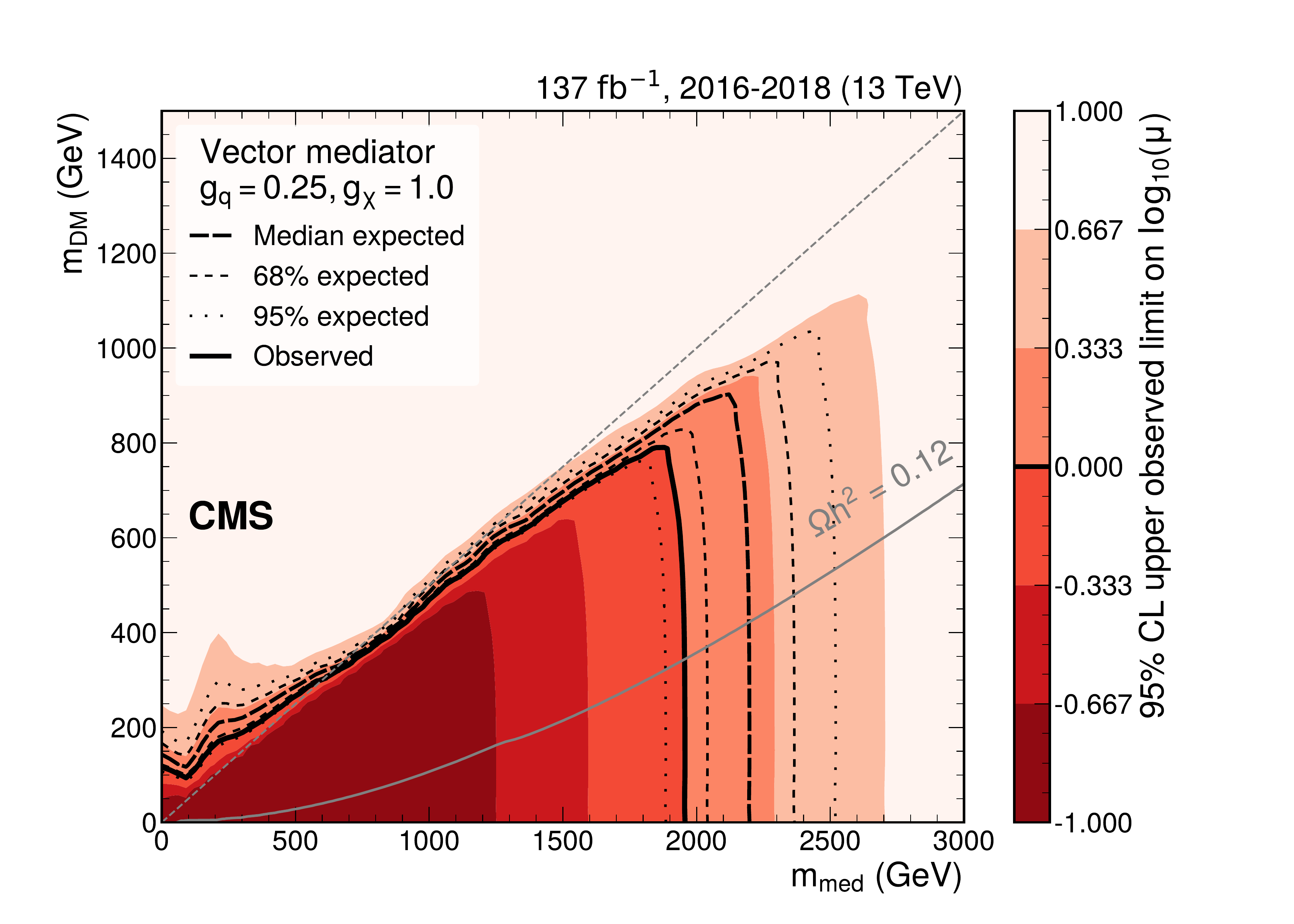}
        \\
        \caption{Exclusion limits at 95\% \CL on the signal strength $\mu=\sigma/\sigma_\text{theo}$ in the \mmed-\mdm plane for coupling values of $\gq=0.25$ and $\gchi=1.0$ for an axial-vector (upper) or vector (lower) mediator. The black solid line indicates the observed exclusion boundary $\mu=1$. The black dashed and dotted lines represent the expected exclusion and the 68 and 95\% \CL intervals around the expected boundary, respectively. Parameter combinations with larger values of $\mu$ (indicated by a darker shade in the color scale) are excluded. The observed exclusion reaches up to $\mmed=1.95\TeV$ (2.2\TeV expected) for low values of $\mdm=1\GeV$ . The gray dashed line indicates the diagonal $\mmed=2\mdm$, above which only off-shell mediator production contributes to the jet+\ptmiss final state. The steep increase of the signal strength limit above the diagonal leads to fluctuations of the exclusion contour, which are due to finite precision in the interpolation method in this region. The gray solid lines represent parameter combinations for which the simplified model reproduces the observed DM relic density in the universe under the assumption of a thermal freeze-out mechanism~\cite{Albert:2017onk,Aghanim:2018eyx}.}
        \label{fig:results_dmsimp_spin1_2d_mass}

\end{figure*}

\begin{figure*}[hbtp]
    \centering
        \includegraphics[width=0.49\textwidth]{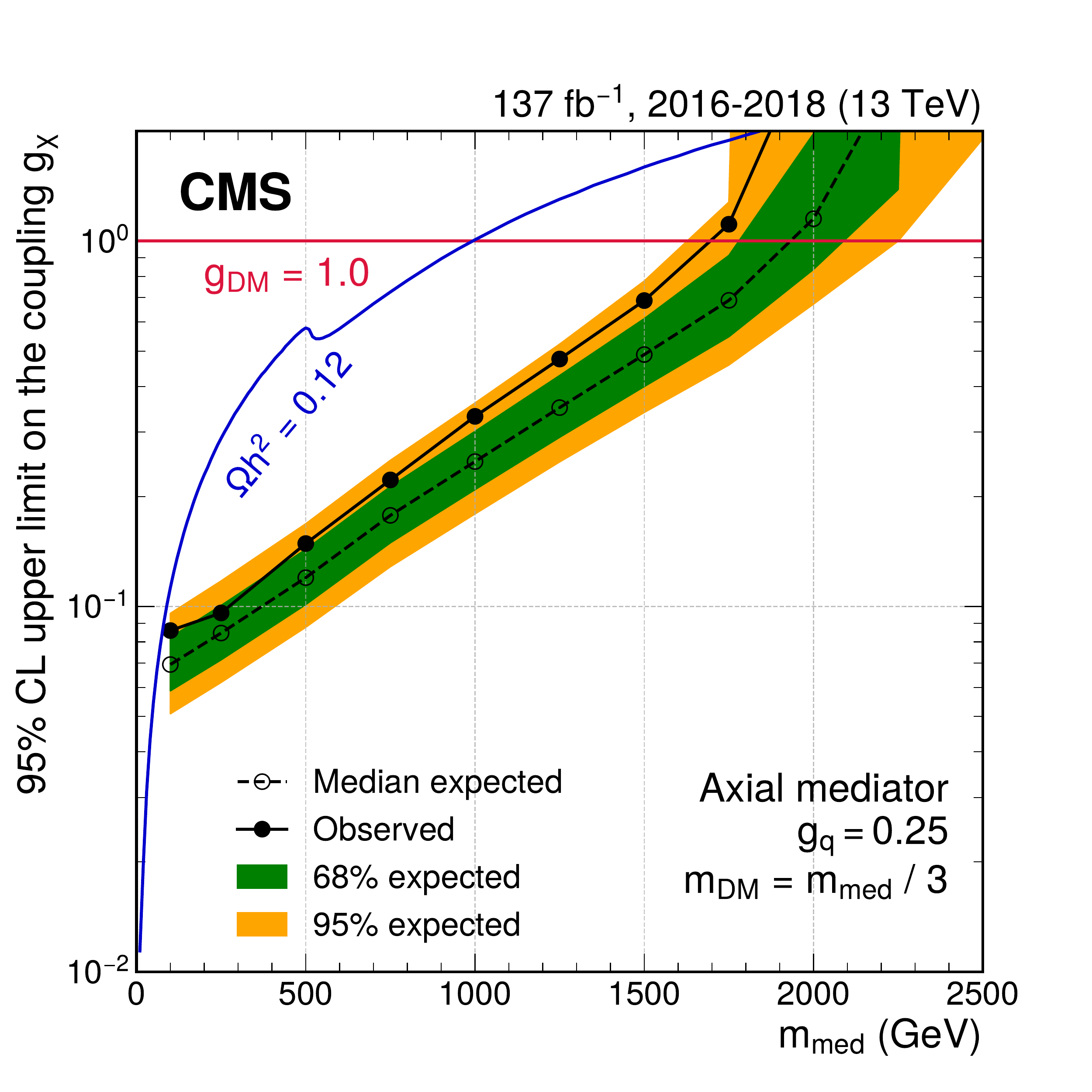}
        \includegraphics[width=0.49\textwidth]{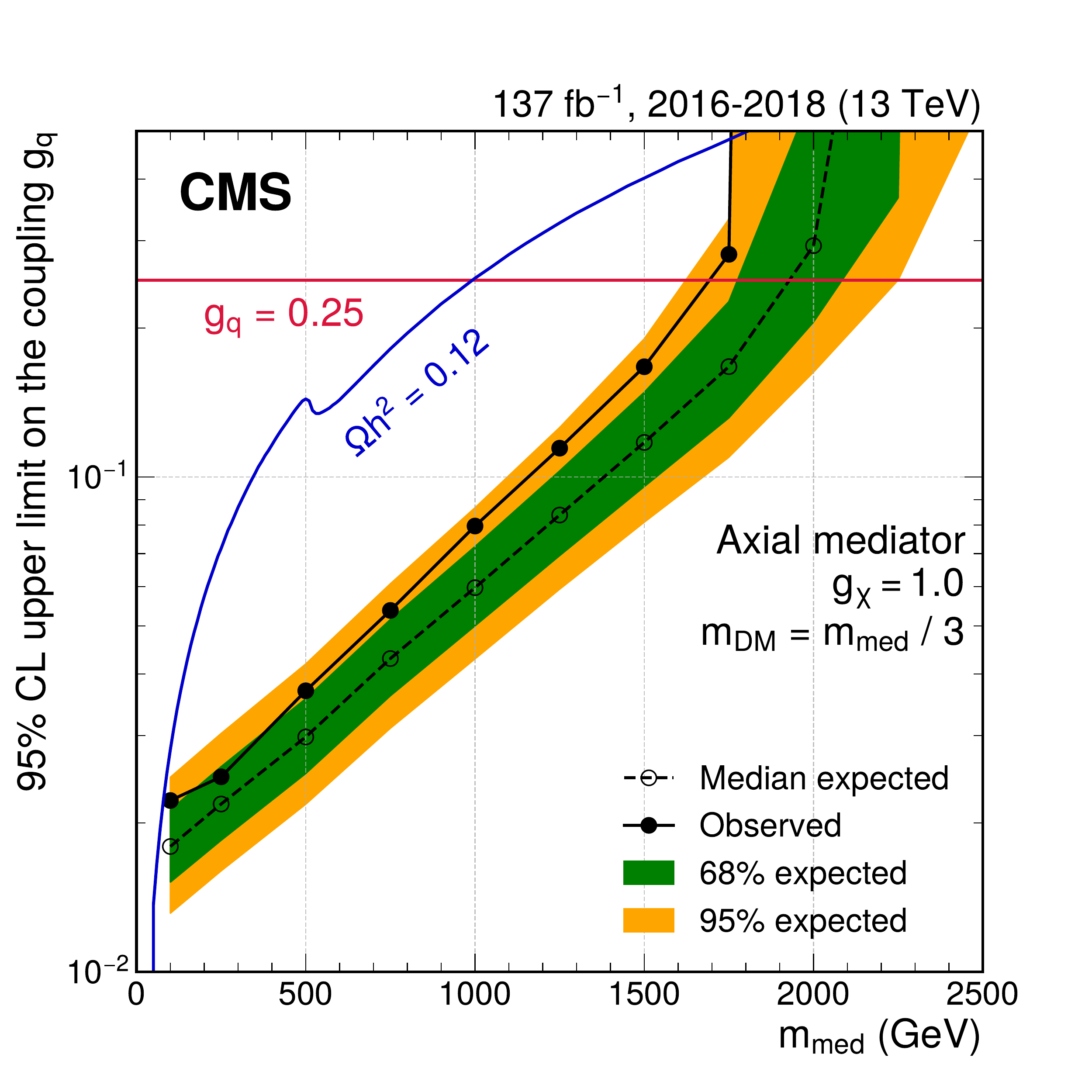}
        \\
        \caption{
            Exclusion limits at 95\% \CL on the couplings \gchi (left) and \gq (right) for an axial-vector mediator. In each panel, the result is shown as a function of the mediator mass $\mmed$, with the mass of the DM candidate fixed to $\mdm=\mmed/3$. In either case, only one coupling is varied, while the other coupling is fixed at its default value ($\gq=0.25$ or $\gchi=1.0$). The blue solid line indicates the parameter combinations for which the simplified model reproduces the observed DM relic density. Around $\mdm\approx m_{\text{top}}$, corresponding to $\mmed\approx500\GeV$, DM annihilation into top quarks becomes possible, leading to a shift in the relic density. The corresponding results for a vector mediator are shown in Fig.~\ref{fig:supp_dmsimp_spin1_vector} in Appendix~\ref{app:appendix_dmsimp_vector_coupling}.
        }
        \label{fig:results_dmsimp_spin1_coupling}

\end{figure*}

The results are further interpreted in terms of simplified models of DM production. In a model with a spin-1 mediator, exclusion limits are calculated in the two-dimensional parameter space of the DM and mediator particle masses, \mdm and \mmed. The coupling between the mediator and the SM quarks is set to a constant value of $\gq=0.25$, the mediator-DM coupling is set to $\gchi=1.0$, and vector and axial-vector type couplings are considered in separate interpretations. The resulting exclusion limits at 95\% \CL on the signal strength $\mu$ are shown in Fig.~\ref{fig:results_dmsimp_spin1_2d_mass}. Values of \mmed up to 1.95\TeV (2.2\TeV expected) are excluded for low \mdm values. The maximum excluded values of \mmed decrease with increasing \mdm, as the branching fraction of the mediator to DM particle decays diminishes.
The dependence of the branching fraction on \mdm is more pronounced in the case of an axial-vector mediator, leading to a reduced maximal exclusion reach in \mdm of 0.7\TeV, as opposed to 1\TeV for the vector case. Compared to the results of Ref.~\cite{Sirunyan:2017jix}, the combined limits improve the maximal exclusion in terms of the mediator mass by approximately 400\GeV, or 20\%. In addition to the constraints in the \mdm-\mmed plane, we also obtain exclusion limits in the planes of \mmed and \gq, as well as \mmed and \gchi, which are shown in Fig.~\ref{fig:results_dmsimp_spin1_coupling} for the case of axial-vector couplings. The coupling value exclusion is derived analytically from the signal strength exclusion at the default coupling values by rescaling the signal cross section according to the production cross section and decay branching fractions of the mediator, using the formalism of Ref.~\cite{Albert:2017onk}. The DM candidate mass \mdm is fixed to $\mmed/3$. For low mediator masses, values of \gq (\gchi) as low as 0.018 (0.070) are excluded, providing significant additional insight into the probed parameter space, compared to the mass exclusion for fixed coupling values. Below $\mmed\approx 750\GeV$, the constraints on \gq are the strongest to date and exceed the sensitivity of searches for mediators decaying to quarks~\cite{Aaboud:2018fzt,Sirunyan:2019vxa}. The coupling exclusion result for the vector mediator is similar to the axial-vector case, and is shown in Fig.~\ref{fig:supp_dmsimp_spin1_vector} in Appendix~\ref{app:appendix_dmsimp_vector_coupling}.

\begin{figure*}[hbtp]
    \centering
        \includegraphics[width=0.49\textwidth]{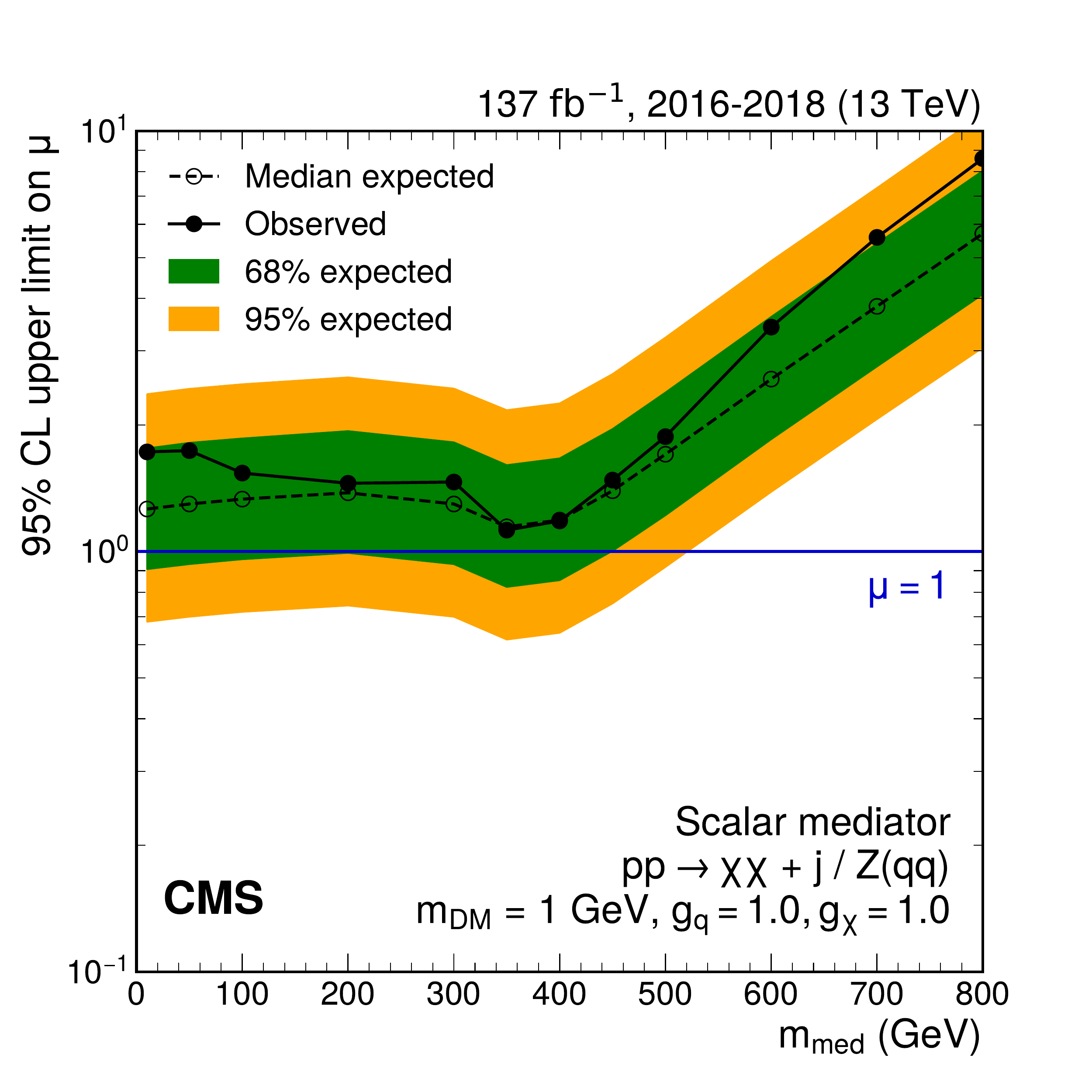}
        \includegraphics[width=0.49\textwidth]{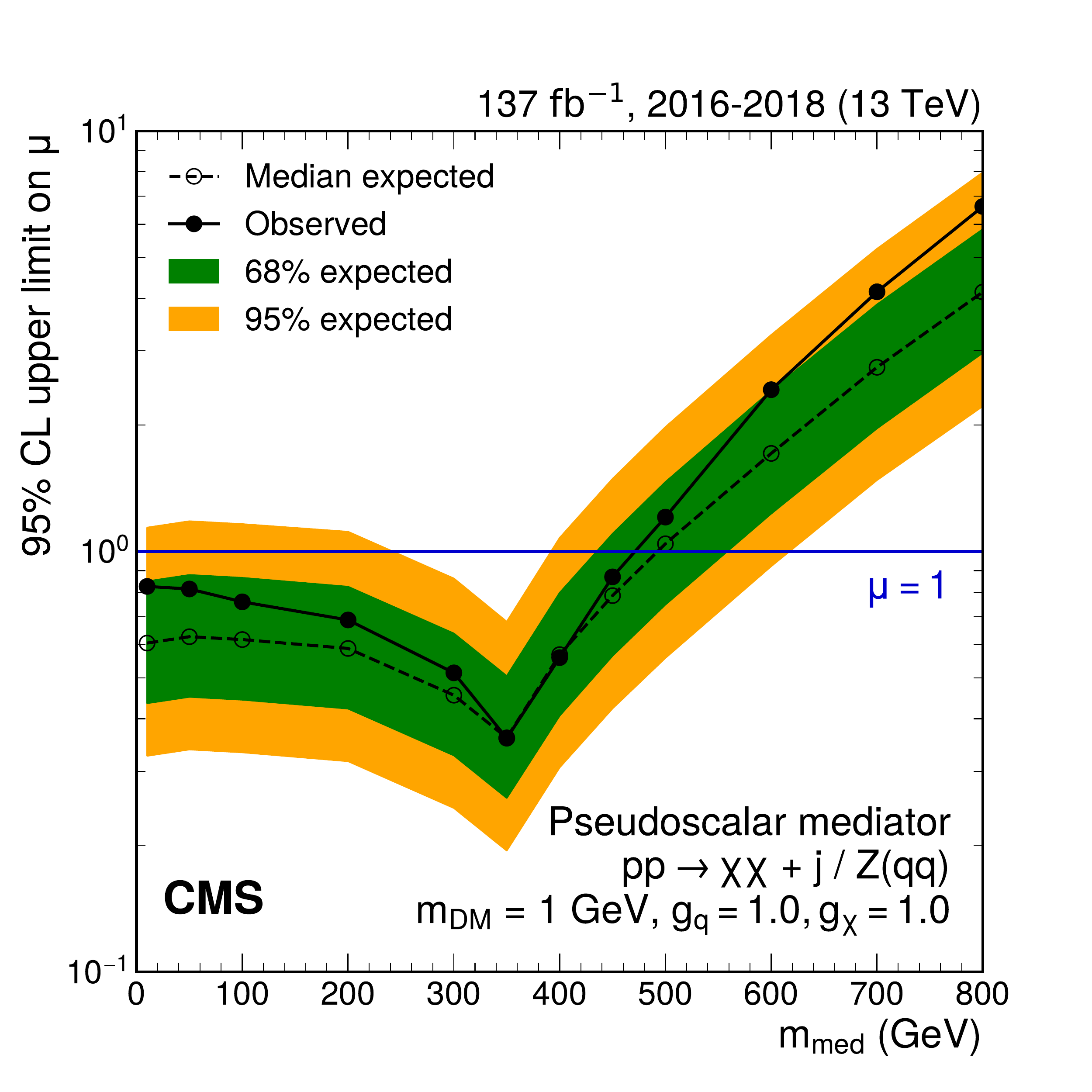}
        \\
        \caption{Upper limits at 95\% \CL on the signal strength $\mu=\sigma/\sigma_\text{theo}$ as a function of \mmed for scenarios with scalar (left) and pseudoscalar (right) mediators and coupling values of $\gq=1.0$, $\gchi=1.0$, for a constant value of $\mdm=1\GeV$. The blue solid line indicates the exclusion boundary $\mu=1$. In the case of a pseudoscalar mediator, \mmed values up to 470\GeV are excluded (490\GeV expected).}
        \label{fig:results_dmsimp_spin0_1d_mass}

\end{figure*}

The expected upper limits on the signal strength in the case of spin-0 mediators are shown in Fig.~\ref{fig:results_dmsimp_spin0_1d_mass}. The mediator couplings are assumed to be $\gq=1.0$ and $\gchi=1.0$, and the DM candidate mass is fixed to 1\GeV. For scalar mediators, signal strengths larger than 1.2 can be excluded at low mediator mass values of $\approx$50\GeV. A pseudoscalar mediator with a mass below $\mmed=470\GeV$ is excluded (490\GeV expected). In both cases, the signal strength limits show distinctive features around the top quark decay threshold of $\mmed=2m_\PQt$. As the mediator is produced via a top quark loop, the signal cross section is enhanced as the mediator mass approaches the threshold from below. Above the threshold, the decay of the mediator into a pair of top quarks becomes possible, leading to a significant suppression of the branching fraction to DM candidates, and therefore the effective signal cross section. A two-dimensional visualization of the pseudoscalar result in the \mmed-\mdm plane is shown in Fig.~\ref{fig:supp_pseudoscalar_2d} in Appendix~\ref{app:appendix_dmsimp_pseudo_2d}. The constraints on the pseudoscalar model presented here are the most stringent to date.

\clearpage

\subsection{Fermion portal interpretation}

\begin{figure*}[hbtp]
    \centering
        \includegraphics[width=0.6\textwidth]{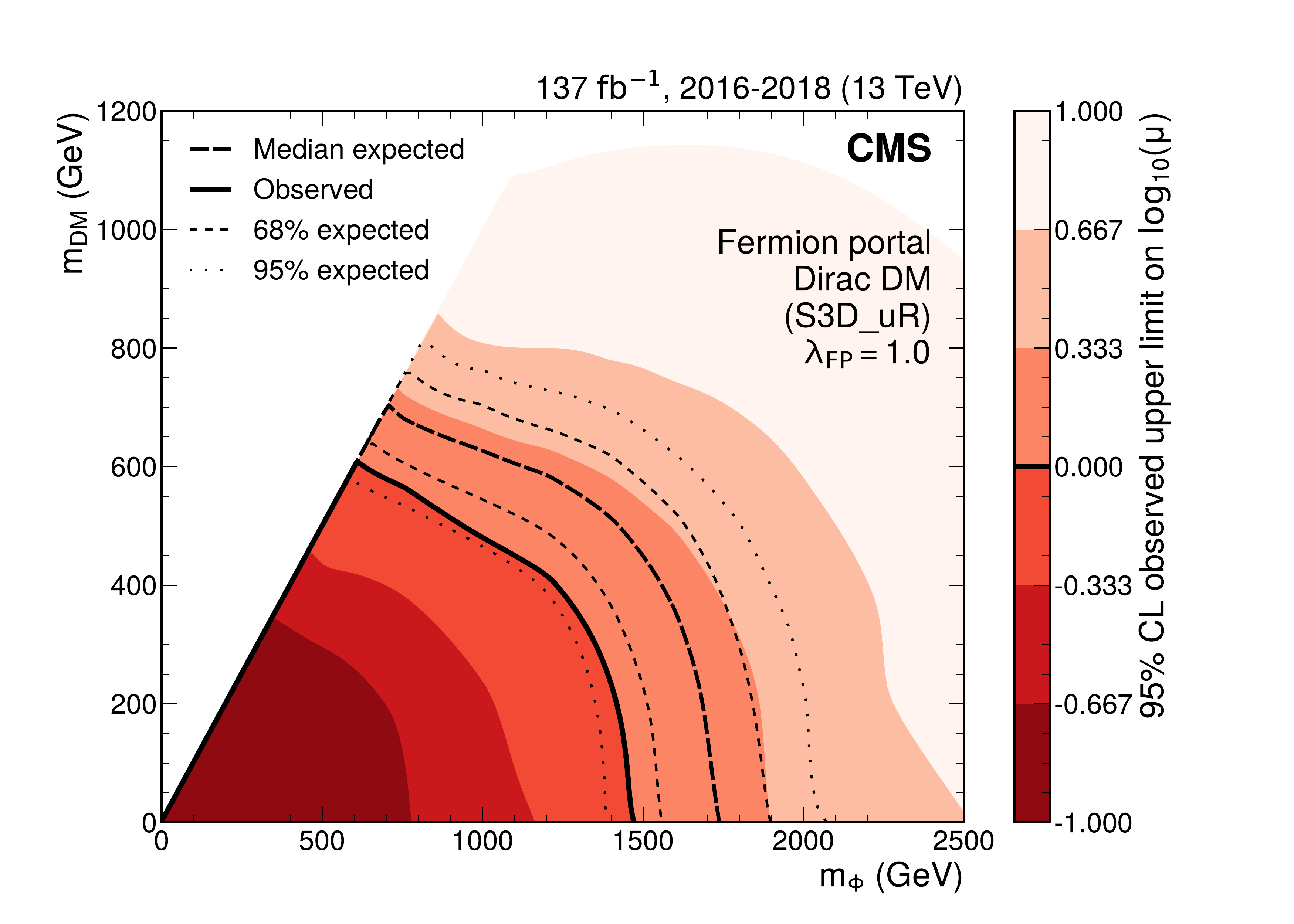}
        \\
        \caption{
            Exclusion limits at 95\% \CL in the plane of the mediator mass \mphi and the DM candidate mass \mdm in the fermion portal model. The black solid line indicates the observed exclusion boundary. The black dashed and dotted lines represent the expected exclusion and the 68 and 95\% \CL intervals around the expected boundary, respectively.
        }
        \label{fig:results_tchan}

\end{figure*}

For the fermion portal model, the results of the analysis are shown in Fig.~\ref{fig:results_tchan} in the plane of the mediator mass $\mphi$ and the DM candidate mass \mdm. The coupling between the mediator, DM candidate and the right-handed up quark is set to a constant value of $\lambdafp=1$. At low \mdm values, mediator masses of up to 1.5\TeV \ are excluded (1.7\TeV expected), which are the most stringent constraints on this model to date.

\subsection{The ADD interpretation}

\begin{figure*}[hbtp]
    \centering
        \includegraphics[width=0.6\textwidth]{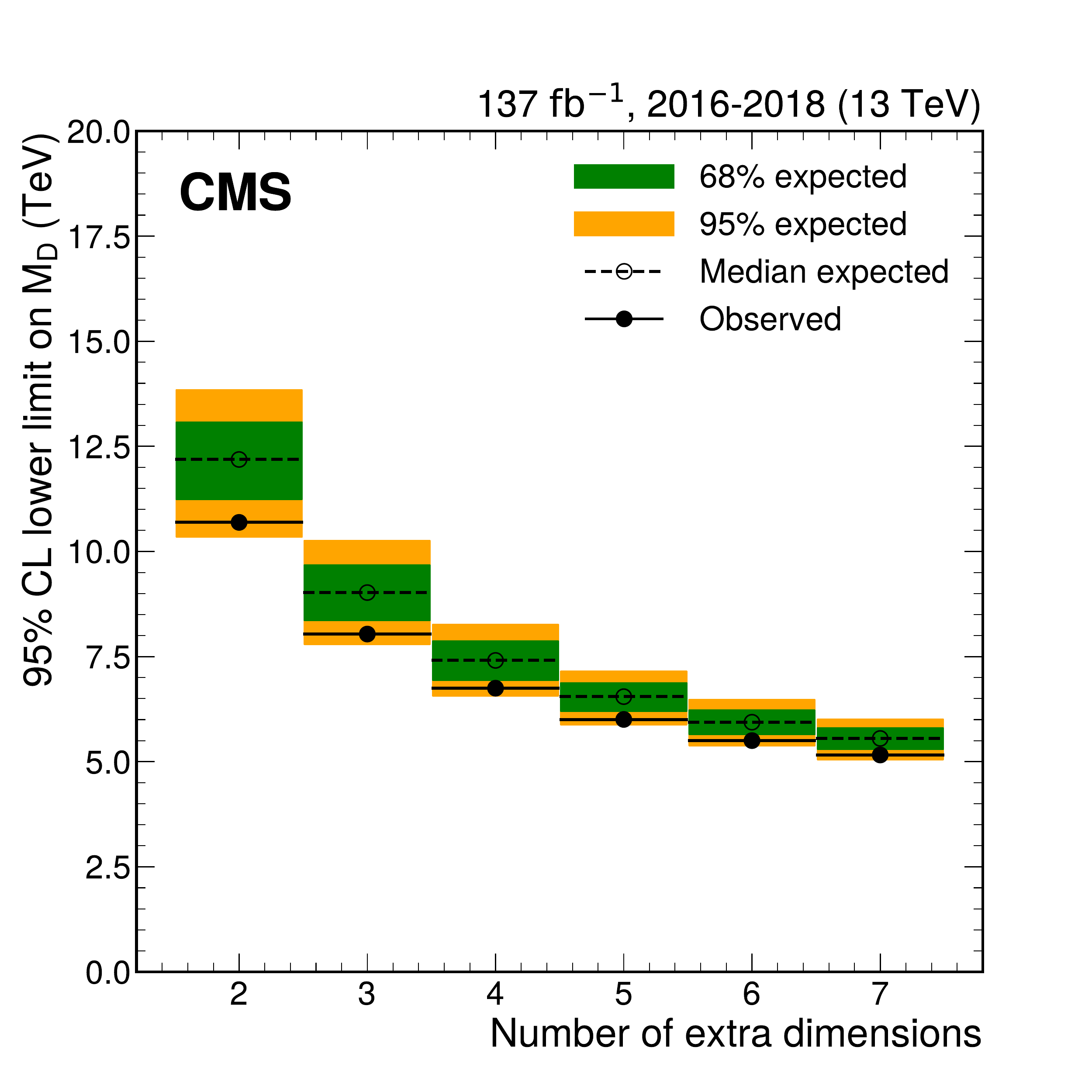}
        \\
        \caption{
            Exclusion limits at 95\% \CL on \MD in the ADD scenario for different values of the number of extra dimensions $d$.
        }
        \label{fig:results_add_md}

\end{figure*}

In the ADD scenario, lower limits on the fundamental Planck scale \MD for the number of extra dimensions $d$ ranging from 2 to 7 are shown in Fig.~\ref{fig:results_add_md}. For the lowest number of extra dimensions considered here, $d=2$, $\MD$ values of up to 10.7\TeV are excluded (12.2\TeV expected). As the number of extra dimensions increases, the probed \MD value is reduced to 5.2\TeV for $d=7$ (5.6\TeV expected). Compared to the result of Ref.~\cite{Sirunyan:2017jix}, these limits represent an improvement of approximately 8\% for low values of $d$ (20\% expected). At larger values, the relative gain in \MD sensitivity is smaller, as a result of the dependence of the signal cross section on \MD, which becomes steeper as $d$ increases. The results are also shown in Table~\ref{tab:supp_add} in Appendix~\ref{app:appendix_add_table}.

\subsection{Leptoquark interpretation}

\begin{figure*}[hbtp]
    \centering
        \includegraphics[width=0.49\textwidth]{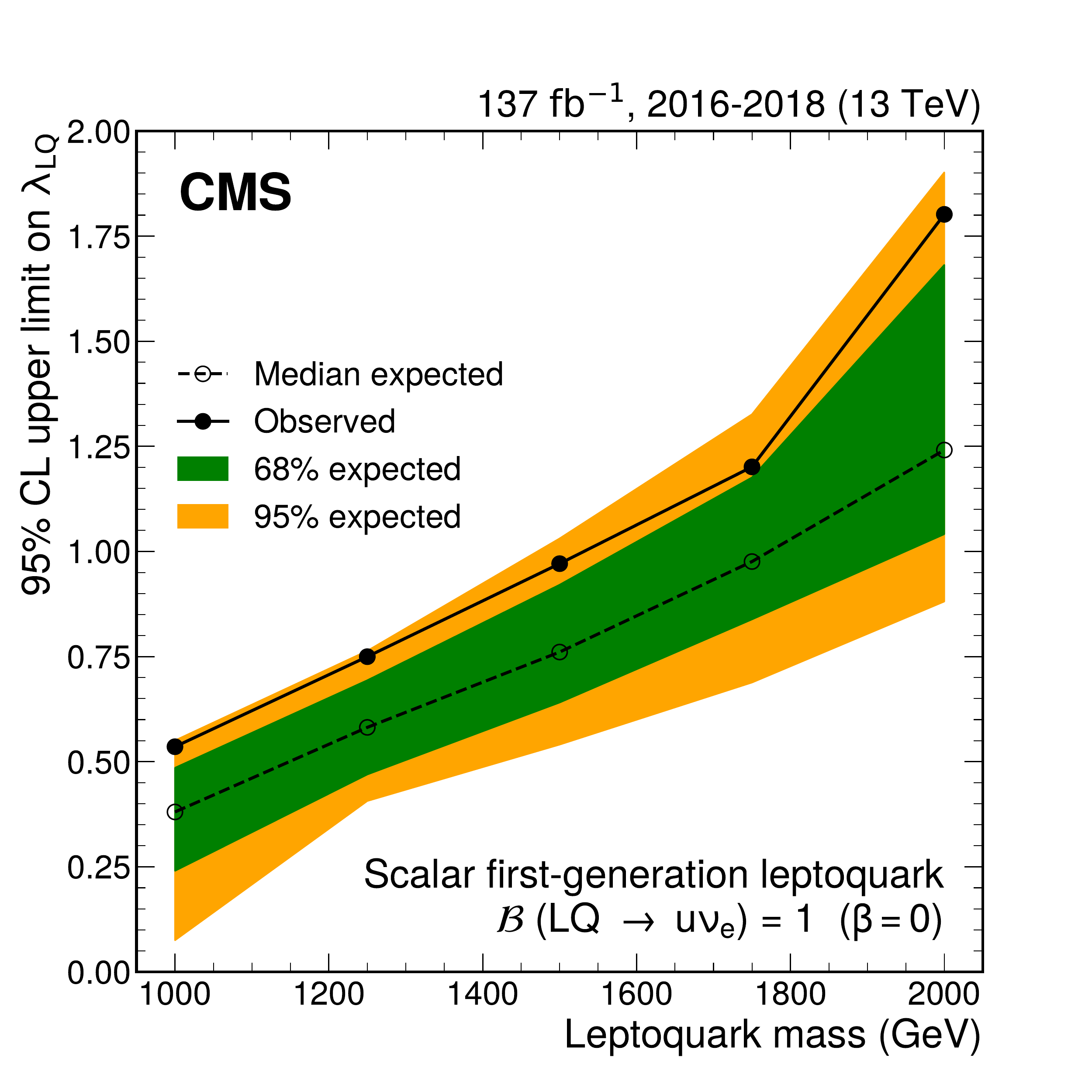}
        \\
        \caption{Upper limits at 95\% \CL on the leptoquark coupling $\lambdalq$ as a function of the leptoquark mass. The branching fraction for the decay of the leptoquark into an electron neutrino and up quark is assumed to be 100\% ($\beta=0$). The dashed line indicates the median expected exclusion contour.}
        \label{fig:results_leptoquark_coupling}

\end{figure*}

Finally, upper limits are placed on the production cross section of LQs coupled to up quarks and neutrinos with a coupling value $\lambdalq$. The branching fraction for the decay of the LQ into an up quark  and an electron neutrino is assumed to be 100\% (also referred to as to $\beta=0$ in the literature). The limits are shown in Fig.~\ref{fig:results_leptoquark_coupling}. Generally, both single and pair LQ production contribute to the signal, with the coupling \lambdalq mainly influencing the single production rate. The pair production dominates at lower LQ masses of $\mlq<1\TeV$, a region which has already been excluded by previous searches~\cite{Sirunyan:2019xwh}. In the higher-mass regime, $\mlq>1\TeV$, the contribution from single production is increased, providing additional sensitivity to the value of \lambdalq. The minimum value of the coupling \lambdalq excluded ranges from about 0.5 at $\mlq=1\TeV$ (0.4 expected) to $\lambdalq=1.0$ at $\mlq=1.5\TeV$ (0.75 expected) and $\lambdalq=1.8$ at 2\TeV (1.25 expected), which are the most stringent constraints from a direct search to date.

\section{Summary}
A search for physics beyond the standard model in events with energetic jets and large missing transverse momentum has been presented. A data set of proton-proton collisions at a center-of-mass energy of 13\TeV, corresponding to an integrated luminosity of 101\fbinv is analyzed, and the analysis results are combined with those of an earlier search using an independent data set collected at the same center-of-mass energy, corresponding to an integrated luminosity of 36\fbinv~\cite{Sirunyan:2017jix}. Separate analysis categories are defined for events with a large-radius jet consistent with a hadronic decay of a \PW\ or a \PZ\ boson, and for events without such a jet. A joint maximum likelihood fit over a combination of signal and control regions is used to constrain standard model (SM) background processes and to extract a possible signal. The data are found to be in good agreement with the fit results, with no evidence for a significant signal contribution. The result is interpreted in terms of exclusion limits at $95\%$ confidence level on the parameters of a number of models of beyond-the-SM physics. We constrain the branching fraction of the Higgs boson decay to invisible particles to be below $27.8\%$. In simplified models of the production of dark matter (DM) candidates via a spin-1 $s$-channel mediator (vector or axial-vector), values of the mediator mass of up to $1.95\TeV$ are excluded, assuming the couplings of $\gq=0.25$ between the mediator and quarks, and $\gchi=1.0$ between the mediator and the Dirac fermion DM particles. Assuming a fixed ratio $\mdm=\mmed/3$, coupling values as low as $\gq=0.018$ and $\gchi=0.070$ can be excluded for $\mmed=100\GeV$. In a similar model with a pseudoscalar spin-0 mediator, \mmed values less than $470\GeV$ are excluded. The fermion portal model, in which a colored scalar mediator couples to a DM candidate and a right-handed up quark, is excluded for mediator mass values up to $1.5\TeV$ at low values of the DM candidate mass $\mdm$, assuming $\lambdafp=1$. In a model of large extra dimensions, values of the fundamental Planck scale below from $10.7$ to $5.2\TeV$ can be excluded, depending on the number of extra dimensions between $2$ and $7$. Finally, the production of leptoquarks decaying into the up quark and the electron neutrino is excluded for coupling values between the leptoquarks and the SM fermions larger than $0.5$ to $1.8$, for leptoquark masses between $1.0$ and $2.0\TeV$. The constraints on \gq in the spin-1 models, on the mediator and dark matter masses in the pseudoscalar and fermion portal models, and on the leptoquark coupling represent the most stringent bounds to date.

\begin{acknowledgments}
    We congratulate our colleagues in the CERN accelerator departments for the excellent performance of the LHC and thank the technical and administrative staffs at CERN and at other CMS institutes for their contributions to the success of the CMS effort. In addition, we gratefully acknowledge the computing centers and personnel of the Worldwide LHC Computing Grid and other centers for delivering so effectively the computing infrastructure essential to our analyses. Finally, we acknowledge the enduring support for the construction and operation of the LHC, the CMS detector, and the supporting computing infrastructure provided by the following funding agencies: BMBWF and FWF (Austria); FNRS and FWO (Belgium); CNPq, CAPES, FAPERJ, FAPERGS, and FAPESP (Brazil); MES (Bulgaria); CERN; CAS, MoST, and NSFC (China); MINCIENCIAS (Colombia); MSES and CSF (Croatia); RIF (Cyprus); SENESCYT (Ecuador); MoER, ERC PUT and ERDF (Estonia); Academy of Finland, MEC, and HIP (Finland); CEA and CNRS/IN2P3 (France); BMBF, DFG, and HGF (Germany); GSRT (Greece); NKFIA (Hungary); DAE and DST (India); IPM (Iran); SFI (Ireland); INFN (Italy); MSIP and NRF (Republic of Korea); MES (Latvia); LAS (Lithuania); MOE and UM (Malaysia); BUAP, CINVESTAV, CONACYT, LNS, SEP, and UASLP-FAI (Mexico); MOS (Montenegro); MBIE (New Zealand); PAEC (Pakistan); MSHE and NSC (Poland); FCT (Portugal); JINR (Dubna); MON, RosAtom, RAS, RFBR, and NRC KI (Russia); MESTD (Serbia); SEIDI, CPAN, PCTI, and FEDER (Spain); MOSTR (Sri Lanka); Swiss Funding Agencies (Switzerland); MST (Taipei); ThEPCenter, IPST, STAR, and NSTDA (Thailand); TUBITAK and TAEK (Turkey); NASU (Ukraine); STFC (United Kingdom); DOE and NSF (USA).
     
    \hyphenation{Rachada-pisek} Individuals have received support from the Marie-Curie program and the European Research Council and Horizon 2020 Grant, contract Nos.\ 675440, 724704, 752730, 758316, 765710, 824093, and COST Action CA16108 (European Union); the Leventis Foundation; the Alfred P.\ Sloan Foundation; the Alexander von Humboldt Foundation; the Belgian Federal Science Policy Office; the Fonds pour la Formation \`a la Recherche dans l'Industrie et dans l'Agriculture (FRIA-Belgium); the Agentschap voor Innovatie door Wetenschap en Technologie (IWT-Belgium); the F.R.S.-FNRS and FWO (Belgium) under the ``Excellence of Science -- EOS" -- be.h project n.\ 30820817; the Beijing Municipal Science \& Technology Commission, No. Z191100007219010; the Ministry of Education, Youth and Sports (MEYS) of the Czech Republic; the Deutsche Forschungsgemeinschaft (DFG), under Germany's Excellence Strategy -- EXC 2121 ``Quantum Universe" -- 390833306, and under project number 400140256 - GRK2497; the Lend\"ulet (``Momentum") Program and the J\'anos Bolyai Research Scholarship of the Hungarian Academy of Sciences, the New National Excellence Program \'UNKP, the NKFIA research grants 123842, 123959, 124845, 124850, 125105, 128713, 128786, and 129058 (Hungary); the Council of Science and Industrial Research, India; the Latvian Council of Science; the Ministry of Science and Higher Education and the National Science Center, contracts Opus 2014/15/B/ST2/03998 and 2015/19/B/ST2/02861 (Poland); the National Priorities Research Program by Qatar National Research Fund; the Ministry of Science and Higher Education, project no. 0723-2020-0041 (Russia); the Programa Estatal de Fomento de la Investigaci{\'o}n Cient{\'i}fica y T{\'e}cnica de Excelencia Mar\'{\i}a de Maeztu, grant MDM-2015-0509 and the Programa Severo Ochoa del Principado de Asturias; the Stavros Niarchos Foundation (Greece); the Rachadapisek Sompot Fund for Postdoctoral Fellowship, Chulalongkorn University and the Chulalongkorn Academic into Its 2nd Century Project Advancement Project (Thailand); the Kavli Foundation; the Nvidia Corporation; the SuperMicro Corporation; the Welch Foundation, contract C-1845; and the Weston Havens Foundation (USA).
\end{acknowledgments}

\bibliography{auto_generated}
\clearpage
\appendix

\section{Additional figures and tables}
\label{app:appendix}

\subsection{Event selection summary tables}
\label{app:appendix_selection_tables}
The event selection criteria for the signal regions of the different analysis categories are summarized in Table~\ref{tab:supp_selection}. The topological selection differences between regions in the same category are shown in Table~\ref{tab:supp_regions}.

\begin{table*}[htb]
    \renewcommand{\arraystretch}{1.5}
    \topcaption{Summary of the common selection requirements for mono-$\PV$ and monojet categories. For the control region selections, the requirements on \ptmiss and \dphijm are replaced by the equivalent selections based on the hadronic recoil, which is calculated as the vectorial sum of the \ptvecmiss and the respective lepton or photon transverse momenta used to define the control region selection. The \dpfcalo and \dphitkpf requirements are always evaluated based on \ptmiss, and not the hadronic recoil.}
    \centering
    \begin{tabular}{c l r}
        {Category}                  & {Variable / Description}                            & {Selection}                                                \\
        \hline
        \multirow{9}{*}{All}        & Electron veto                                       & $\pt>10\GeV$ and $\abs{\eta} < 2.5$                        \\
                                    & Muon veto                                           & $\pt>10\GeV$ and $\abs{\eta} < 2.4$                        \\
                                    & \PGt lepton veto                                    & $\pt>18\GeV$ and $\abs{\eta} < 2.3$                        \\
                                    & Photon veto                                         & $\pt>15\GeV$ and $\abs{\eta} < 2.5$                        \\
                                    & \PQb jet veto                                       & DeepCSV ``medium'', $\pt > 20\GeV,~\abs{\eta} < 2.4$       \\
                                    & \ptmiss                                             & $>$250\GeV                                                 \\
                                    & \dpfcalo                                            & $<$0.5                                                     \\
                                    & \dphitkpf                                           & $<$2                                                       \\
                                    & \dphijm                                             & $>$0.5                                                     \\\\
        \multirow{4}{*}{All (2018)} & \multirow{2}{*}{Calorimeter failure mitigation (I)} & no AK4 jet with $\pt>30\GeV,$                              \\
                                    &                                                     & $-1.57<\phi<-0.87$, $-3.0<\eta<-1.3$                       \\\\
                                    & Calorimeter failure mitigation (II)                 & $\phi(\ptvecmiss)\notin[-1.62,-0.62]$ if $\ptmiss<470\GeV$ \\\\
        Monojet                     & Leading AK4 jet                                     & $\pt>100\GeV$ and $\abs{\eta} < 2.4$                       \\
        \\
        \multirow{2}{*}{Mono-V}     & \multirow{2}{*}{Leading AK8 jet}                    & $\pt>250\GeV$, $\abs{\eta}<2.4, 65 <\msd<120\GeV$          \\
                                    &                                                     & Subcategorization based on \textsc{DeepAK8} score          \\
    \end{tabular}
    \label{tab:supp_selection}

\end{table*}

\begin{table*}[htb]
    \renewcommand{\arraystretch}{1.5}
    \topcaption{Summary of the topological selections used for different regions in the same category. Note that the trigger-level \ptmiss calculation does not take into account muons, which makes the \ptmiss based trigger equally suitable for the signal region and muon-based control regions.}
    \centering
    \begin{tabular}{l r r r c r}
        Region type     & \multicolumn{3}{c}{Multiplicities} & Trigger & Special selection                                                       \\
                        & \Pe                                & \PGm    & \PGg              &                   & (relative to signal)            \\
        \hline
        Signal          & 0                                  & 0       & 0                 & \ptmiss           & \NA                             \\
        Single electron & 1                                  & 0       & 0                 & Electron / photon & $\mT<160\GeV$, $\ptmiss>60\GeV$ \\
        Single muon     & 0                                  & 1       & 0                 & \ptmiss           & $\mT<160\GeV$                   \\
        Dielectron      & 2                                  & 0       & 0                 & Electron / photon & $60<m(\ell\ell)<120\GeV$        \\
        Dimuon          & 0                                  & 2       & 0                 & \ptmiss           & $60<m(\ell\ell)<120\GeV$        \\
        Photon          & 0                                  & 0       & 1                 & Photon            & \NA                             \\
    \end{tabular}
    \label{tab:supp_regions}

\end{table*}

\clearpage

\subsection{Hadronic recoil distributions in the control regions}
\label{app:appendix_recoil_cr}

\begin{figure*}[hbtp]
    \centering
        \includegraphics[width=0.35\textwidth]{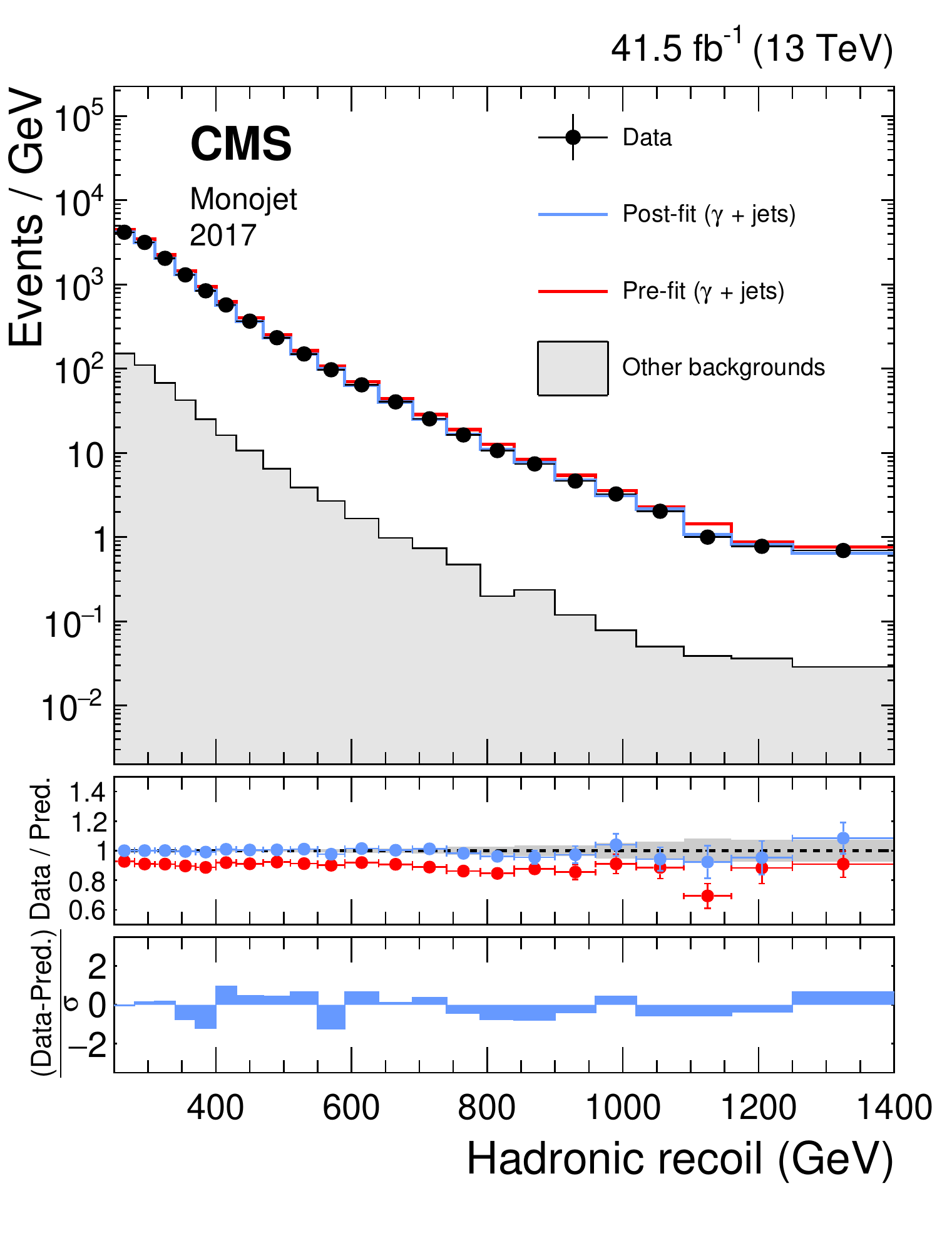}
        \includegraphics[width=0.35\textwidth]{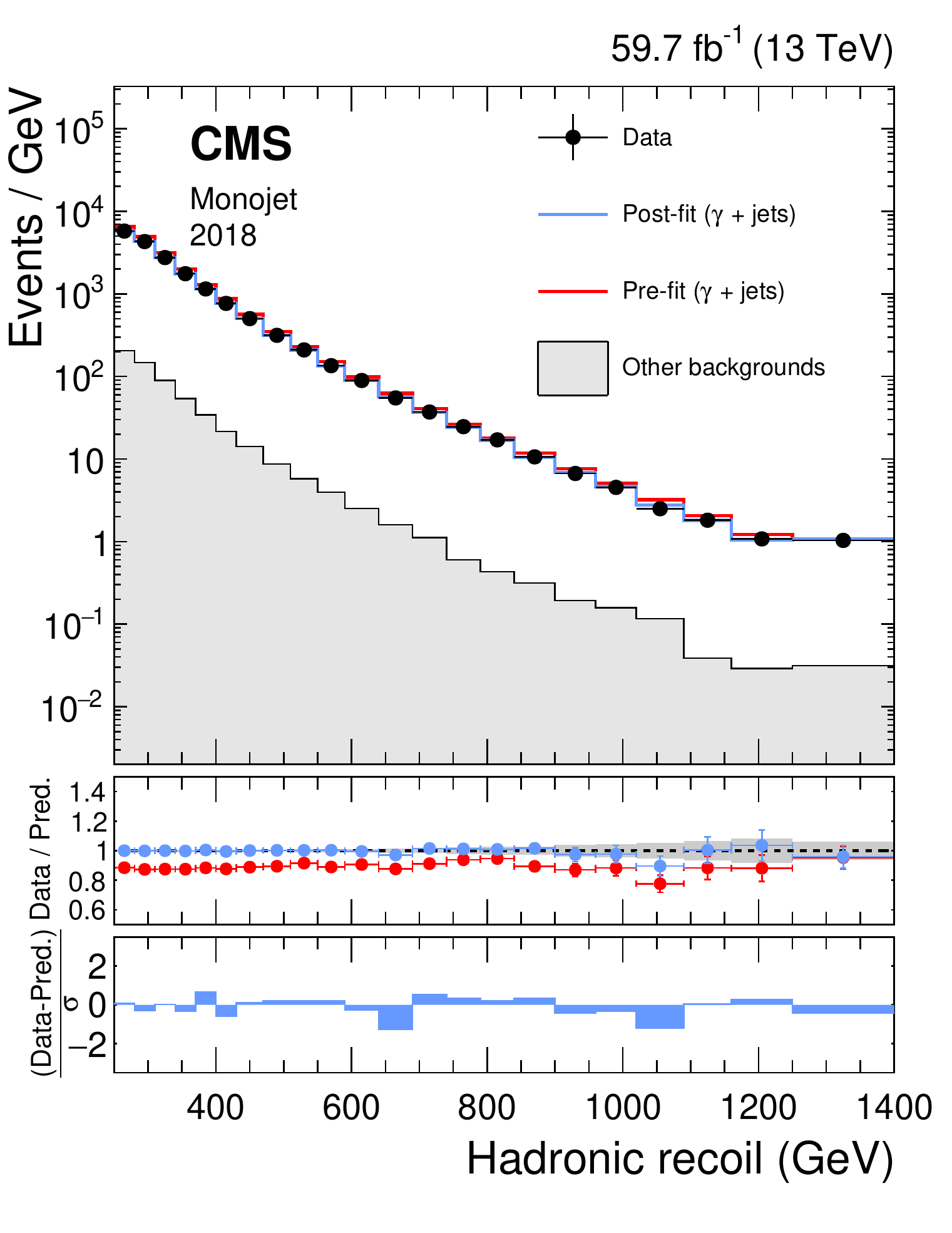}
        \\
        \includegraphics[width=0.35\textwidth]{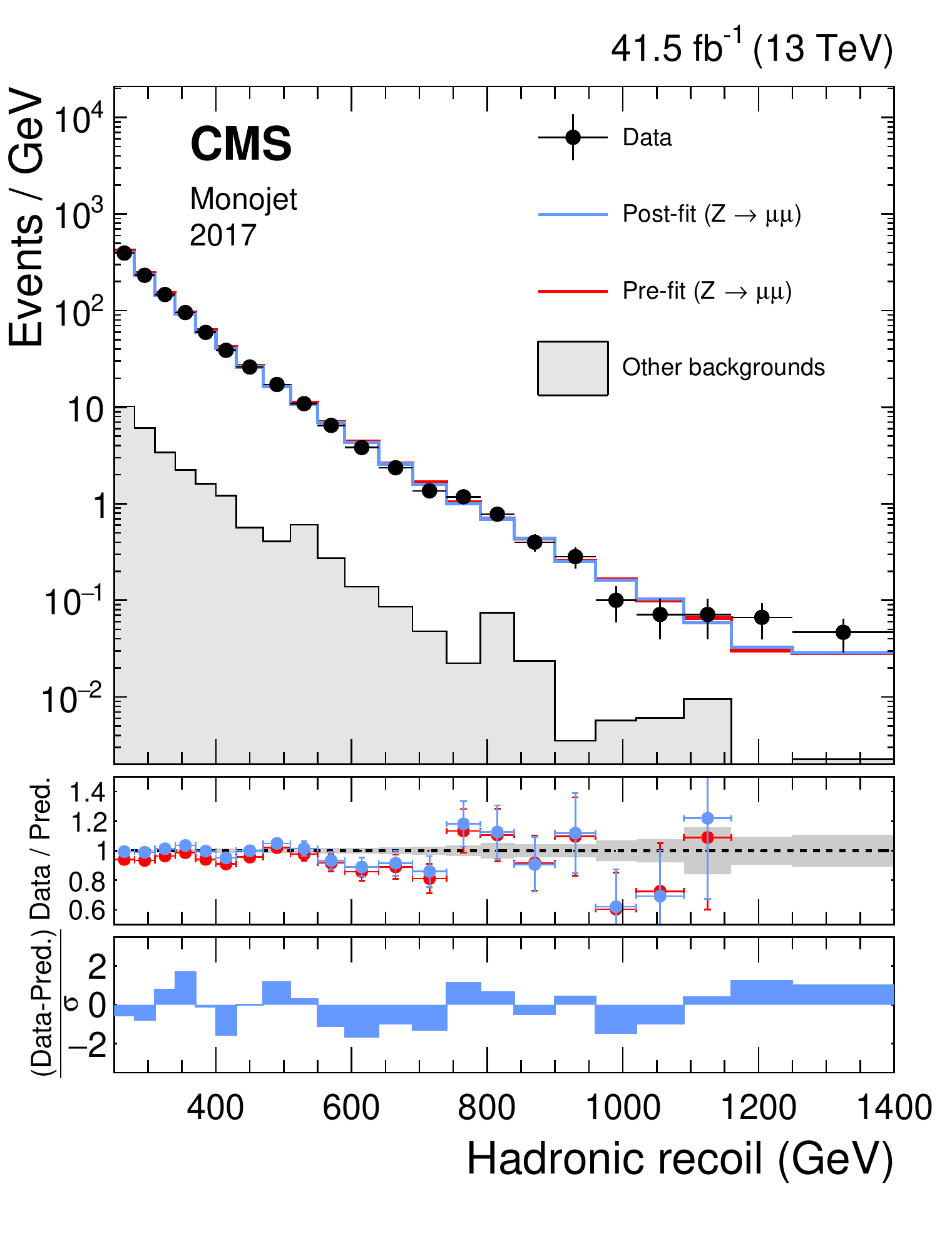}
        \includegraphics[width=0.35\textwidth]{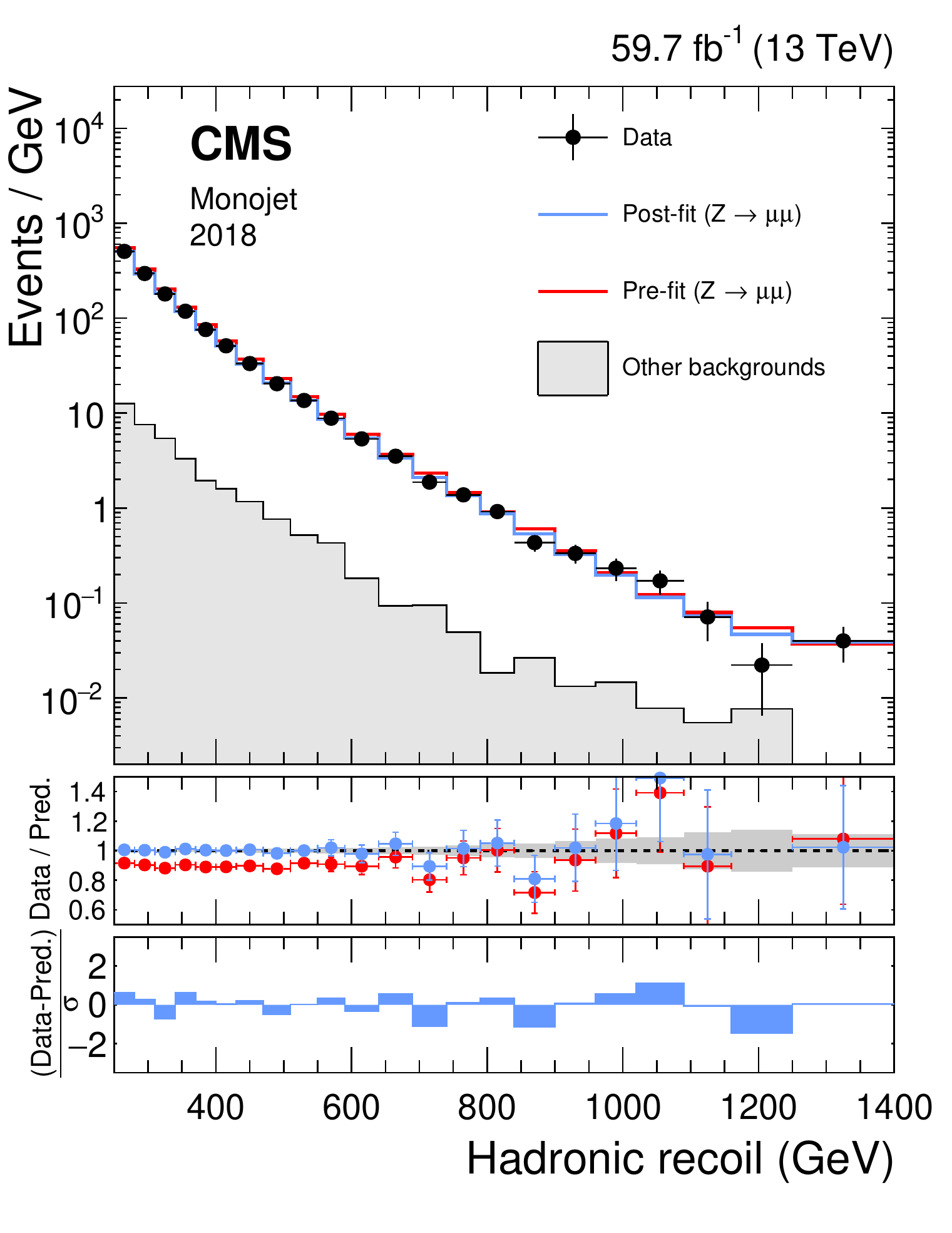}
        \\
        \includegraphics[width=0.35\textwidth]{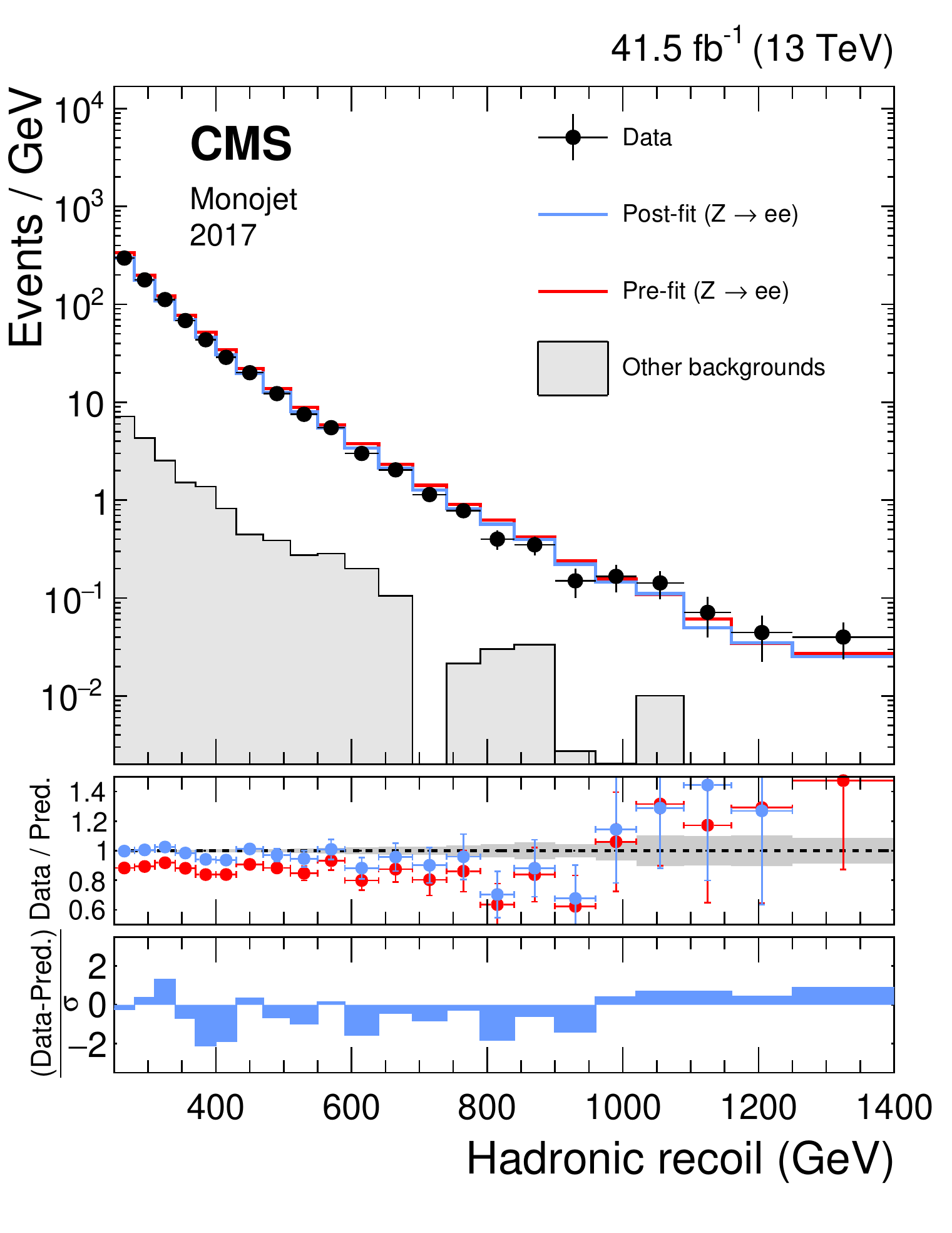}
        \includegraphics[width=0.35\textwidth]{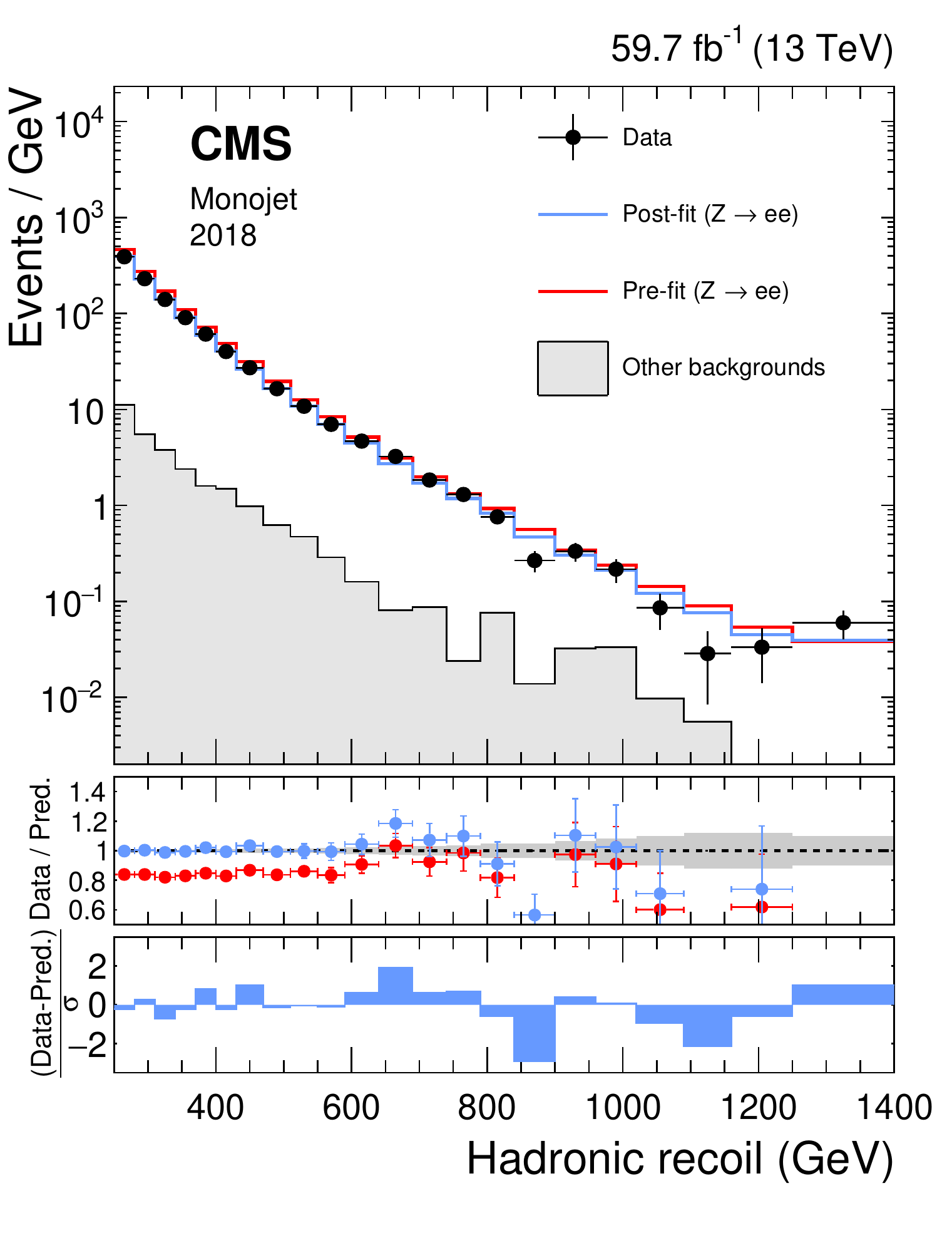}
        \caption{
           Hadronic recoil distributions in the photon (upper), dimuon (middle) and dielectron control regions (lower) in the monojet category. The ``Other backgrounds" include QCD multijet production (photon control region), and top quark, diboson, and \Wjets processes (dimuon and dielectron control regions).
        }
        \label{fig:postfit_zll_photon_monojet}

\end{figure*}

\begin{figure*}[hbtp]
    \centering
        \includegraphics[width=0.40\textwidth]{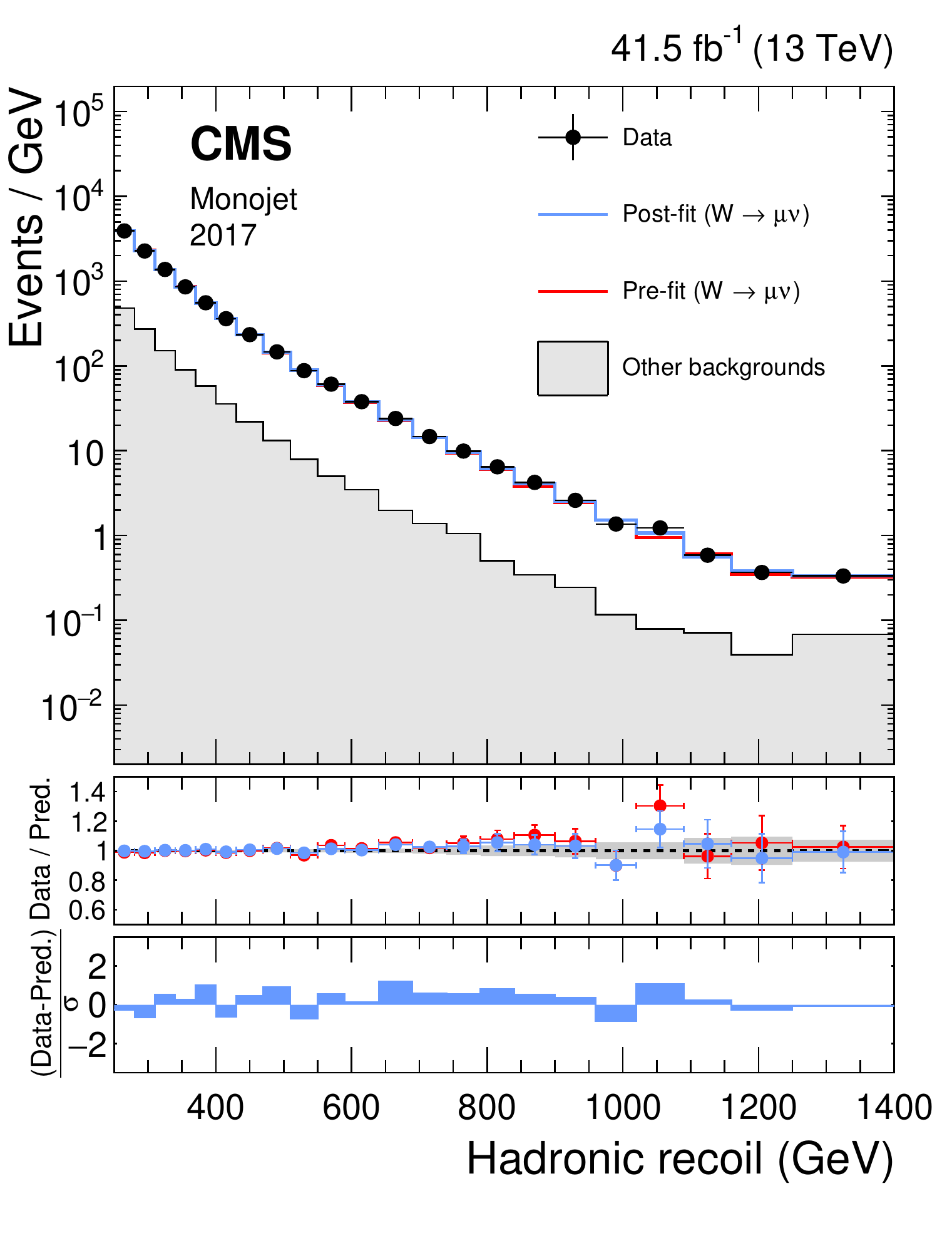}
        \includegraphics[width=0.40\textwidth]{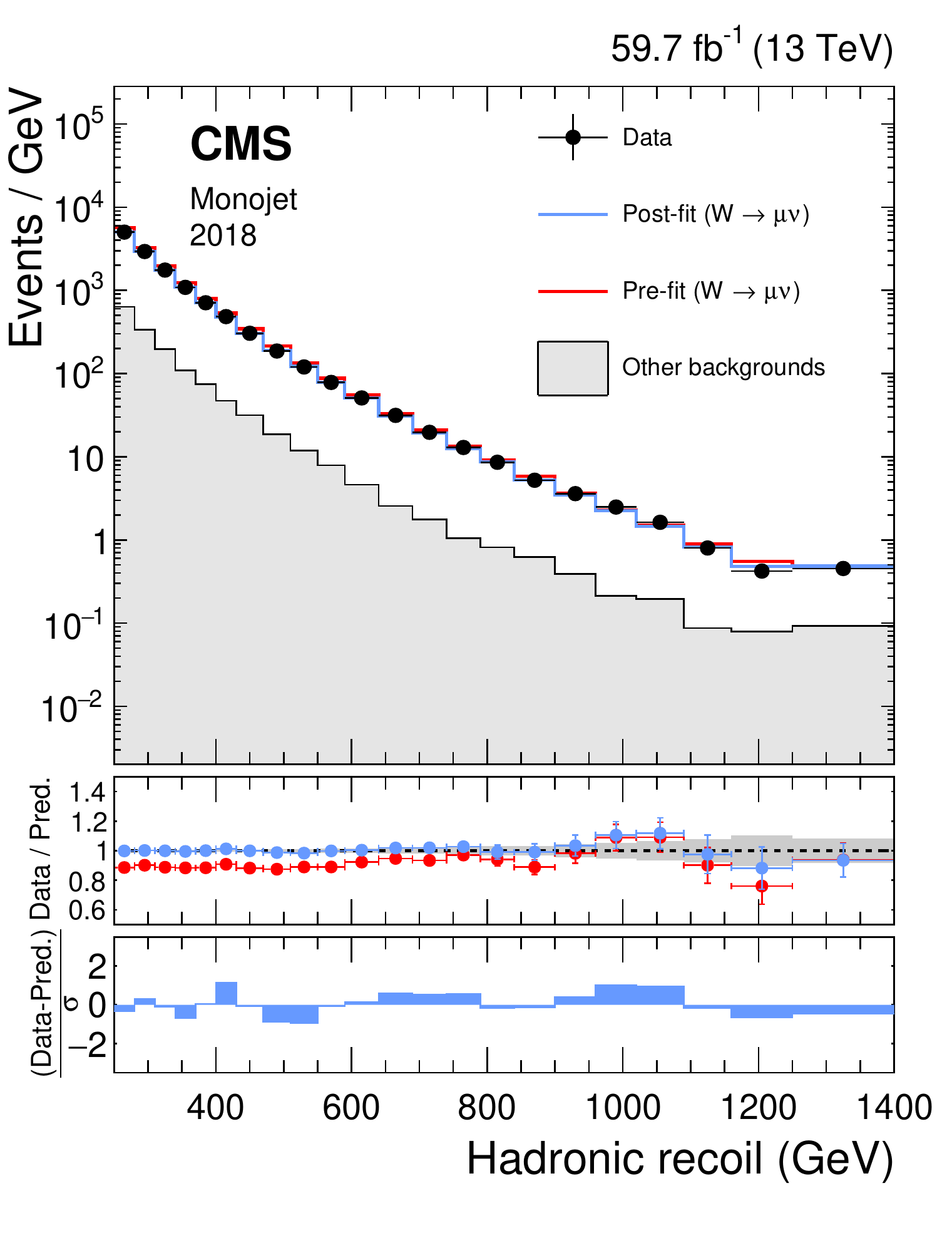}
        \\
        \includegraphics[width=0.40\textwidth]{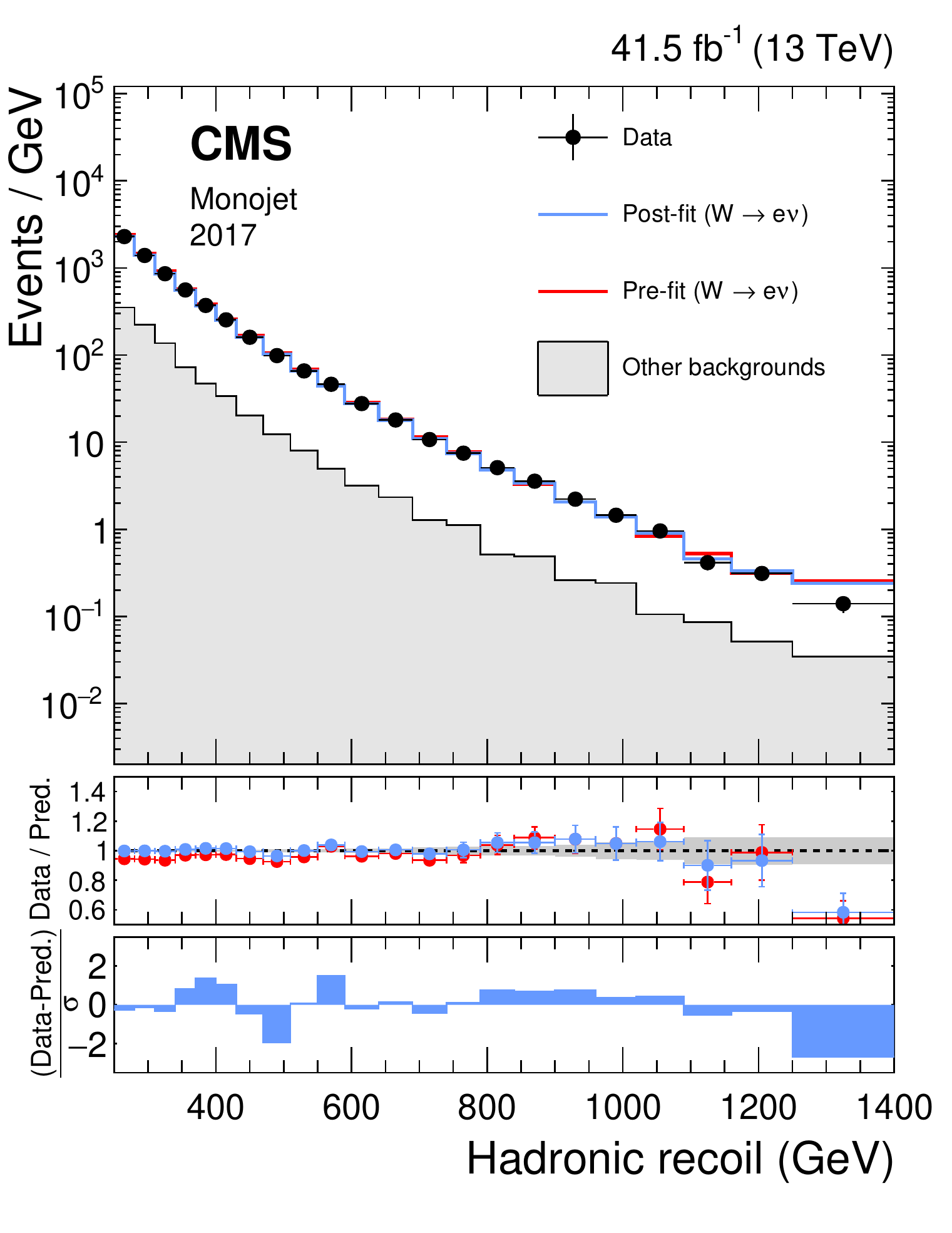}
        \includegraphics[width=0.40\textwidth]{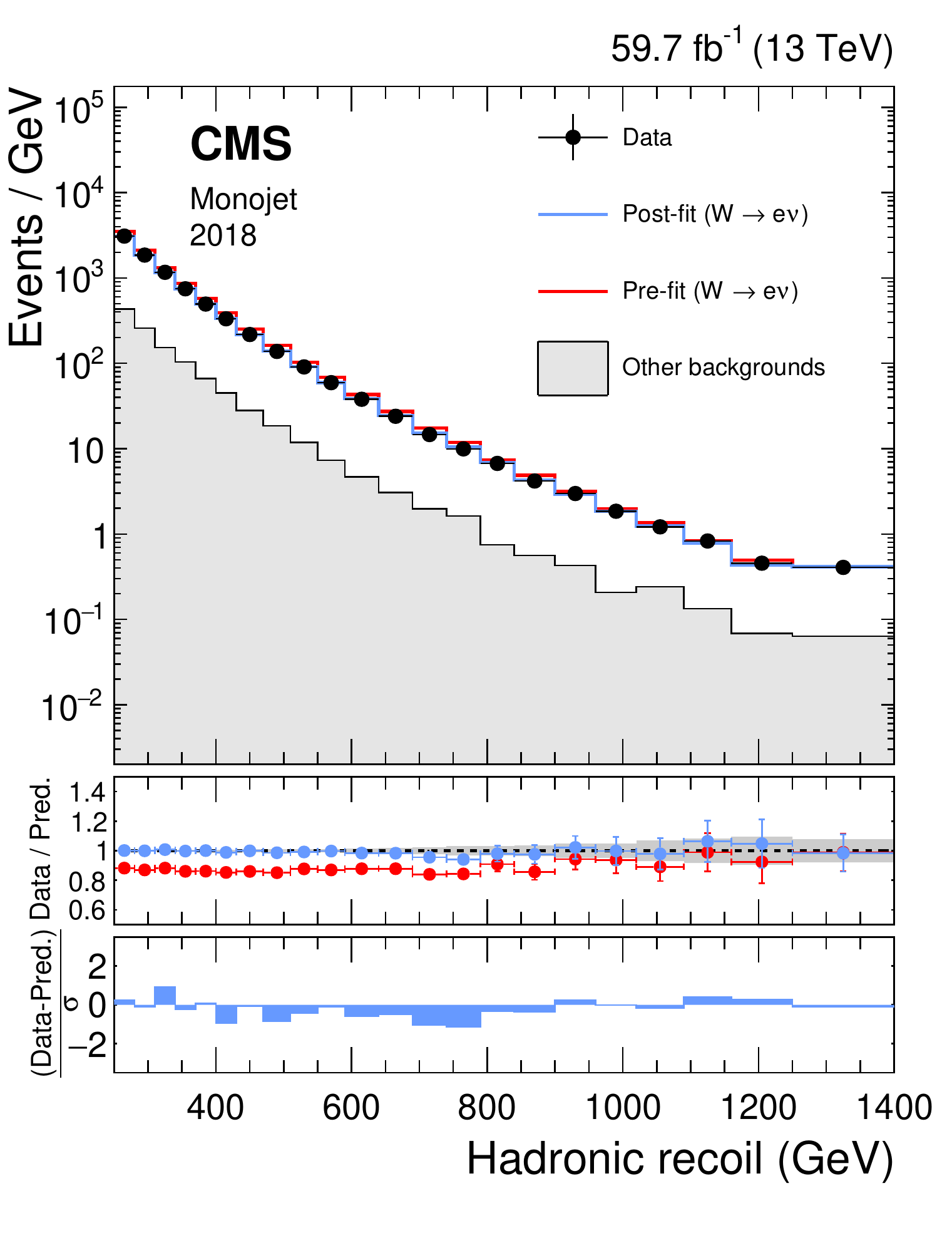}
        \caption{
            Hadronic recoil distributions in the single muon (upper), and single electron regions (lower) in the monojet category.
            The ``Other backgrounds" include top quark, diboson, and QCD multijet processes.
        }
        \label{fig:postfit_wln_monojet}

\end{figure*}

\begin{figure*}[hbtp]
    \centering
        \includegraphics[width=0.35\textwidth]{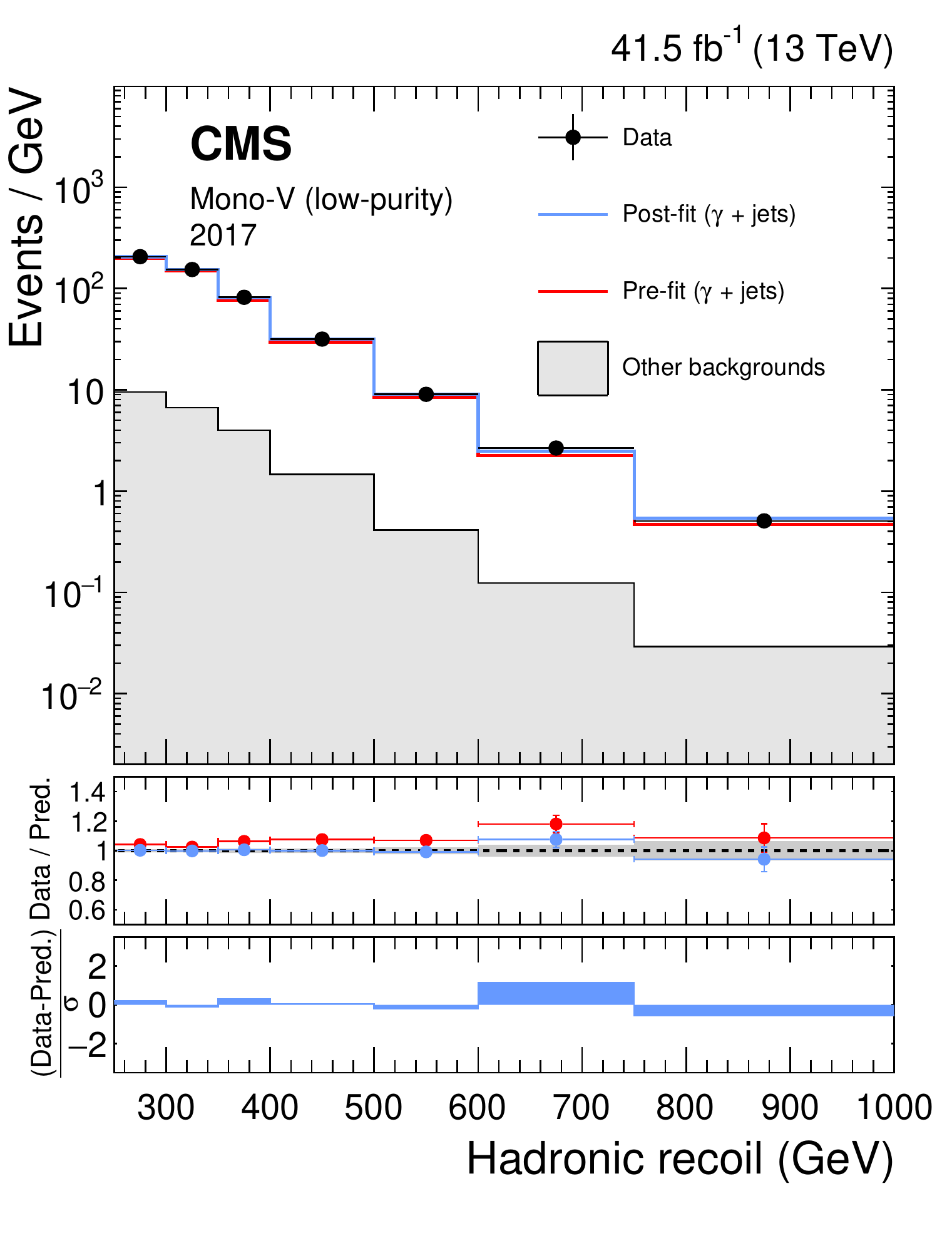}
        \includegraphics[width=0.35\textwidth]{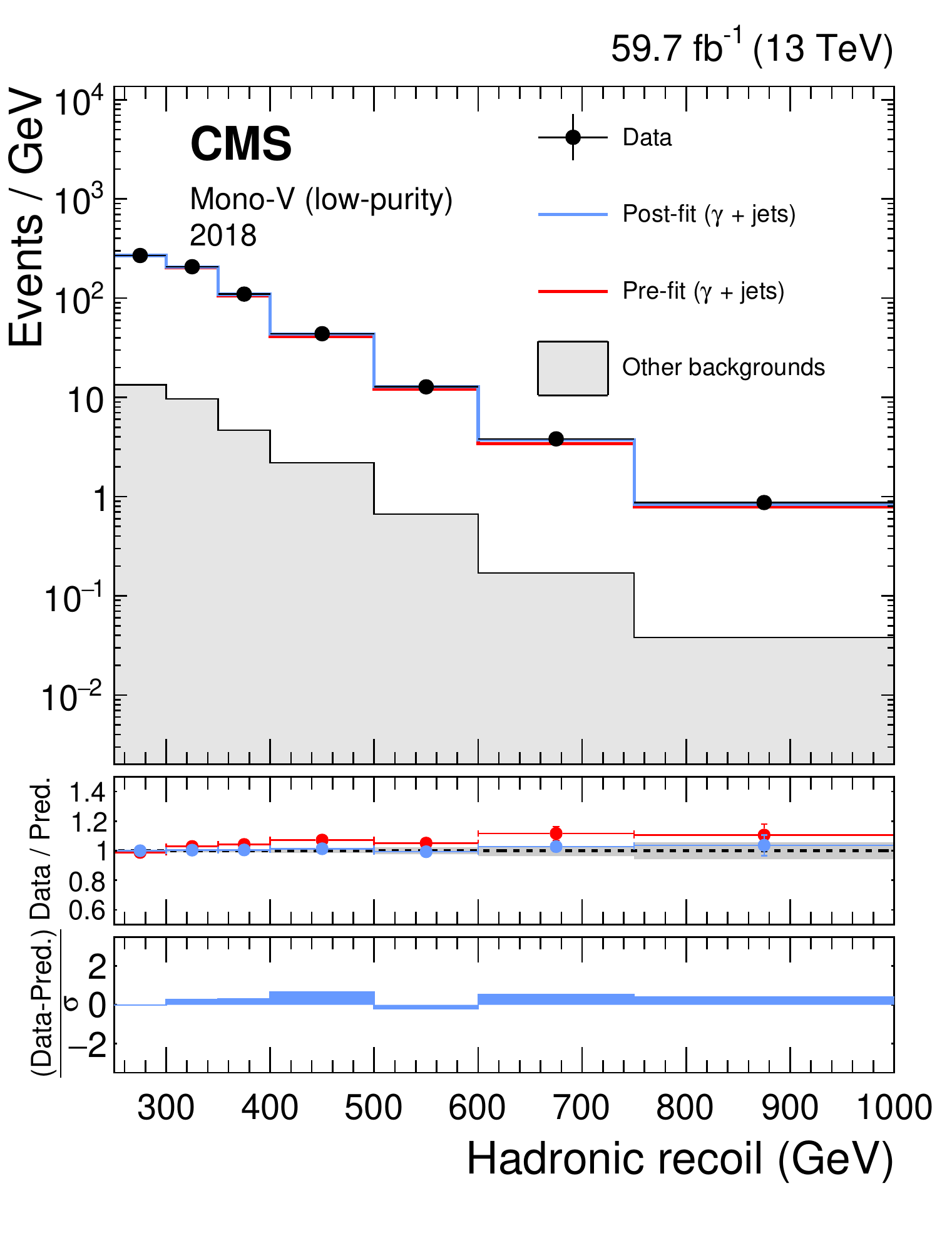}
        \\
        \includegraphics[width=0.35\textwidth]{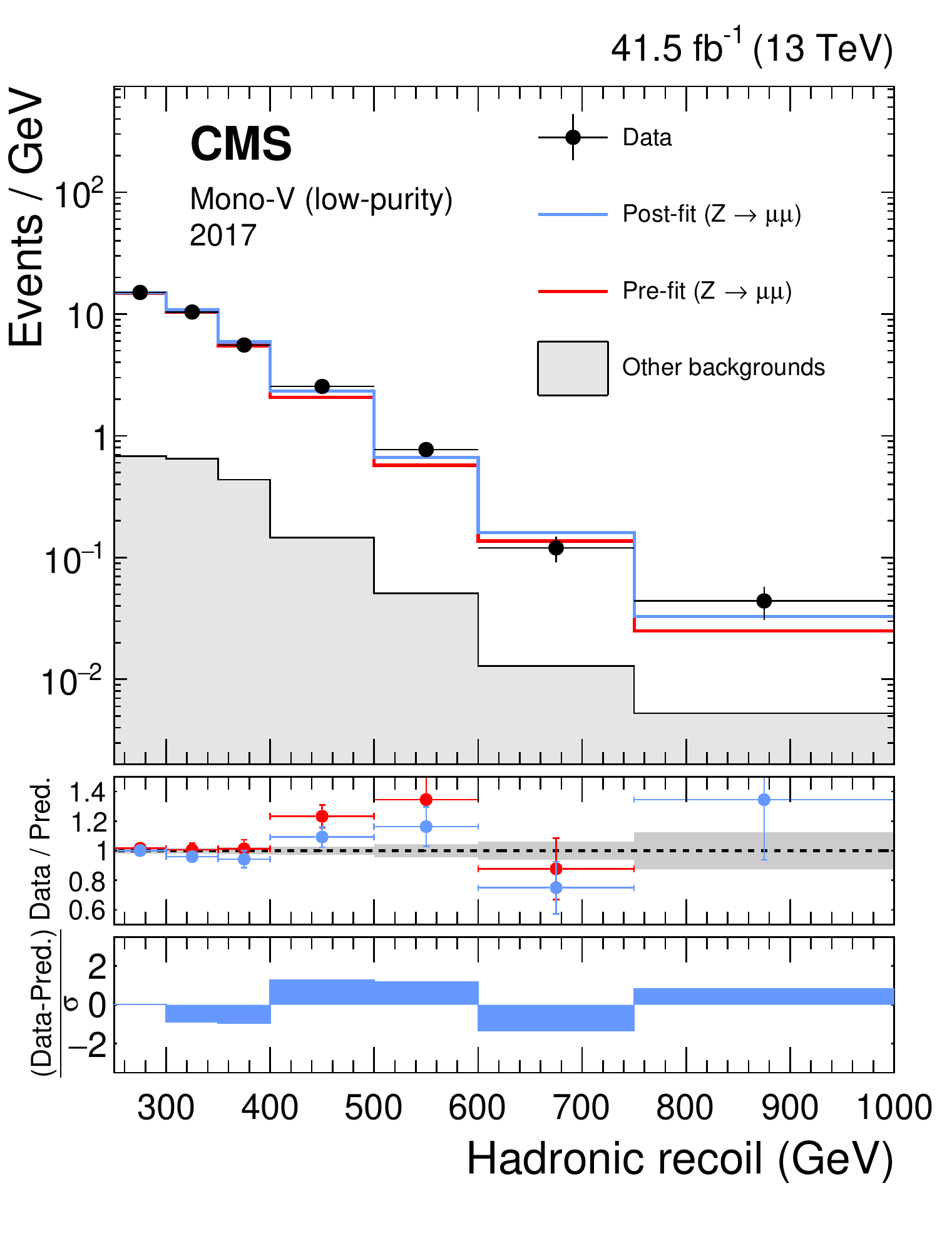}
        \includegraphics[width=0.35\textwidth]{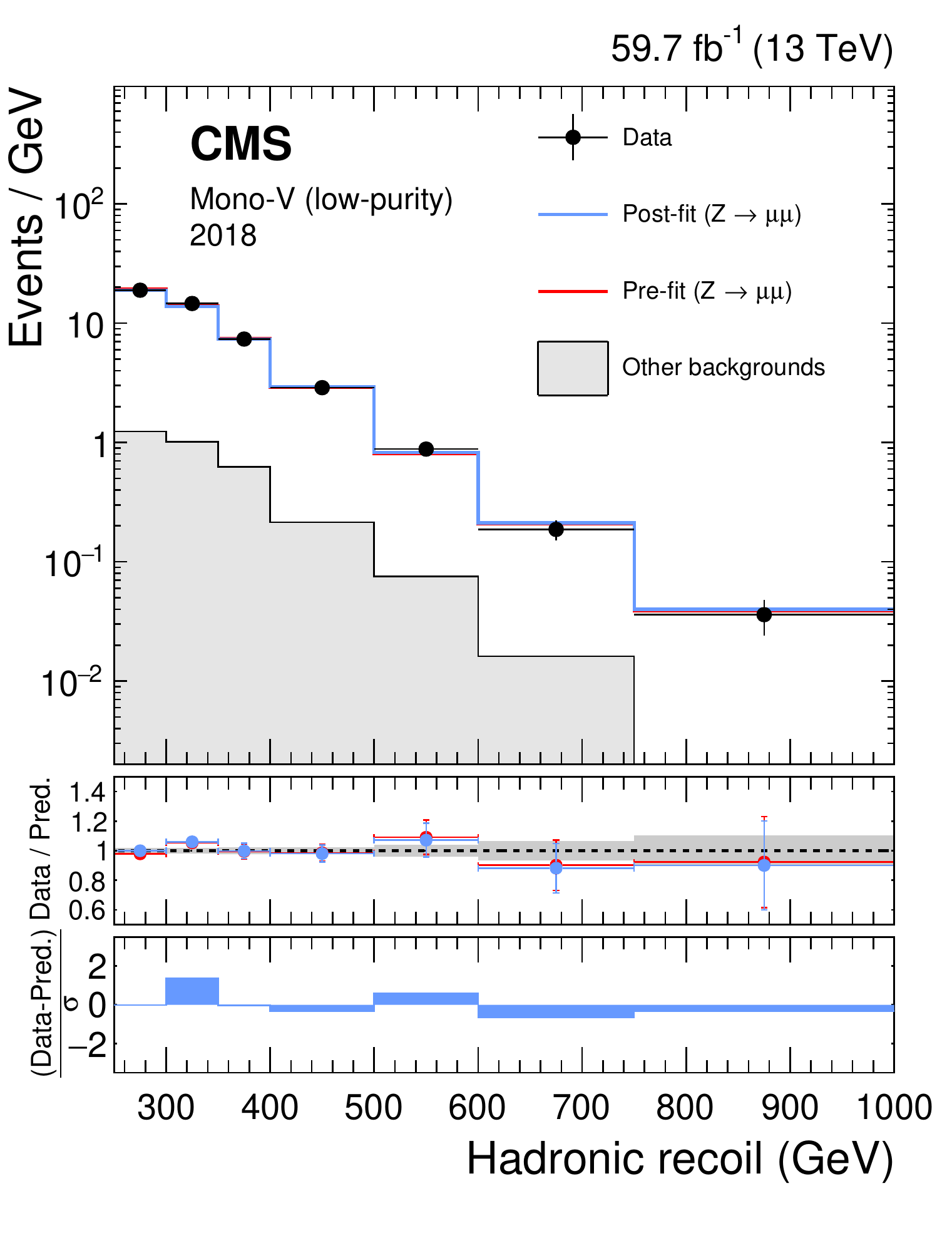}
        \\
        \includegraphics[width=0.35\textwidth]{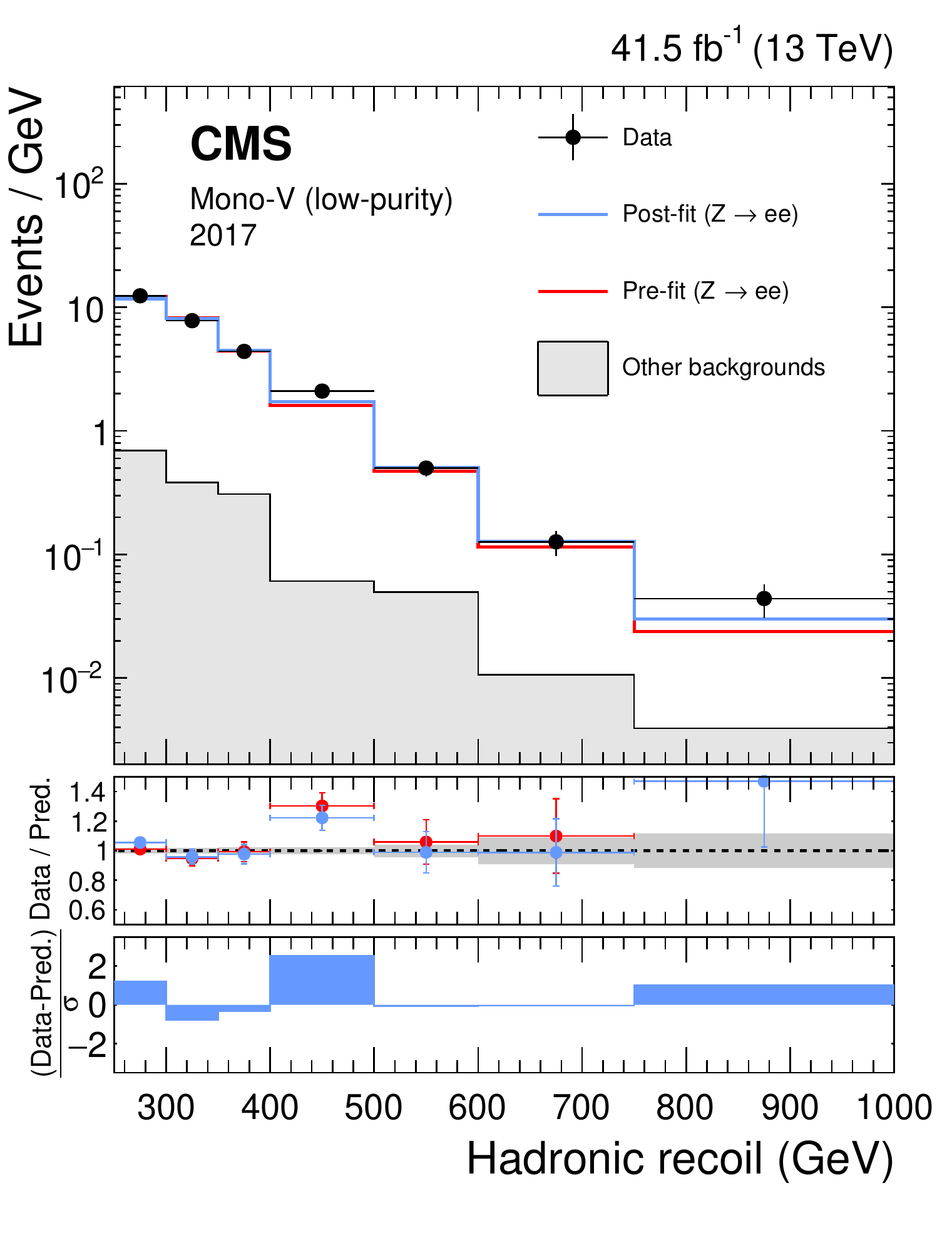}
        \includegraphics[width=0.35\textwidth]{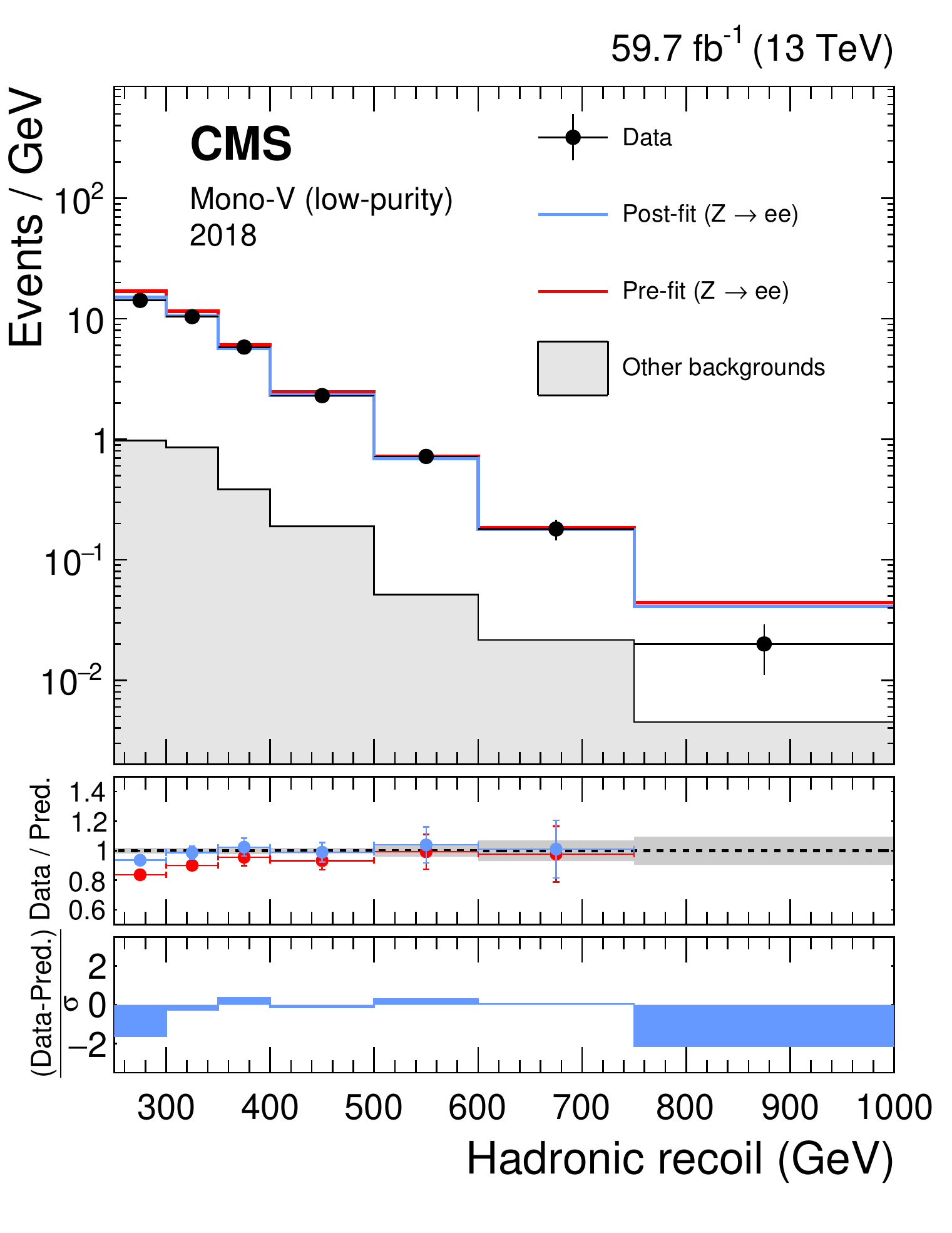}
        \caption{
           Hadronic recoil distributions in the photon (upper), dimuon (middle) and dielectron control regions (lower) in the low-purity mono-V category. The ``Other backgrounds" include QCD multijet production (photon control region), and top quark, diboson, and \Wjets processes (dimuon and dielectron control regions).
        }
        \label{fig:postfit_cr_zll_monovloose}

\end{figure*}

\begin{figure*}[hbtp]
    \centering
        \includegraphics[width=0.40\textwidth]{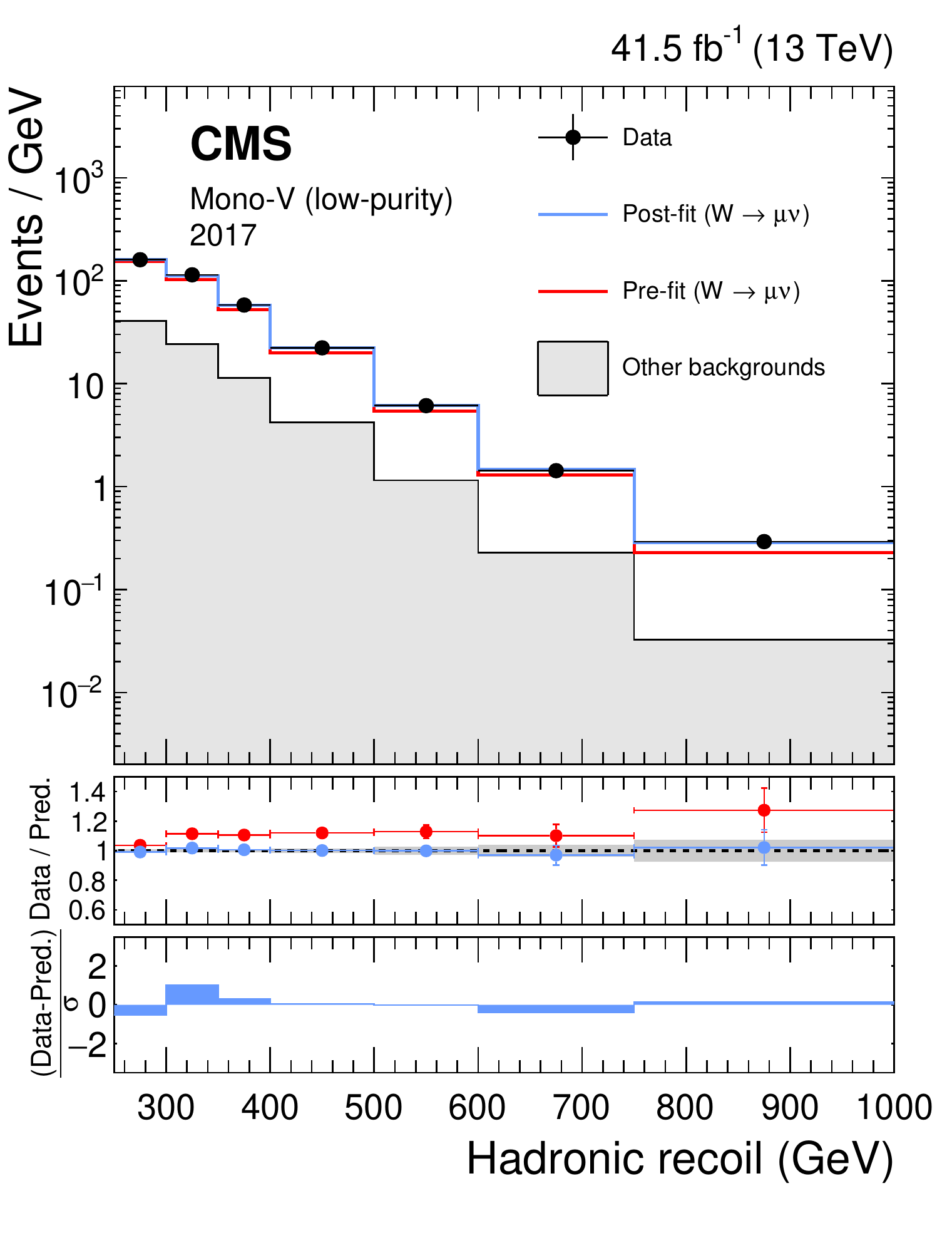}
        \includegraphics[width=0.40\textwidth]{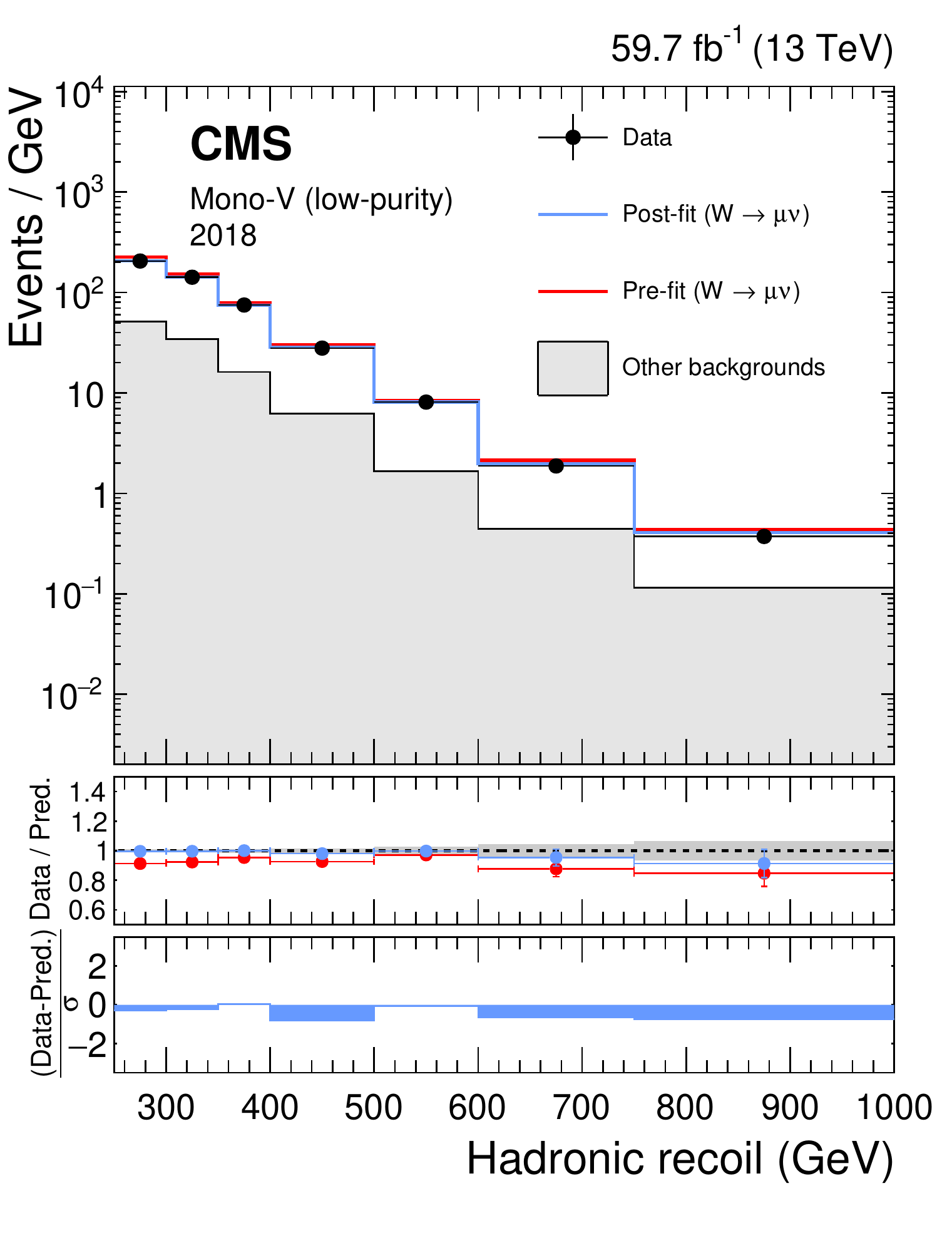}
        \\
        \includegraphics[width=0.40\textwidth]{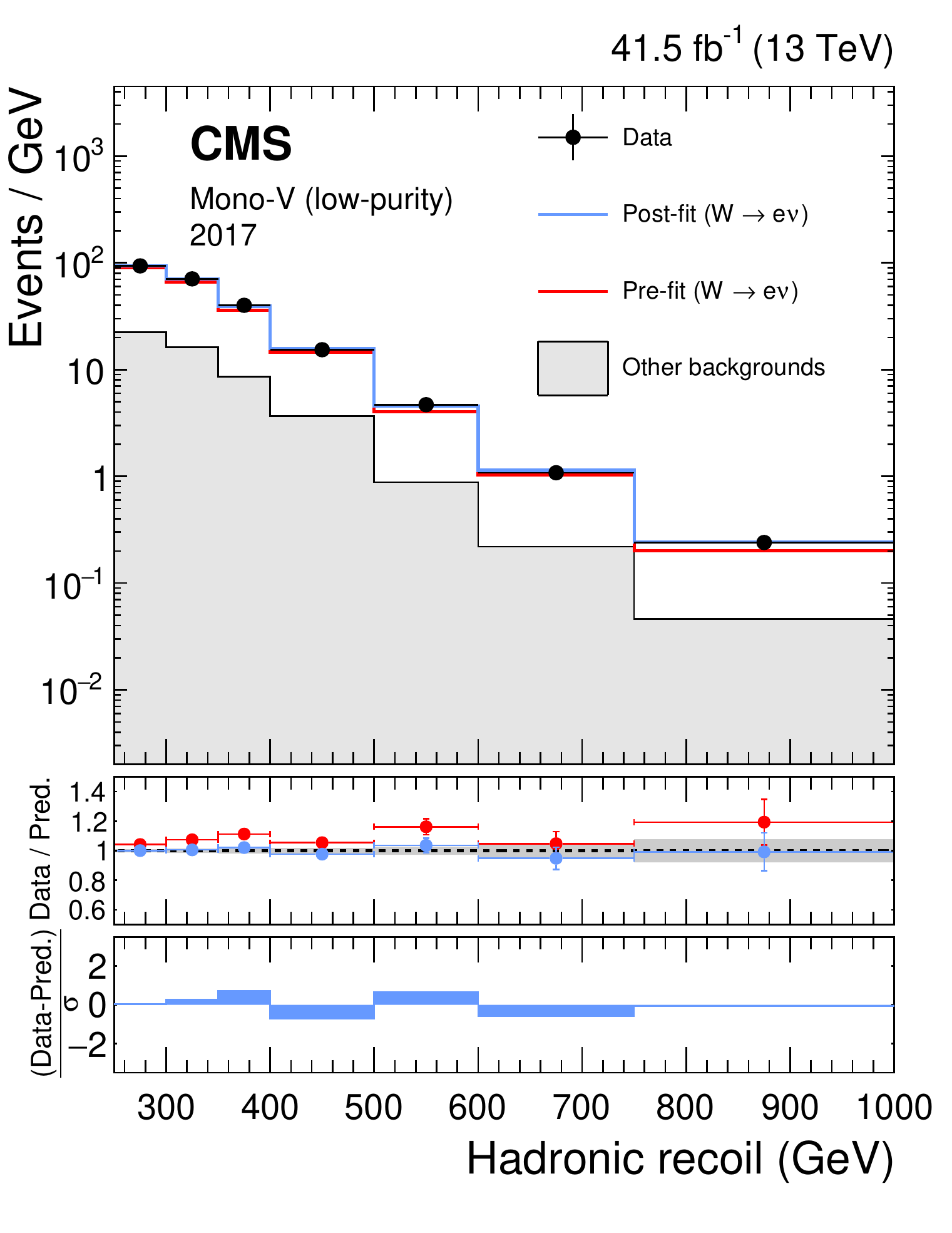}
        \includegraphics[width=0.40\textwidth]{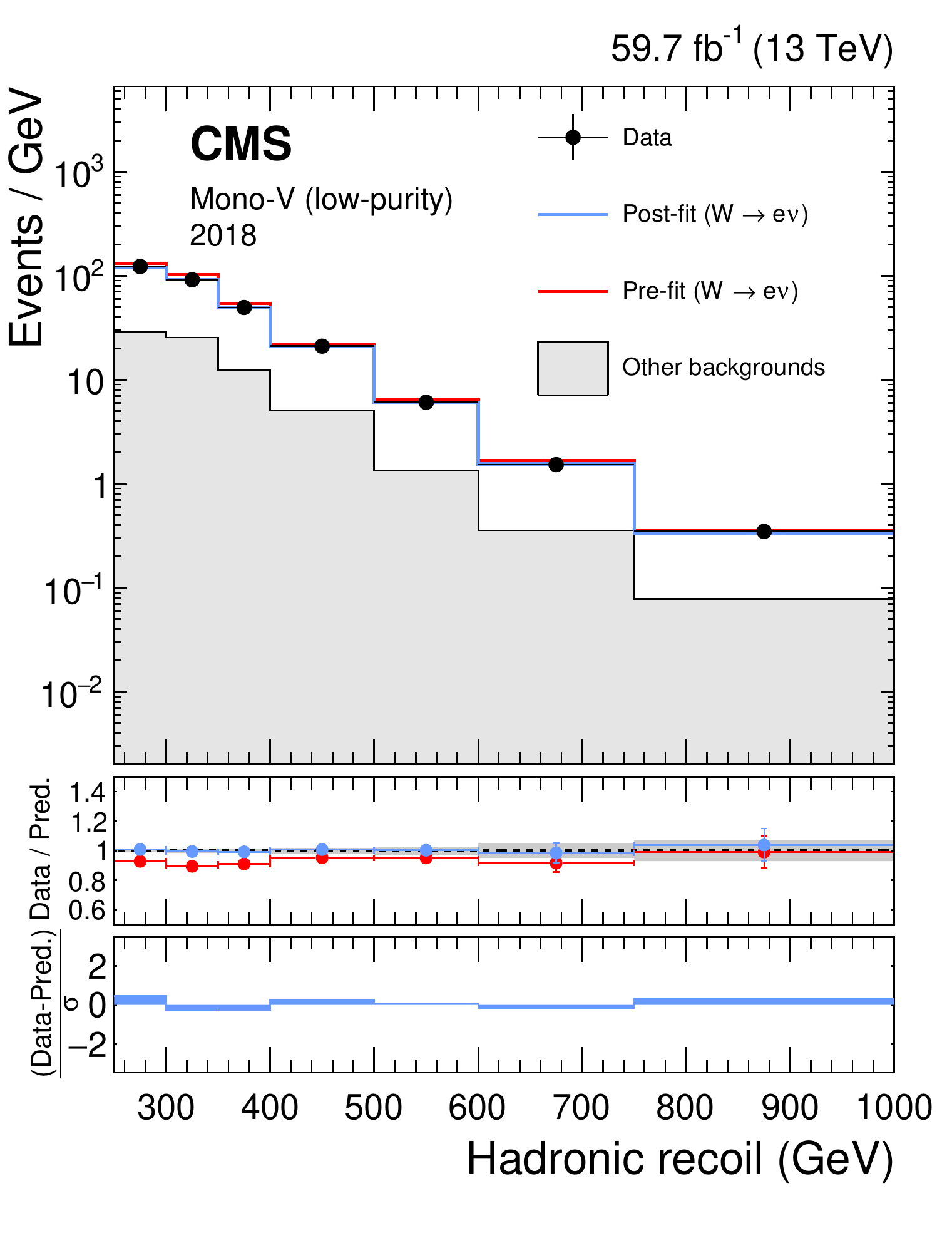}
        \caption{
            Hadronic recoil distributions in the single muon (upper), and single electron regions (lower) in the low-purity mono-V category.
            The ``Other backgrounds" include top quark, diboson, and QCD multijet processes.
        }
        \label{fig:postfit_cr_wln_monovloose}

\end{figure*}

\begin{figure*}[hbtp]
    \centering
        \includegraphics[width=0.35\textwidth]{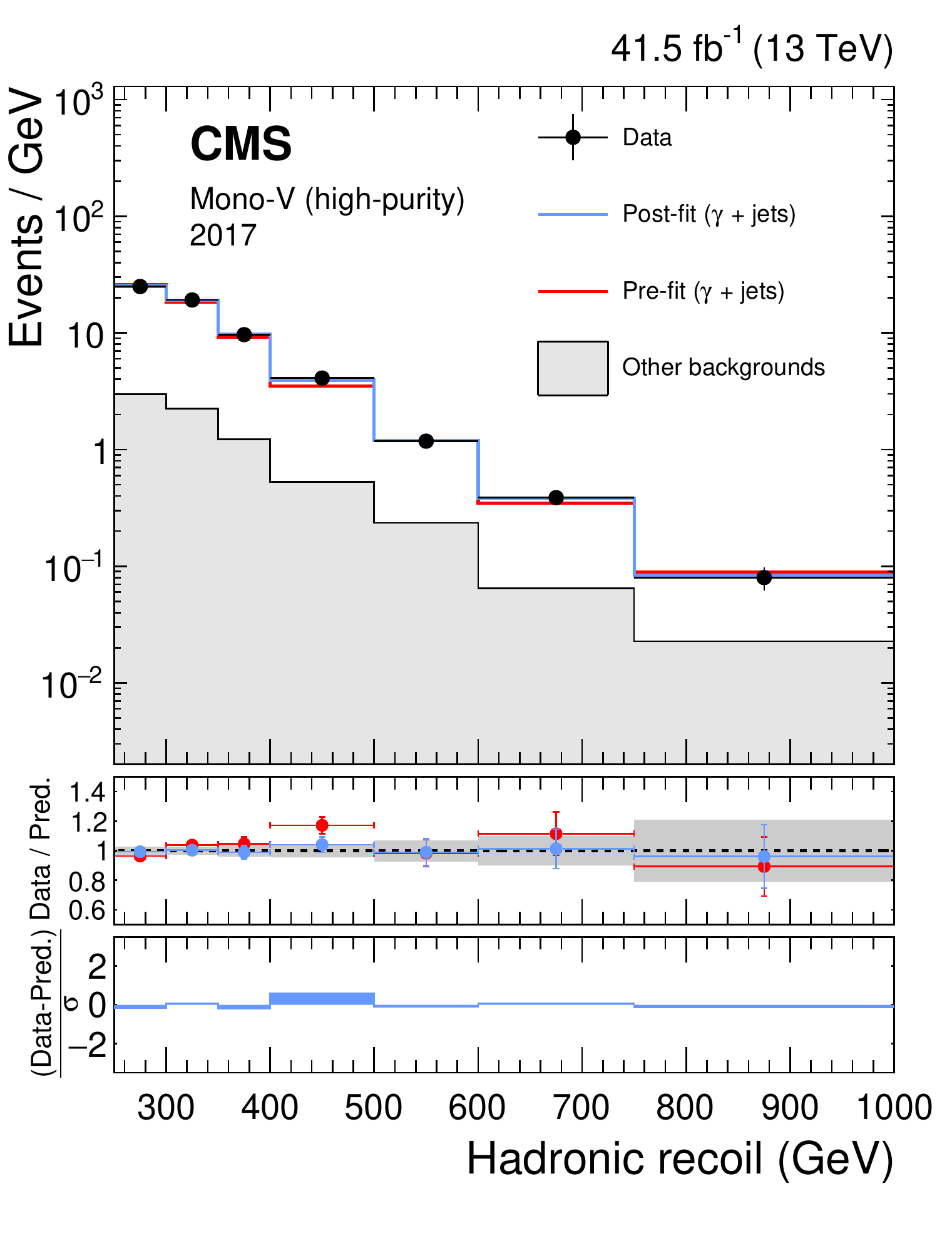}
        \includegraphics[width=0.35\textwidth]{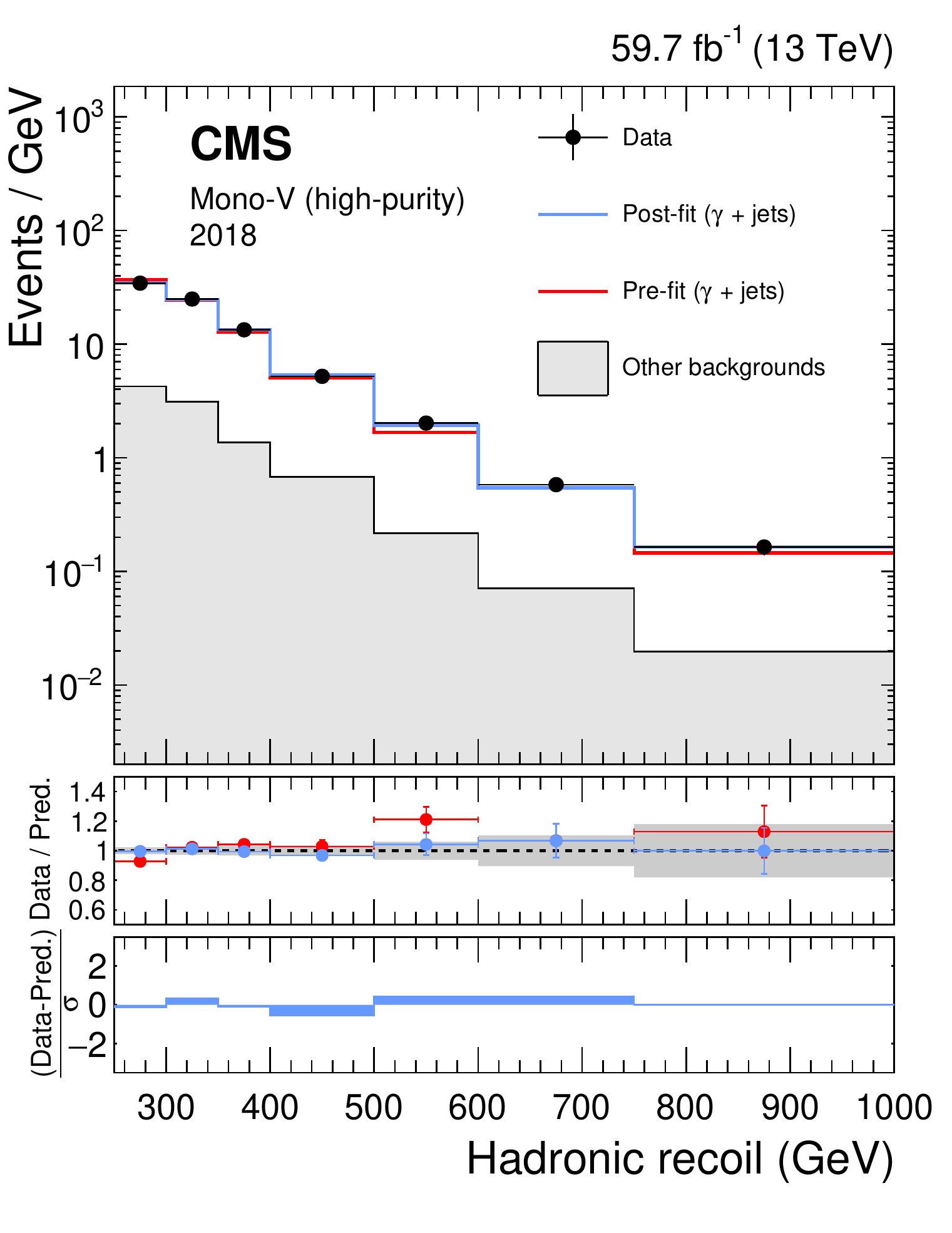}
        \\
        \includegraphics[width=0.35\textwidth]{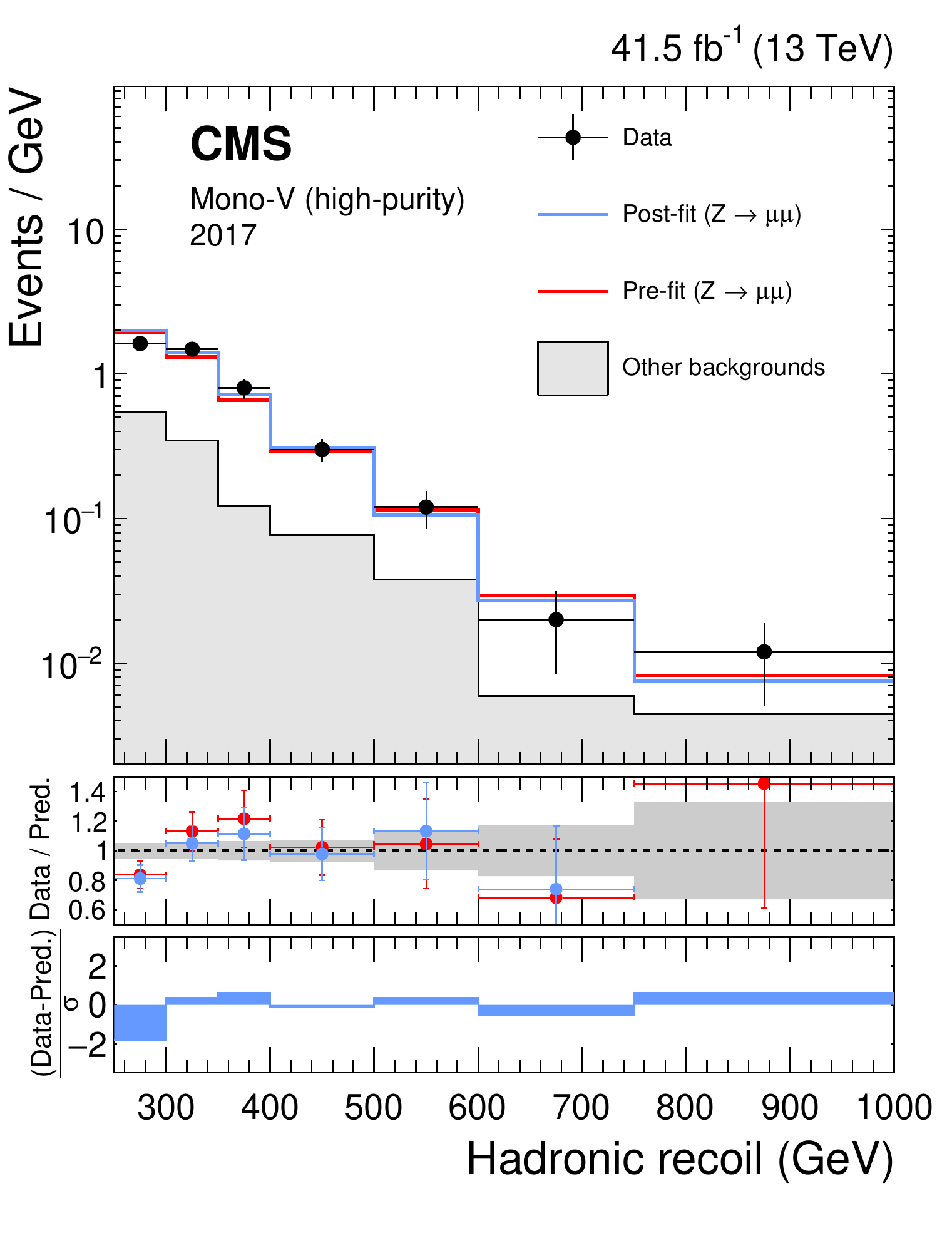}
        \includegraphics[width=0.35\textwidth]{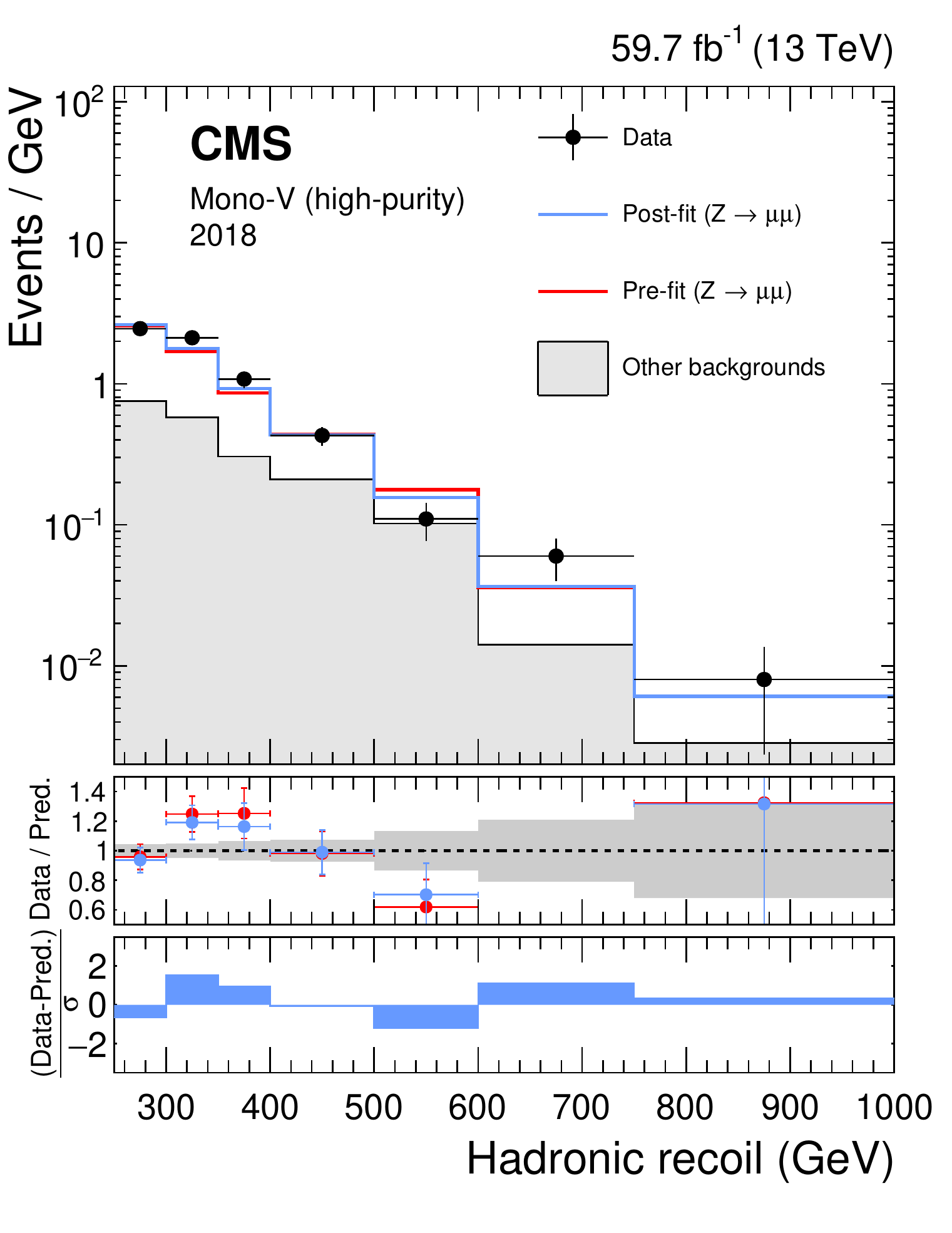}
        \\
        \includegraphics[width=0.35\textwidth]{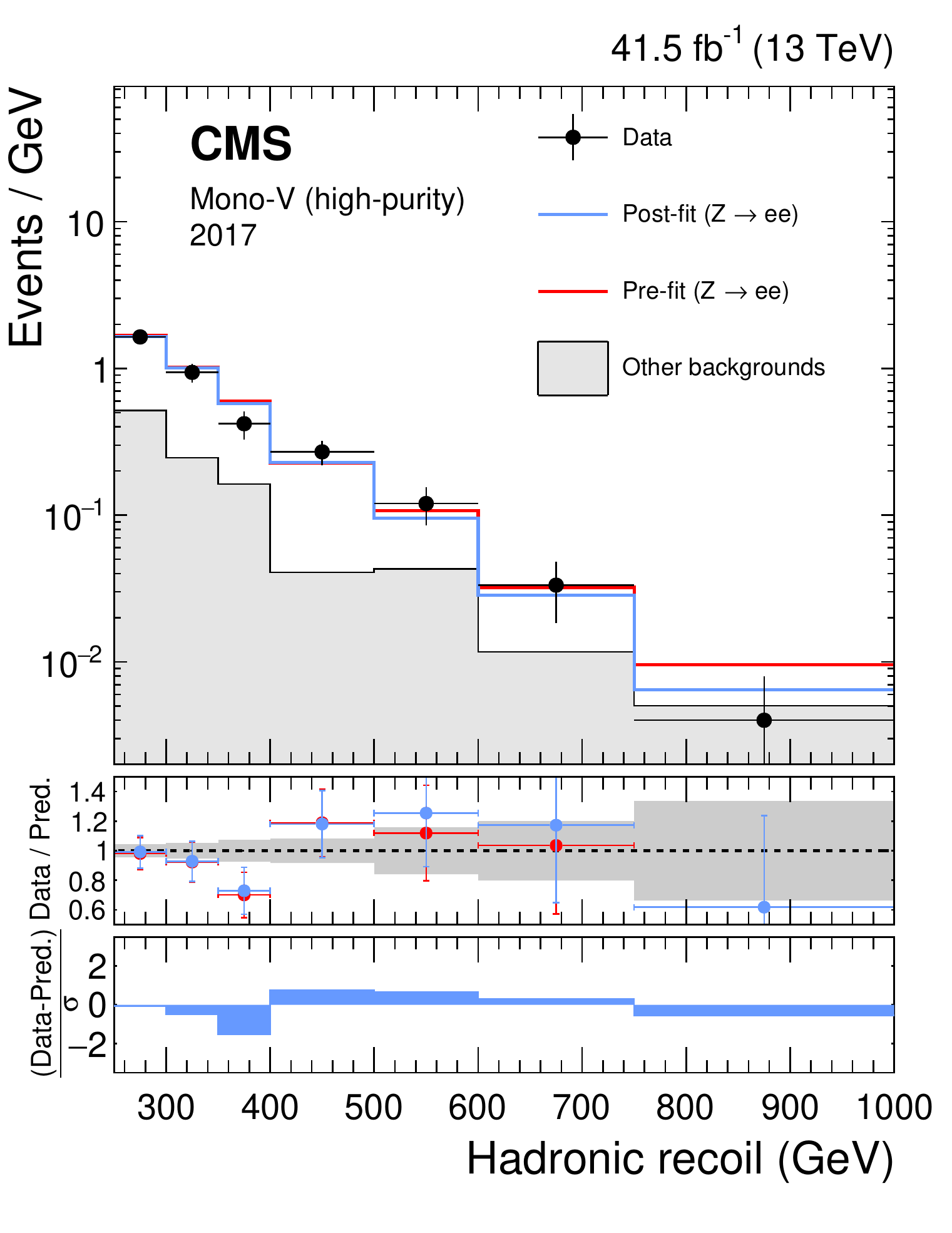}
        \includegraphics[width=0.35\textwidth]{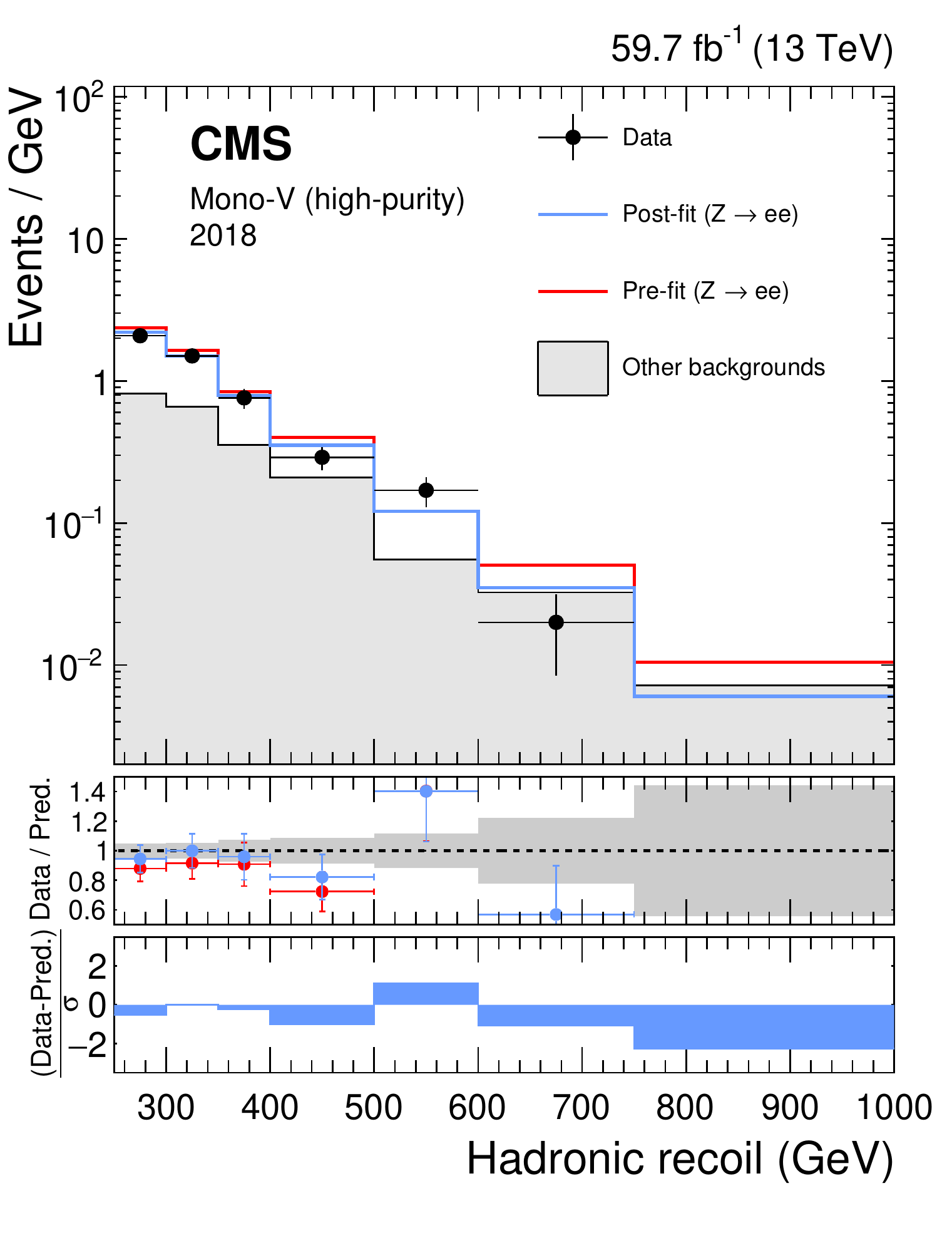}
        \caption{
            Hadronic recoil distributions in the photon (upper), dimuon (middle) and dielectron control regions (lower) in the high-purity mono-V category. The ``Other backgrounds" include QCD multijet production (photon control region), and top quark, diboson, and \Wjets processes (dimuon and dielectron control regions).
         }        \label{fig:postfit_cr_zll_monovtight}

\end{figure*}

\begin{figure*}[hbtp]
    \centering
        \includegraphics[width=0.40\textwidth]{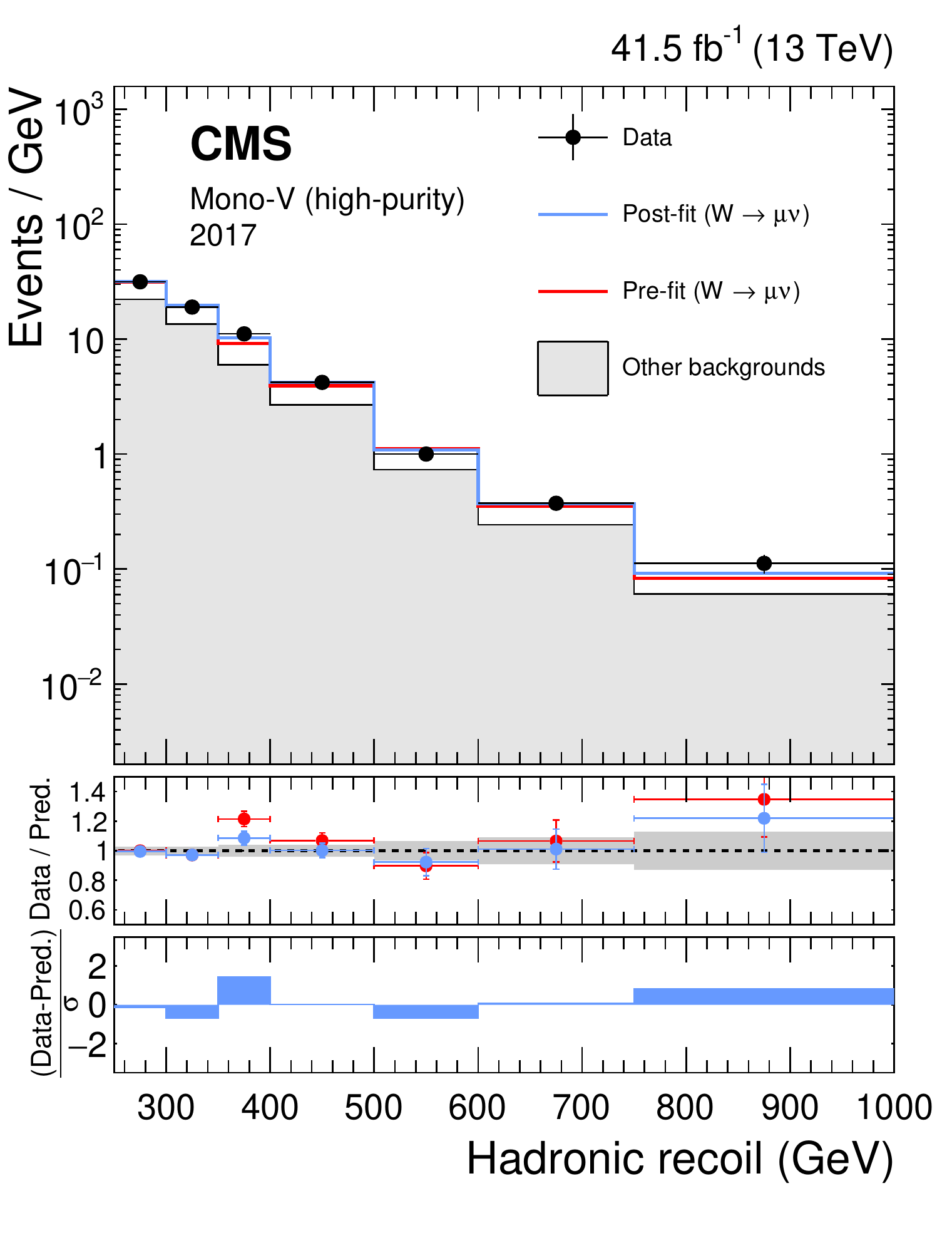}
        \includegraphics[width=0.40\textwidth]{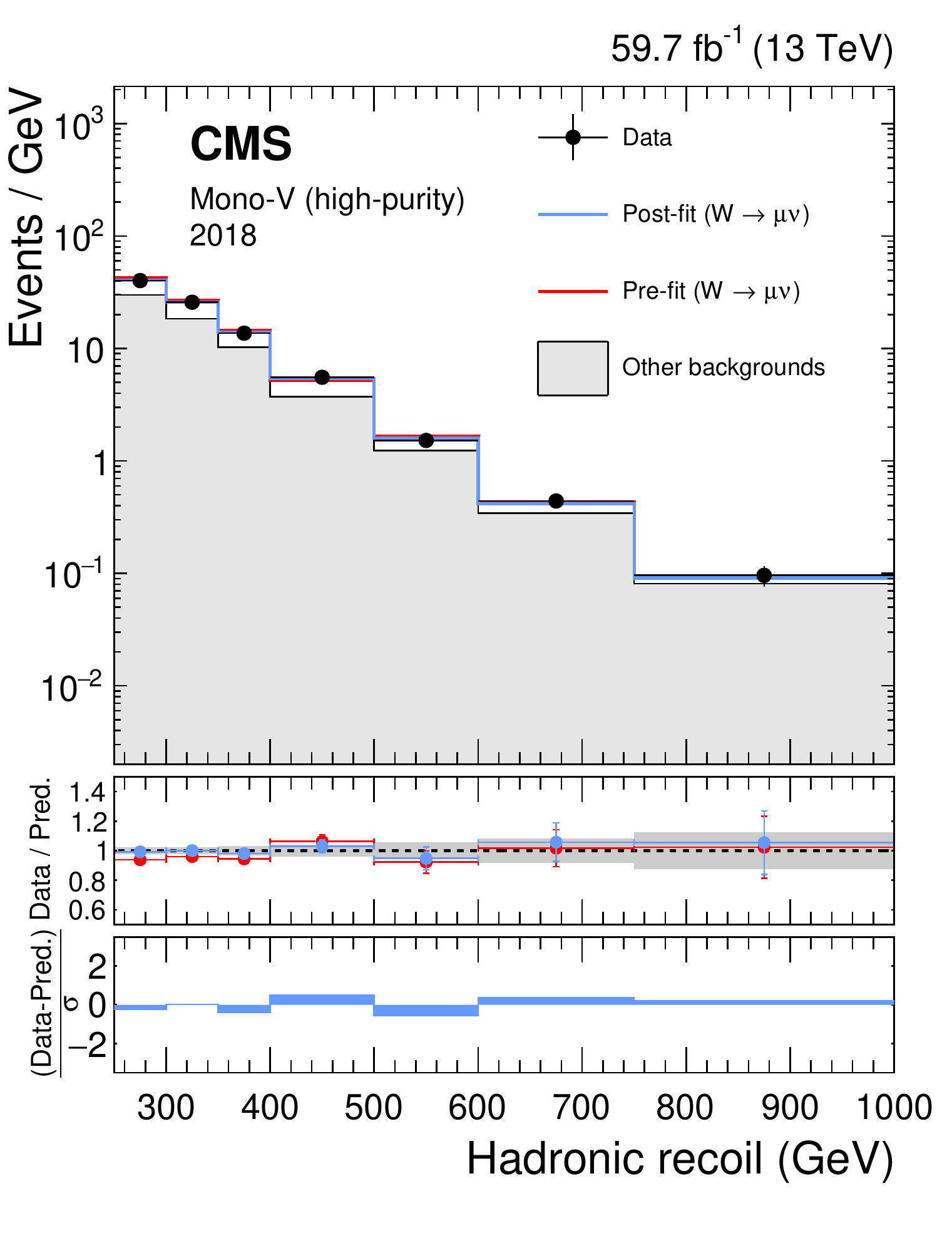}
        \\
        \includegraphics[width=0.40\textwidth]{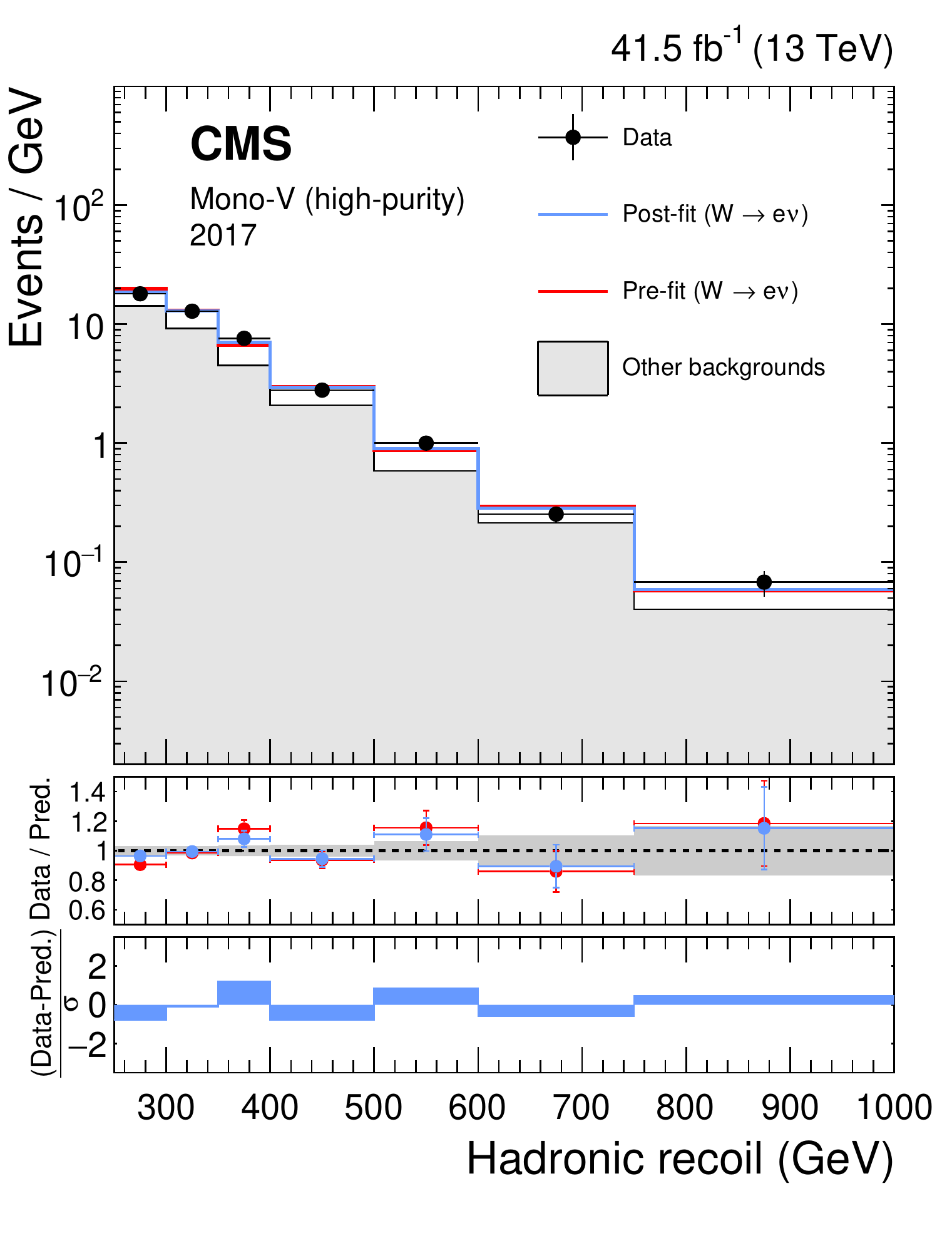}
        \includegraphics[width=0.40\textwidth]{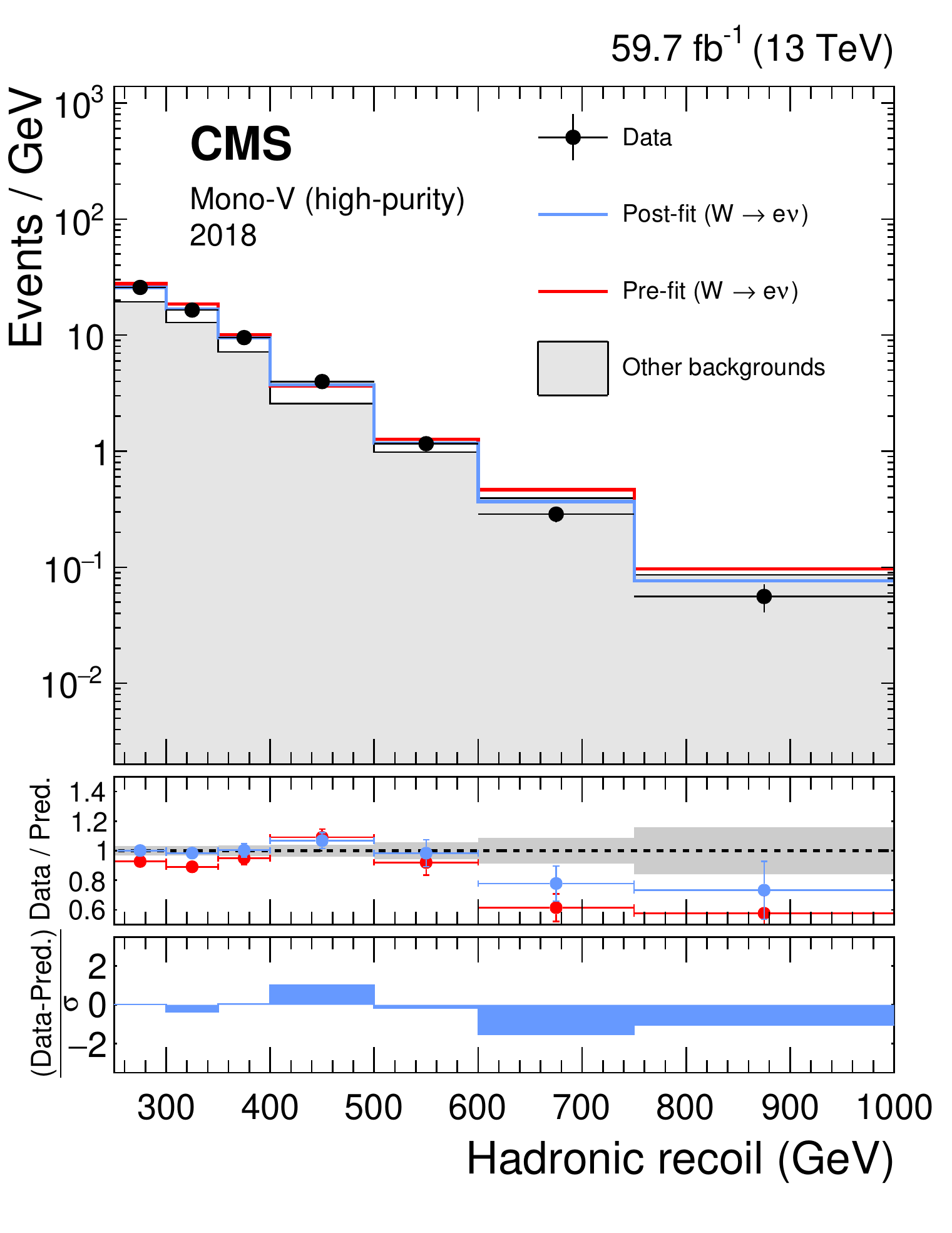}
        \caption{
            Hadronic recoil distributions in the single muon (upper), and single electron regions (lower) in the high-purity mono-V category.
            The ``Other backgrounds" include top quark, diboson, and QCD multijet processes.
        }        \label{fig:postfit_cr_wln_monovtight}

\end{figure*}

The maximum likelihood fit used to determine signal and background contributions is performed including control regions in data. In each of the control regions, the hadronic recoil, defined as the vectorial sum of \ptvecmiss and the transverse components of the selected lepton or photon momenta, is used as a proxy for \ptmiss in the signal region. The recoil distributions for all control regions in all categories are shown in Figs.~\ref{fig:postfit_zll_photon_monojet}--\ref{fig:postfit_cr_wln_monovtight}.

\clearpage

\subsection{Exclusion in the Higgs portal interpretation split by data taking year}
\label{app:appendix_hinv_split}

\begin{figure*}[hbtp]
    \centering
        \includegraphics[width=0.75\textwidth]{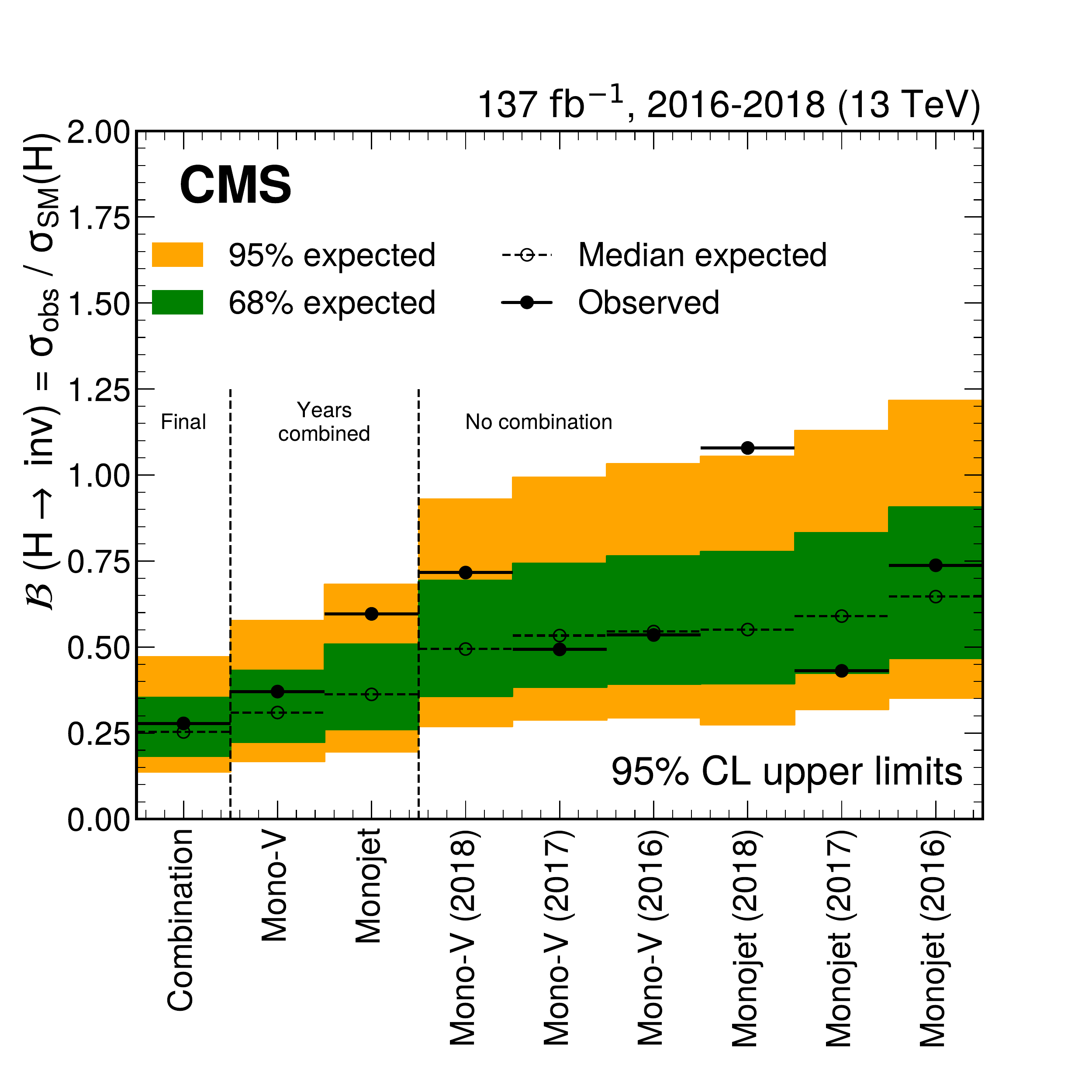}
        \caption{
                Exclusion limits at 95\% \CL on the branching fraction of the Higgs boson to invisible particles.
                The result is shown separately for the monojet and mono-V categories in each data taking year, as well as their combination.
                The final combined limit is $27.8\%$ ($25.3\%$ expected).
                }
        \label{fig:supp_hinv}

\end{figure*}

The constraints placed on decays of the Higgs boson to invisible particles in each data taking year and category are summarized in Fig.~\ref{fig:supp_hinv}. For each individual category and year, separate ML fits are performed, leading to independent best fit values of the nuisance parameters, as well as the signal strength.

\clearpage

\subsection{Coupling limits in a simplified DM model with a vector mediator}
\label{app:appendix_dmsimp_vector_coupling}
\begin{figure*}[hbtp]
    \centering
        \includegraphics[width=0.49\textwidth]{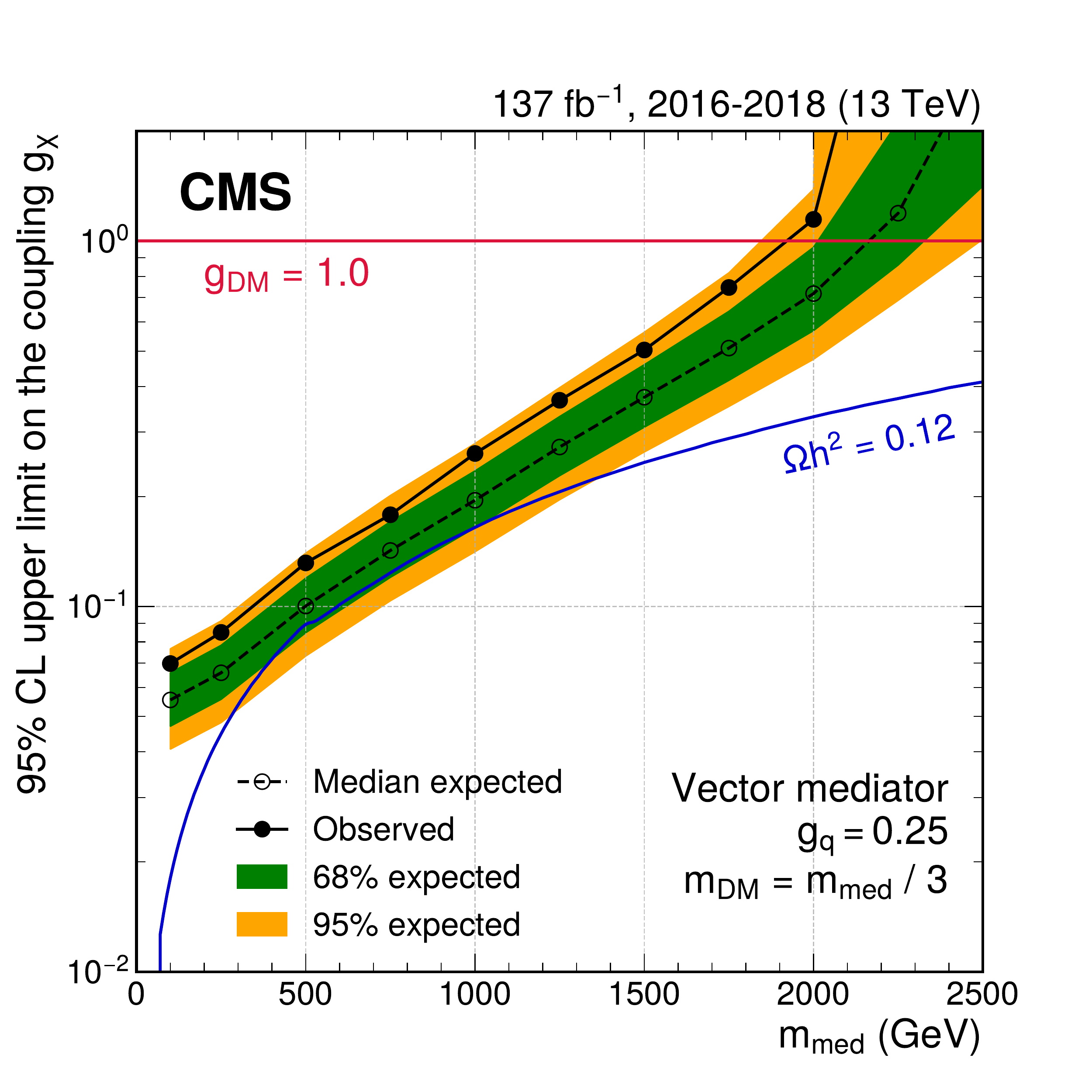}
        \includegraphics[width=0.49\textwidth]{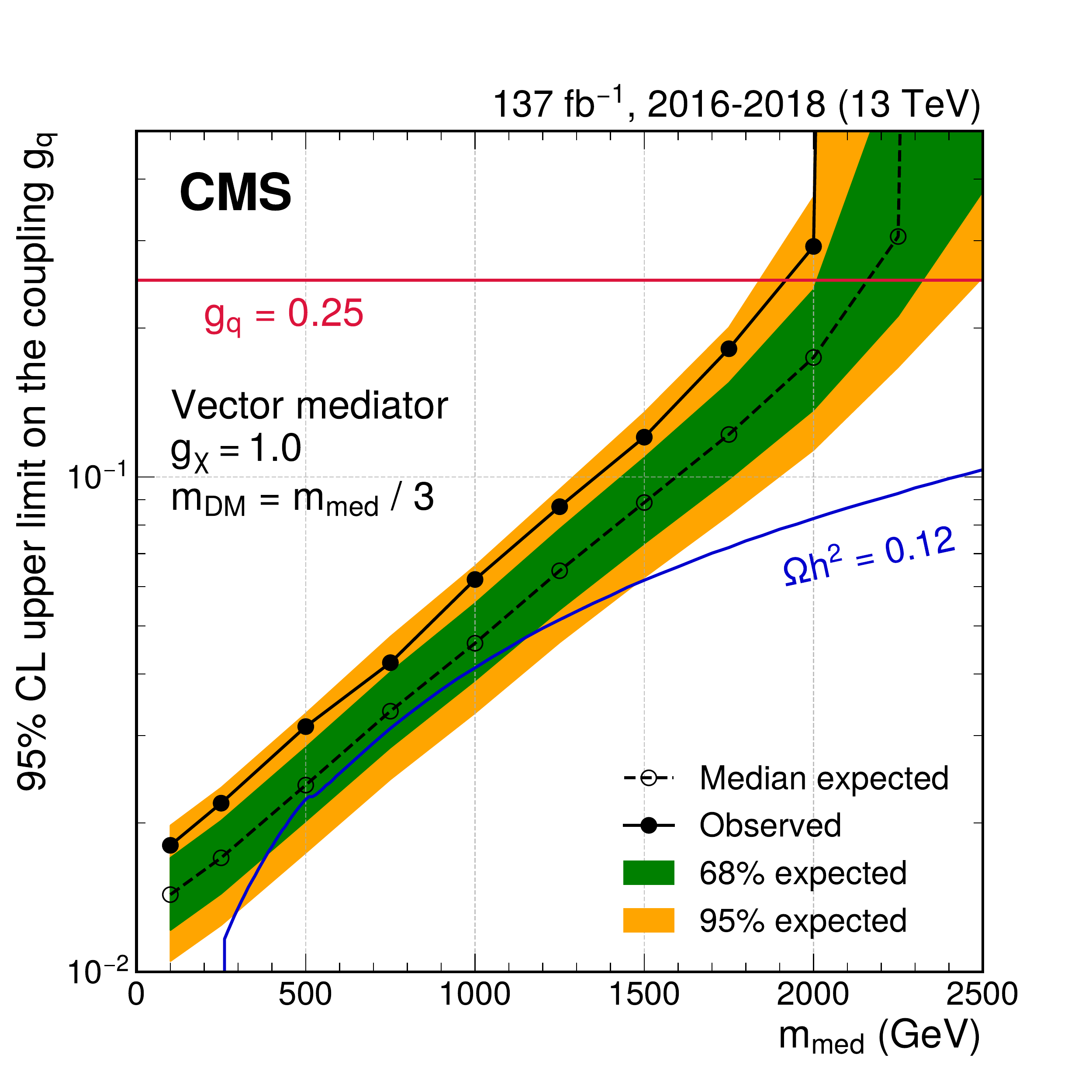}
        \\
        \caption{
            Exclusion limits at 95\% \CL on the couplings \gchi (left) and \gq (right) for a vector mediator. In each panel, the result is shown as a function of the mediator mass $\mmed$, and the mass of the DM candidate is fixed to $\mdm=\mmed/3$. In either case, only one coupling is varied, and the respective other coupling is fixed at its default value ($\gq=0.25$, $\gchi=1.0$). The blue solid line indicates the parameter combinations for which the simplified model reproduces the observed DM relic density. Around $\mdm\approx m_{\text{top}}$, corresponding to $\mmed\approx500\GeV$, DM annihilation into top quarks becomes possible, leading to a shift in the relic density.
        }
        \label{fig:supp_dmsimp_spin1_vector}

\end{figure*}

Coupling limits for a vector mediator are derived in the same manner as the result for an axial mediator shown in Fig.~\ref{fig:results_dmsimp_spin1_coupling}. The  result is shown in Fig.~\ref{fig:supp_dmsimp_spin1_vector}.

\clearpage

\subsection{Two-dimensional exclusion in the simplified DM model with pseudoscalar mediator}
\label{app:appendix_dmsimp_pseudo_2d}
\begin{figure*}[hbtp]
    \centering
        \includegraphics[width=0.75\textwidth]{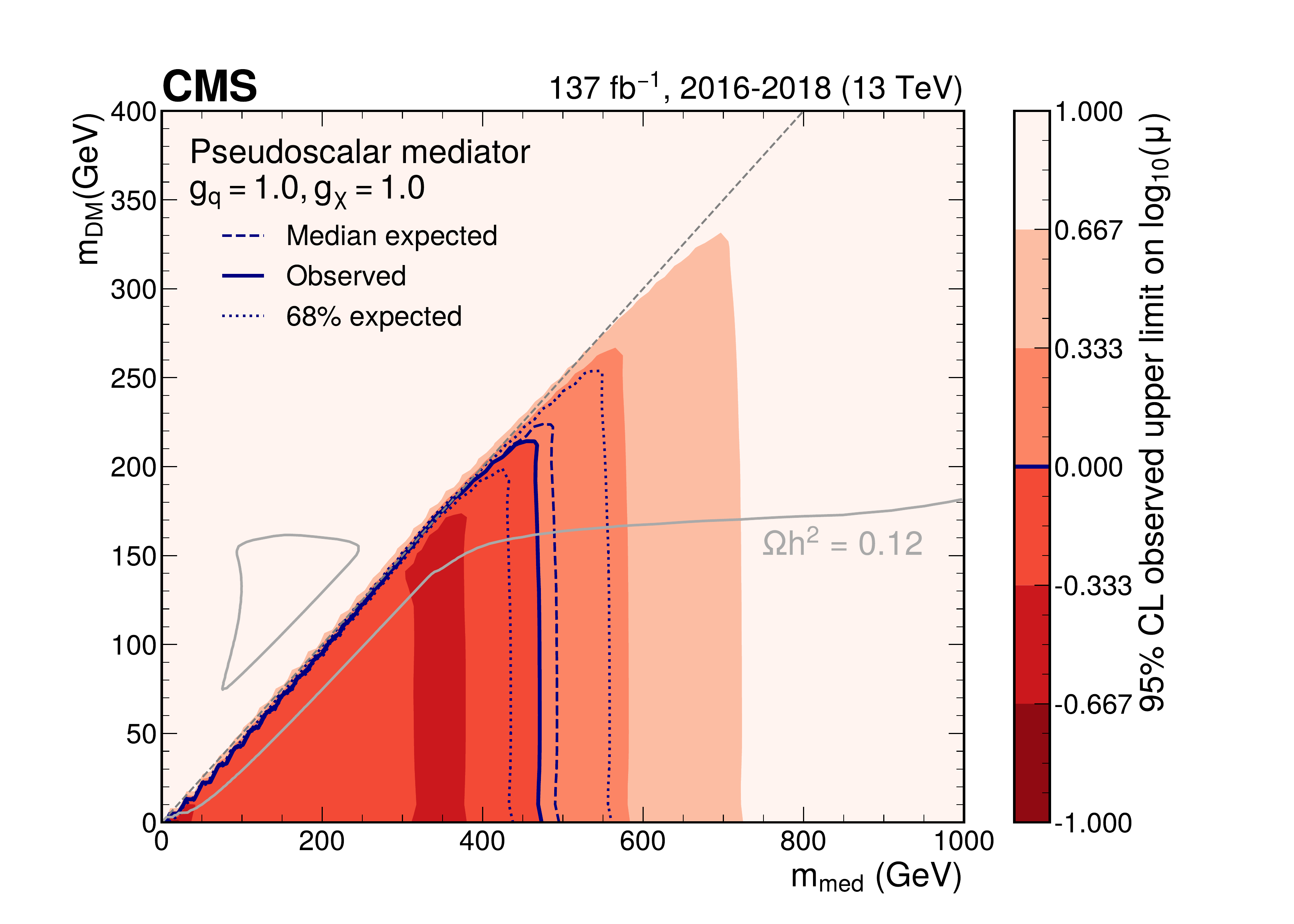}
        \caption{
                Exclusion limits at 95\% \CL on the signal strength $\mu=\sigma/\sigma_\text{theo}$ in the \mmed-\mdm plane for coupling values of $\gq=\gchi=1.0$ and a pseudoscalar mediator.
                The blue solid line indicates the observed exclusion boundary $\mu=1$. The blue dashed and dotted lines represent expected exclusion and the 68\% \CL interval of the expected boundary, respectively. Parameter combinations with larger values of $\mu$ (indicated by a darker shade in the color scale) are excluded. The gray dashed line indicates the diagonal $\mmed=2\mdm$, above which only off-shell mediator production contributes to the jet+\ptmiss final state. The gray solid lines represent parameter combinations for which the simplified model reproduces the observed DM relic density in the universe under the assumption of a thermal freeze-out mechanism~\cite{Albert:2017onk,Aghanim:2018eyx}.
                }
        \label{fig:supp_pseudoscalar_2d}

\end{figure*}

The exclusion limits in the \mmed-\mdm plane for the simplified model with a pseudoscalar mediator are shown Fig.~\ref{fig:supp_pseudoscalar_2d}.

\clearpage

\subsection{Table of exclusion limits in the ADD model}
\label{app:appendix_add_table}
The lower limits on the fundamental Planck mass $\MD$ as a function of the number of extra dimensions are shown in Table~\ref{tab:supp_add}.

\begin{table*}[htb]
    \topcaption{Lower limits at $95\%$ \CL on the fundamental Planck mass $\MD$ in TeV as functions of the number of extra dimensions $d$.}
    \centering
        \begin{tabular}{r r r r}
            $d$ &  & \multicolumn{2}{c}{Lower limit on \MD (\TeVns) }            \\
                         &  & Expected                                                   & Observed \\
\hline
            2            &  &  12.2                                                          & 10.7     \\
            3            &  &  9.0                                                           & 8.0      \\
            4            &  &  7.4                                                           & 6.8      \\
            5            &  &  6.6                                                           & 6.0      \\
            6            &  &  5.9                                                           & 5.5      \\
            7            &  &  5.6                                                           & 5.2      \\
        \end{tabular}
        \label{tab:supp_add}

\end{table*}

\ifthenelse{\boolean{cms@external}}{}{
\clearpage
\numberwithin{table}{section}
\numberwithin{figure}{section}
\section{Supplemental material\label{app:suppMat}}
\subsection{Comparison with direct-detection experiments}

The constraints placed on the $s$-channel simplified models imply bounds on the interaction cross section between DM candidates and nuclei. The fixed-coupling exclusion curves in the \mmed-\mdm plane are translated point-by-point using the formulae described in Ref.~\cite{Boveia:2016mrp}, which depend on the coupling choices $\gq=0.25$ and $\gchi=1.0$ and on the specific signal model. The resulting curves in the \mdm-$\sigma_\text{DM-nucleon}$ plane are compared to the results from direct-detection  (DD) experiments in Fig.~\ref{fig:supp_dmsimp_spin1_dd}. Qualitatively, the results from this search depend on \mdm only weakly (as long as $\mdm<\mmed/2$), leading to stringent constraints also at low values of \mdm. The sensitivity of most DD experiments is limited in this regime as the small value of \mdm translates into a reduced signal-to-noise ratio relative to the case of more massive DM. Depending on the mediator type, the resulting couplings between DM particles and nuclei are either spin dependent (axial-vector) or independent (vector). In the spin-dependent case, the sensitivity of DD experiments is limited relative to collider searches as the DM-nucleus scattering is no longer coherent.

\begin{figure*}[hbtp]
    \centering
        \includegraphics[width=0.47\textwidth]{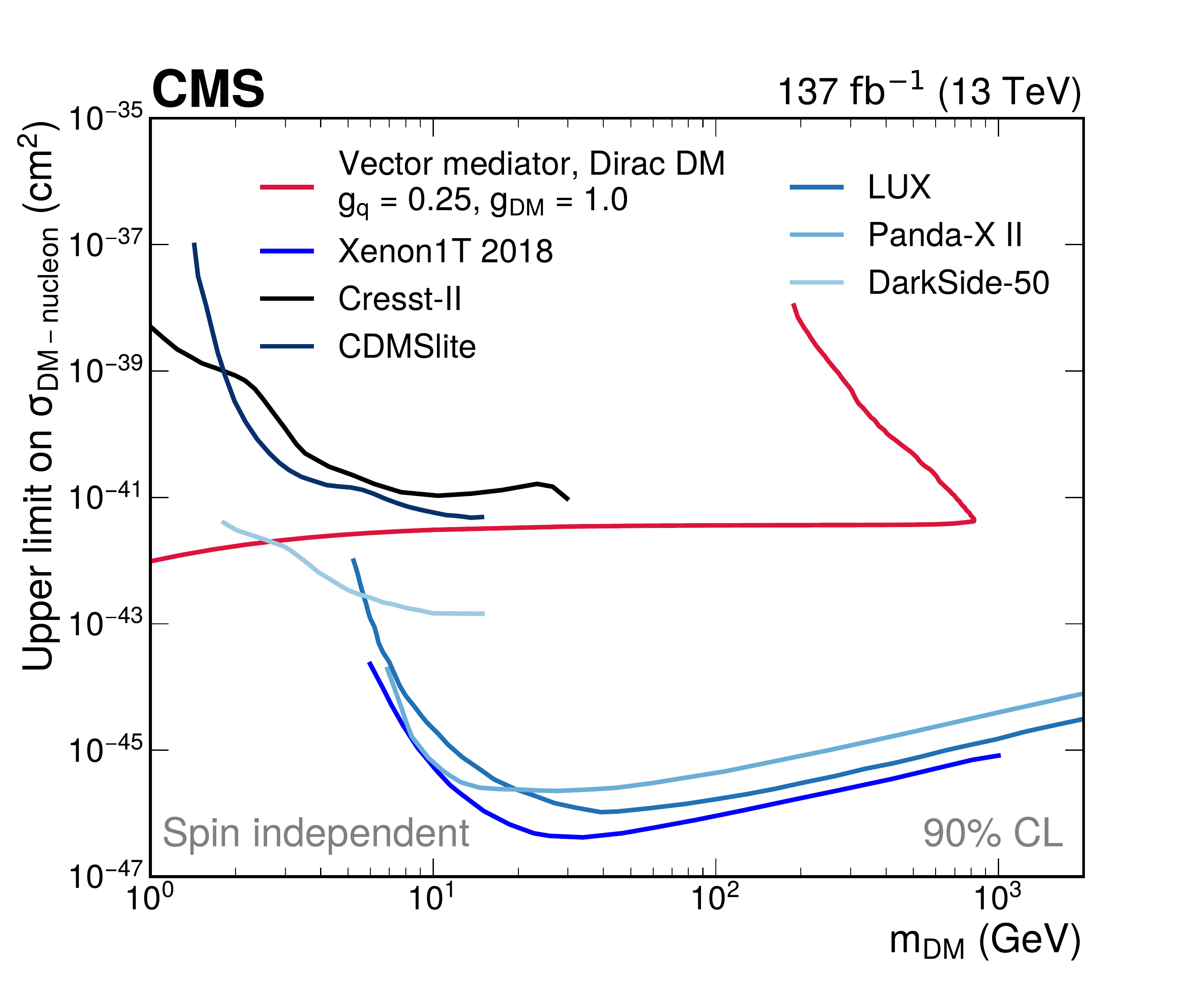}
        \includegraphics[width=0.47\textwidth]{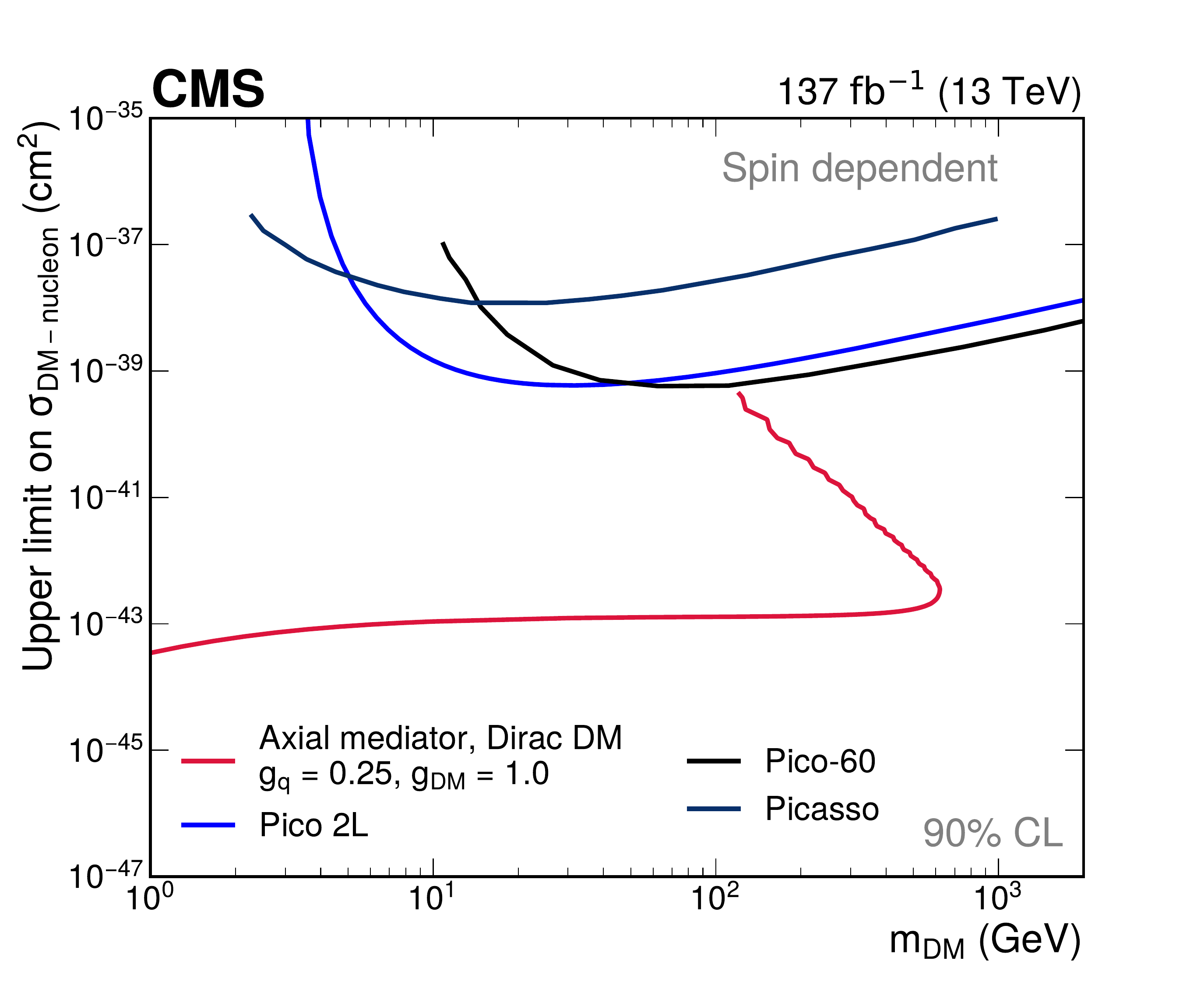}

        \caption{
                Comparison of the simplified model constraints from this search (red line) to results from direct-detection experiments (blue lines).
                The comparison is shown separately for the vector (left) and axial-vector (right) mediators, which translate into spin-independent and
                spin-dependent DM-nucleon couplings, respectively. In the case of spin-independent couplings,
                results from CRESST-II~\cite{Angloher:2015ewa}, CDMSlite~\cite{Agnese:2015nto}, LUX~\cite{Akerib:2016vxi}, DarkSide-50~\cite{Agnes:2018ves}, XENON1T~\cite{Aprile:2018dbl}, and Panda-X II~\cite{Wang:2020coa} are shown for comparison. For spin-dependent couplings, PICO-2L~\cite{Amole:2016pye}, PICASSO~\cite{Behnke:2016lsk}, and PICO-60~\cite{Amole:2019fdf} limits are displayed.
                }
        \label{fig:supp_dmsimp_spin1_dd}

\end{figure*}

\clearpage

\subsection{Distributions of jet tagging variables}

The identification of the \Vqq candidate large-radius jets relies on the SD-corrected mass of a given jet, as well as on the classifier score from the \textsc{DeepAK8} neural network. The ability of these quantities to separate genuine \Vqq candidates from background with jets originating from QCD radiation is demonstrated in Fig.~\ref{fig:supp_ak8}.

\begin{figure*}[hbtp]
    \centering
        \includegraphics[width=0.7\textwidth]{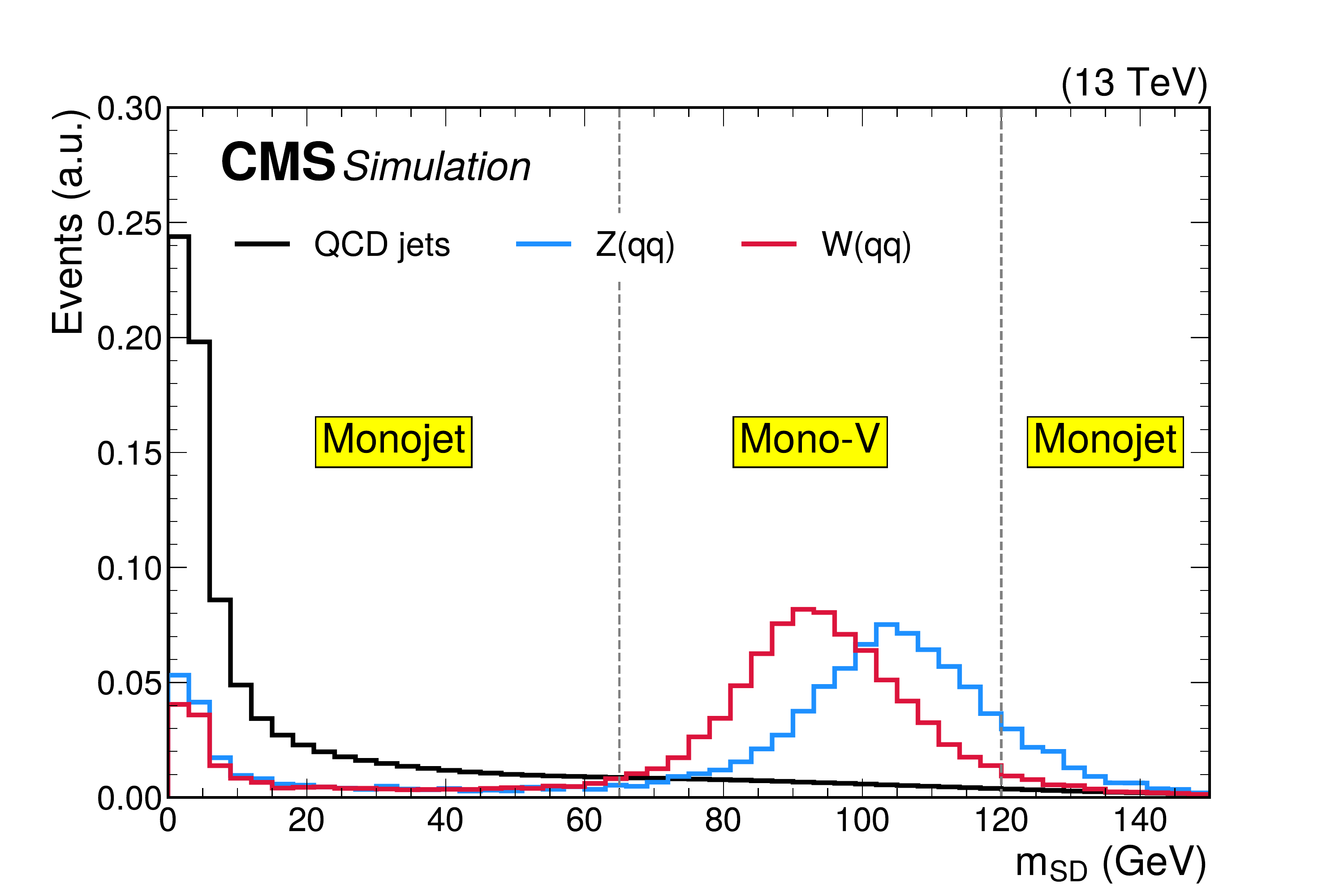}
        \includegraphics[width=0.7\textwidth]{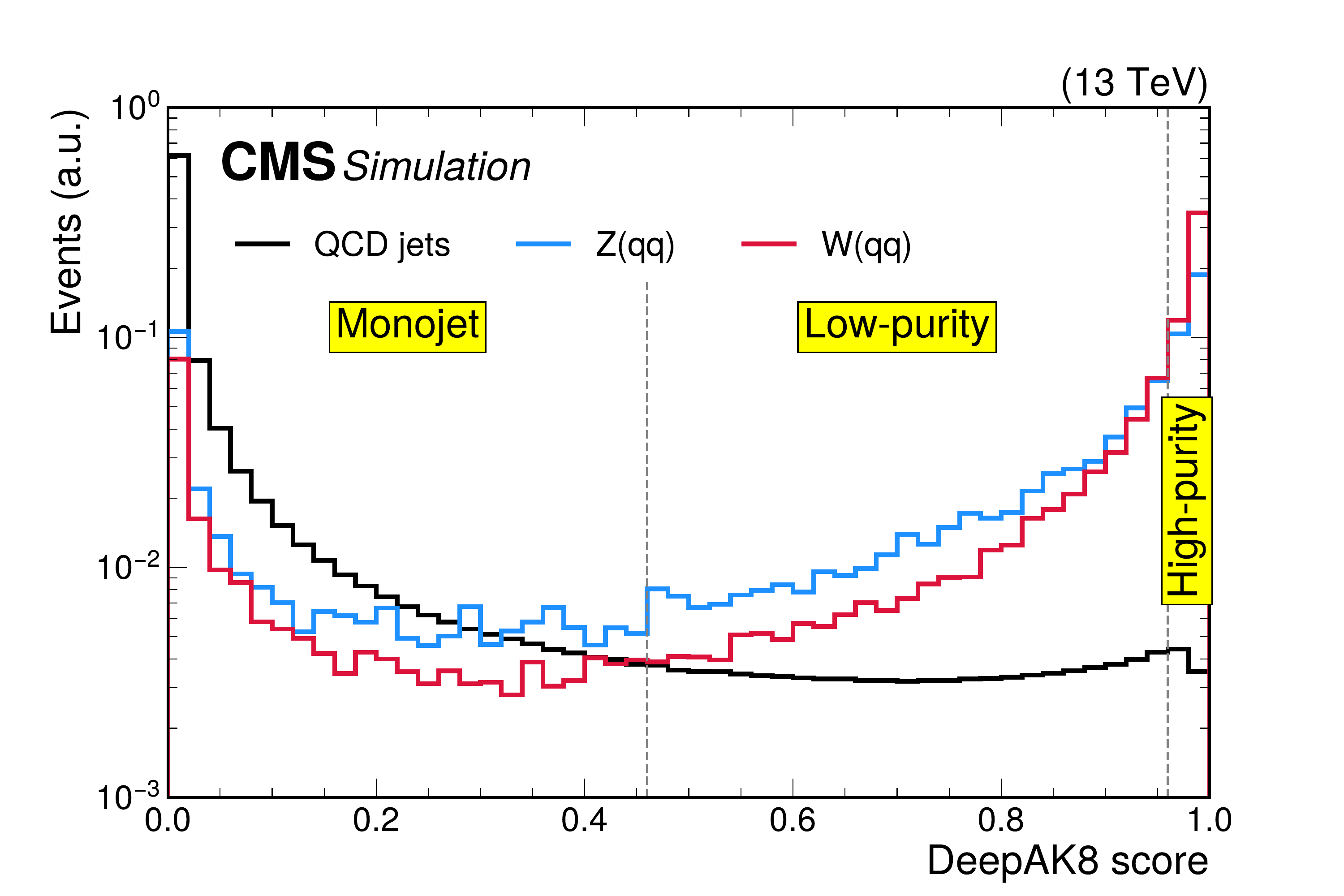}

        \caption{
                Distributions of the variables used for the identification of \Vqq candidate jets. The upper and lower panels show the SD-corrected jet mass and the \textsc{DeepAK8} classifier value, respectively. In each panel, the distributions are shown for the \Zvv background, as well as the $\PW\PH$(inv.) and $\PZ\PH$(inv.) signals. The distributions are shown after applying the mono-V signal region selection, with the exception of the requirements on the two variables shown here. Vertical dashed lines indicate the acceptance boundaries of different regions.
                }
        \label{fig:supp_ak8}

\end{figure*}

\clearpage

\subsection{Large-radius jet tagging efficiencies for reinterpretation}

To aid reinterpretation, the efficiency is calculated for the combination of the SD mass and \textsc{DeepAK8} tagging requirements.
The efficiency is calculated in simulated events passing the full signal mono-V region selections, with the exception of the requirements on \msd and the tagging score. Correction factors accounting for the differences between data and simulation are included. The efficiencies are shown in Fig.~\ref{fig:supp_ak8_eff}.

Efficiencies are provided for the low- and high-purity tagging requirements. We note that for the low-purity tagger, the overlap removal with the high-purity category is already done.

The efficiency is calculated separately for the AK8 jets matching a generator-level $\PZ$ boson, $\PW$ boson, or not matching either (``QCD jet''). A jet is considered to be matching a boson if their angular separation $\Delta R = \sqrt{\smash[b]{(\Delta\eta)^2+(\Delta\phi)^2}}$ is less than 0.8. In order to apply the efficiencies to simulated events, one should first apply all other selection criteria, except for the ones based on the jet mass or other substructure variables. Depending on the matching status of the jet, the respective efficiency evaluated at the \pt of the jet should be then applied as an event weight.

\begin{figure*}[hbtp]
    \centering
        \includegraphics[width=0.45\textwidth]{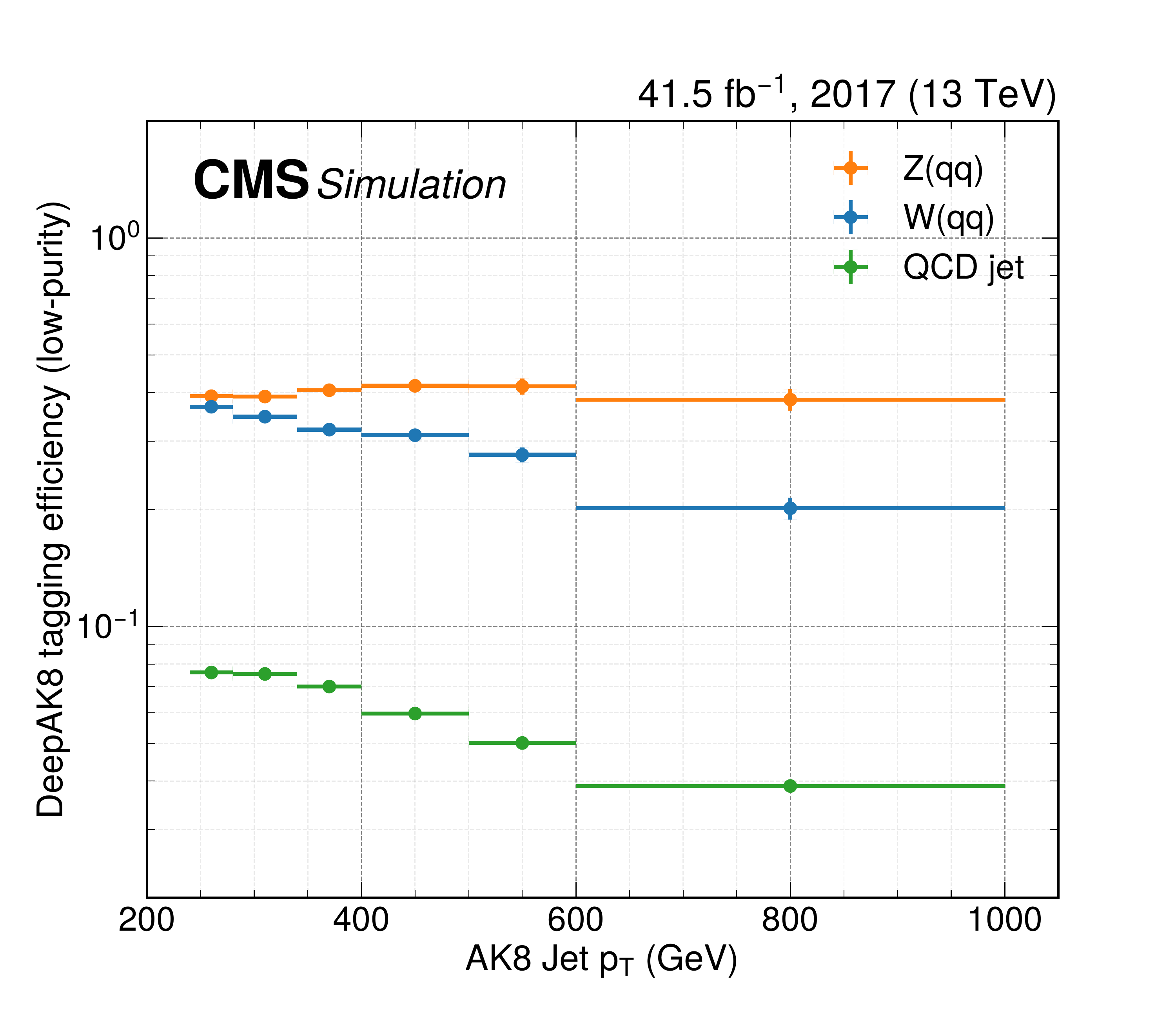}
        \includegraphics[width=0.45\textwidth]{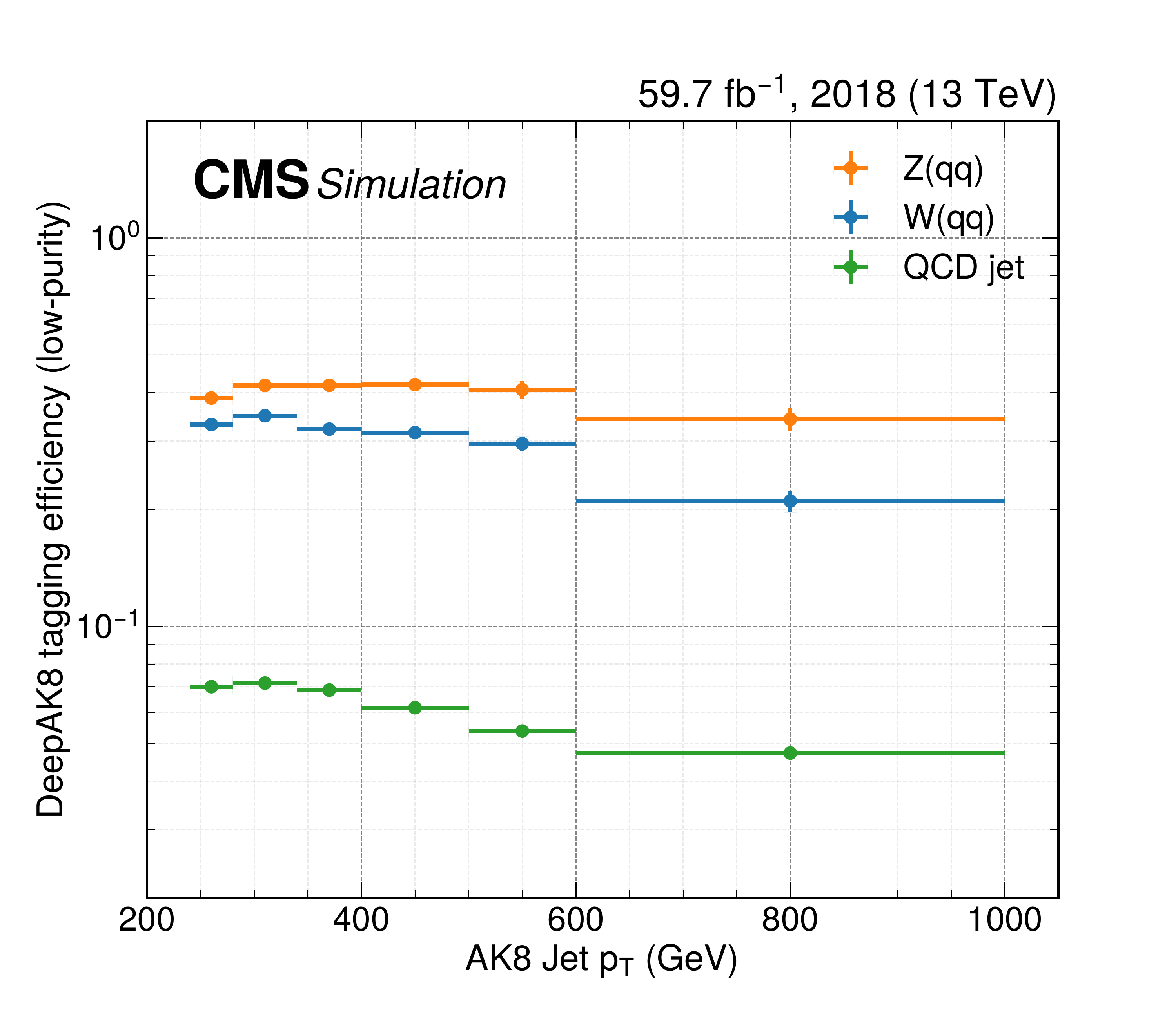}
        \includegraphics[width=0.45\textwidth]{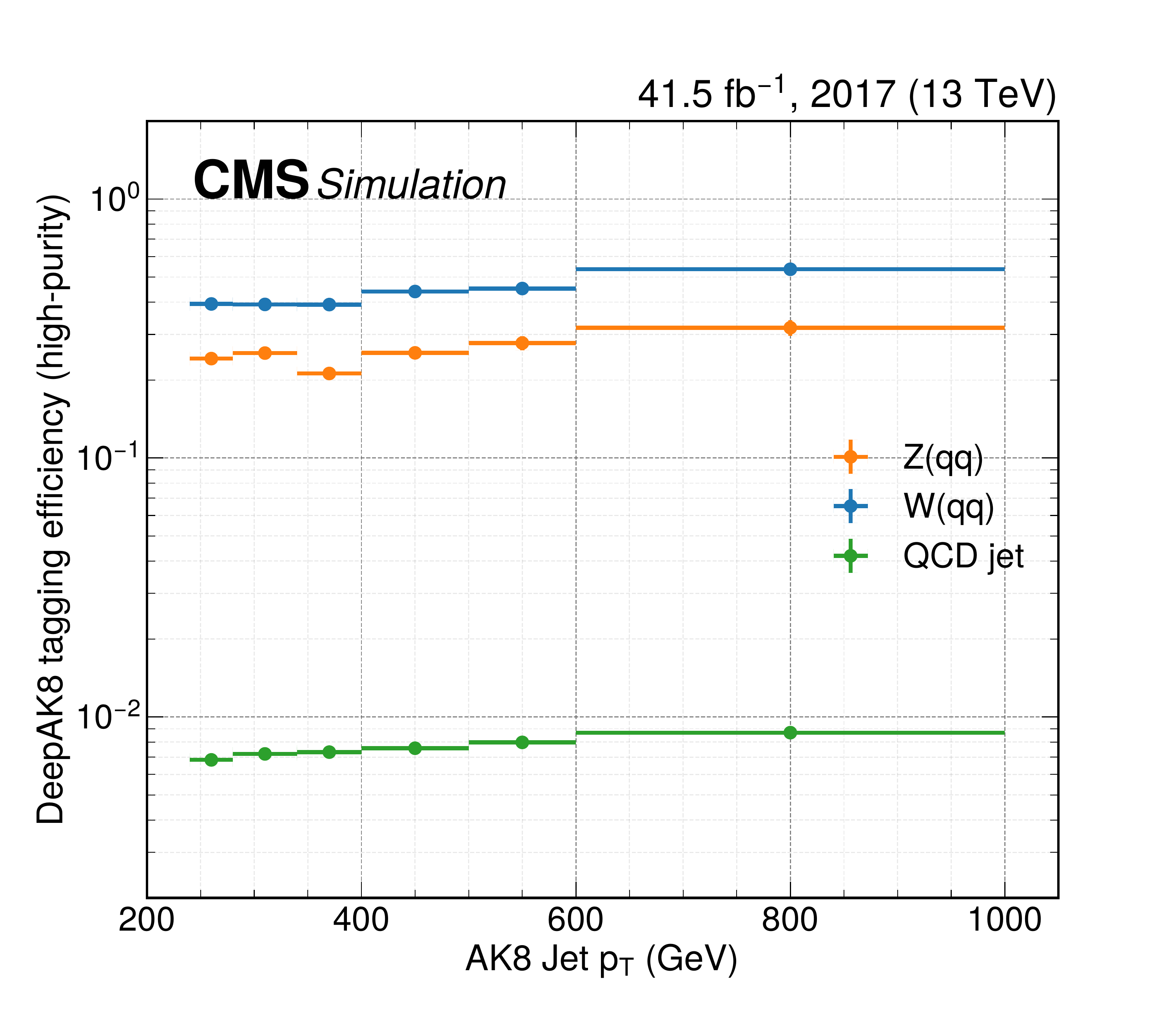}
        \includegraphics[width=0.45\textwidth]{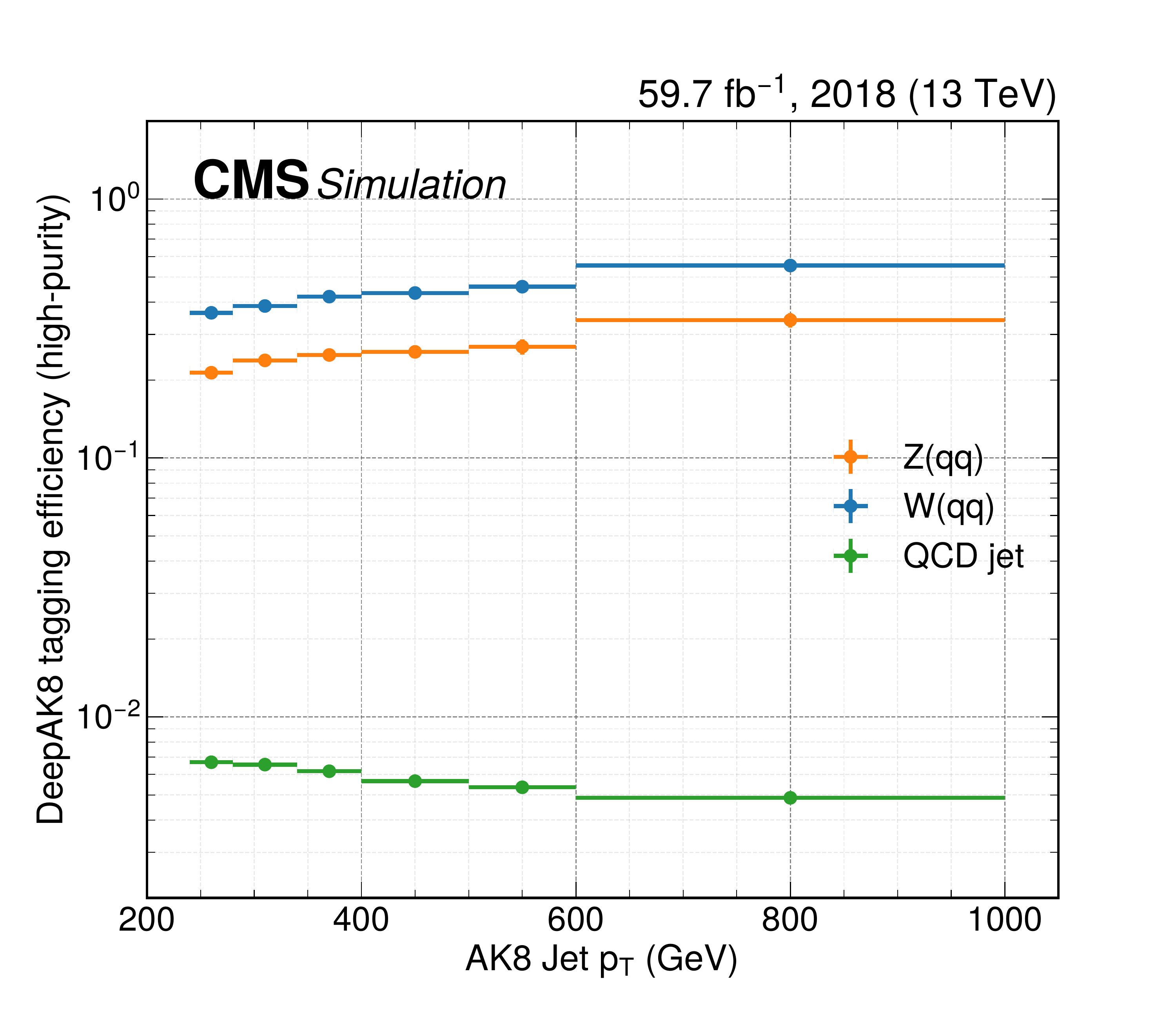}

        \caption{
                Large-radius jet tagging efficiencies for use in reinterpretation of the results. The efficiencies are shown separately for the low- and high-purity selections in the upper and lower panels, respectively, and for the 2017 (left) and 2018 (right) data taking periods. The efficiencies include the effect of the \textsc{DeepAK8} tagger, as well as the SD-corrected mass requirement. In each panel, individual curves represent the efficiency for different types of jets, based on whether the jets are matched to a generator-level $\PW$ boson, $\PZ$ boson, or neither (``QCD jet''). Simulation-to-data corrections are included.
                }
        \label{fig:supp_ak8_eff}

\end{figure*}

\clearpage
\subsection{Monojet \ptmiss distribution for the full data set}

In the statistical analysis described in this paper, data from different data taking periods are sorted into separate bins. In Fig.~\ref{fig:supp_monojet_ptmiss}, the total \ptmiss distribution for all data taking years is shown, which is the bin-by-bin sum of the distributions in the individual years.

\begin{figure*}[hbtp]
    \centering
        \includegraphics[width=0.75\textwidth]{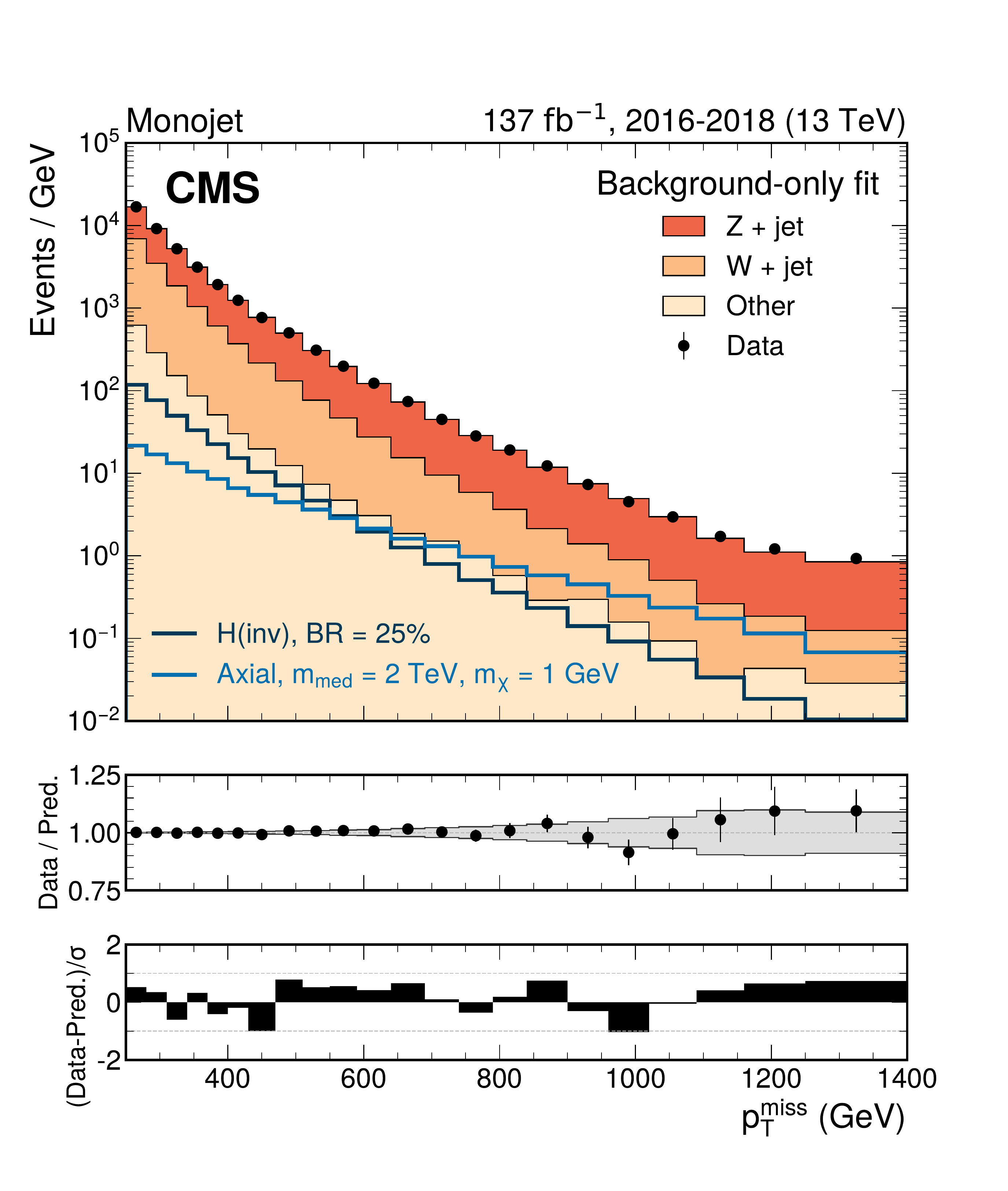}
        \caption{
                Distribution of \ptmiss in the monojet category. The distribution is shown including the contributions from all data taking years. It does not directly represent the input to the statistical method, which instead relies on distributions separated by data taking year. The background estimate is obtained from the background-only fit to all data taking periods, regions, and categories, including mono-V. The total uncertainty in the background estimate, shown as a gray band in the middle panel, takes into account all relevant correlations. The signal templates from the Higgs portal and axial-vector mediator hypotheses are overlaid (solid lines). In both cases, contributions from all production modes are taken into account.
                }
        \label{fig:supp_monojet_ptmiss}

\end{figure*}

\clearpage

\subsection{Analysis implementation in \textsc{MadAnalysis}}

The \madanalysis package is a framework for the reinterpretation of existing analyses in terms of arbitrary new physics models~\cite{Conte:2014zja}.
The framework provides the infrastructure for the implementation of event selections that can be run over simulated signal events.
Once an implementation is available, it is indexed in a public database that allows users to automatically download and execute it~\cite{Dumont:2014tja}.

In order to promote this analysis for reinterpretation, we implement the selection for the monojet category of this analysis in \madanalysis.
A total of 66 analysis regions are defined, with each of the regions representing one recoil bin in one data taking year.
The selections applied for the 2016 and 2017 data sets are identical, and additional criteria are applied to the 2018 data set,
where mitigation requirements are used because of a localized problem in the hadron calorimeter.

In order to validate the implementation, generator-level information from the simulated signal samples is fed into the \delphes framework,
which performs fast parameterized event simulation~\cite{deFavereau:2013fsa}. The \madanalysis implementation is then run based on the \delphes output,
and the final yields per signal region bin are compared to the signal prediction obtained from the CMS analysis framework.

The comparison is made using signal samples for the ADD interpretation, which are generated using \PYTHIA, and are therefore relatively easy to reproduce.
The resulting comparison of the final signal templates is shown in Fig.~\ref{fig:reinterpretation_madanalysis_closure}
for a representative choice of parameter points. It is found that the \delphes/\madanalysis-based result agrees with the CMS result to better than $20\%$ in every bin.
In most bins, the agreement is at the $10\%$ or better level. While only a few parameter points are shown here, it has been verified
that the agreement is similar for the full range of parameters. The level of agreement observed here is sufficiently good to enable reliable reinterpretation.

\begin{figure*}[hbtp]
    \centering
        \includegraphics[width=0.49\textwidth]{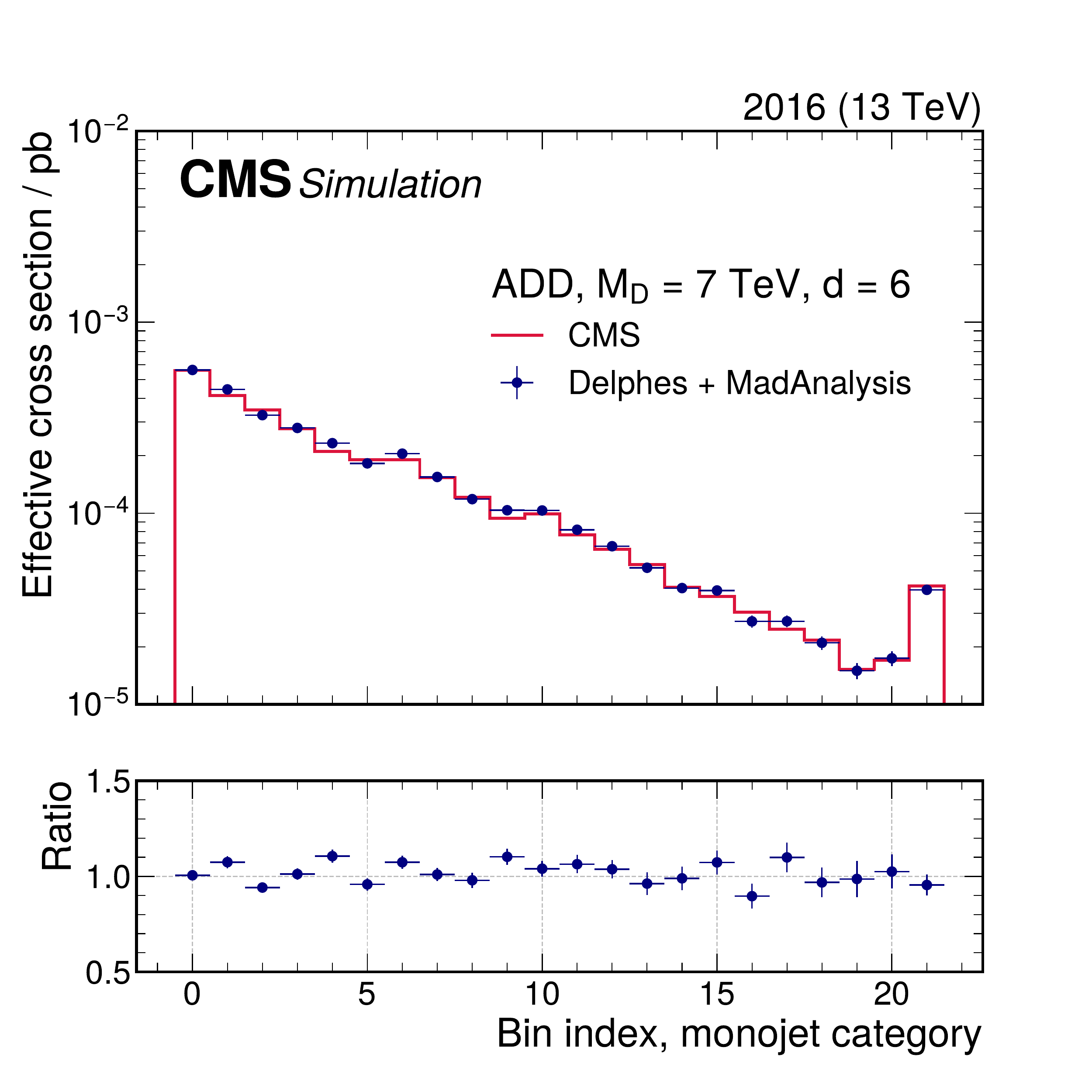}
        \includegraphics[width=0.49\textwidth]{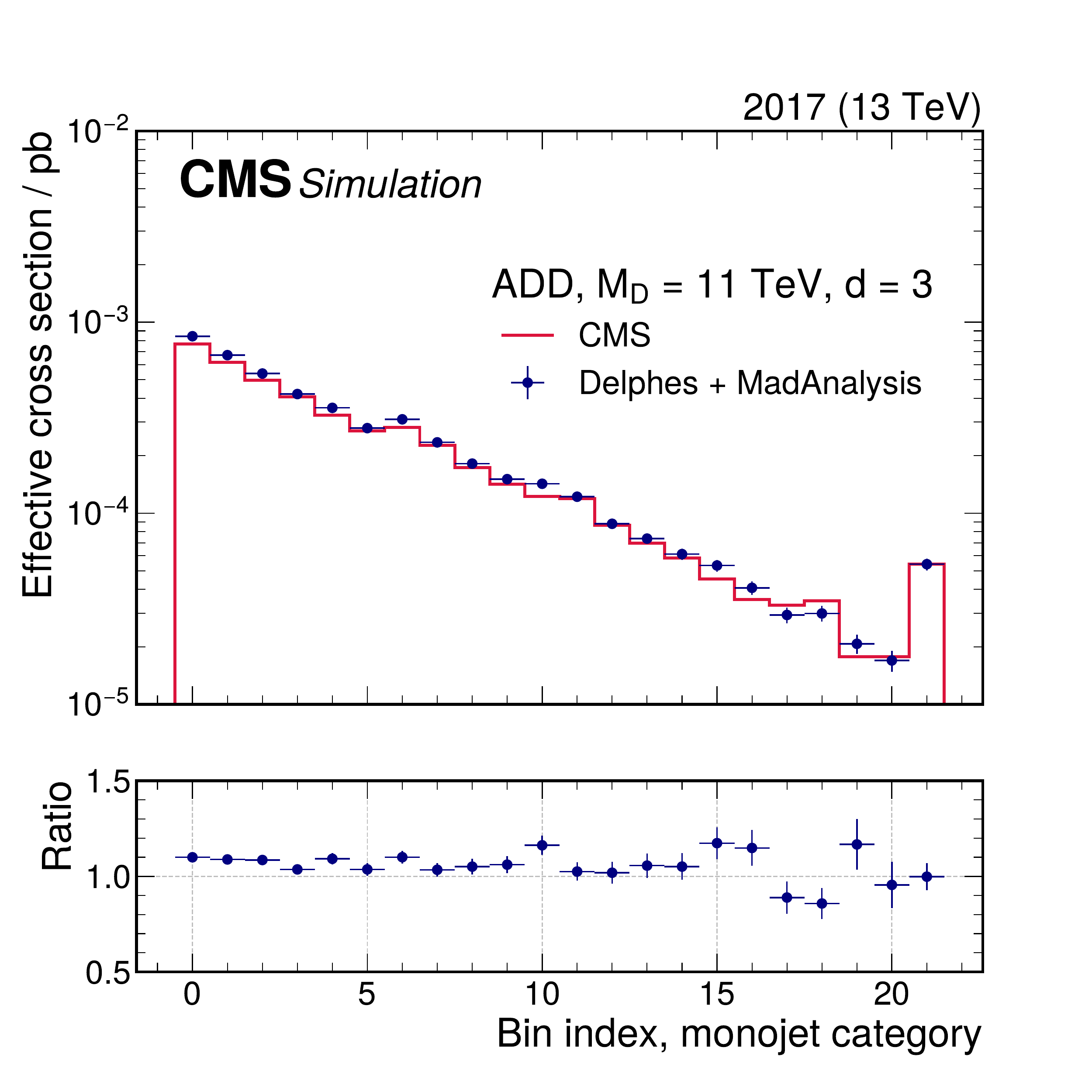}
        \includegraphics[width=0.49\textwidth]{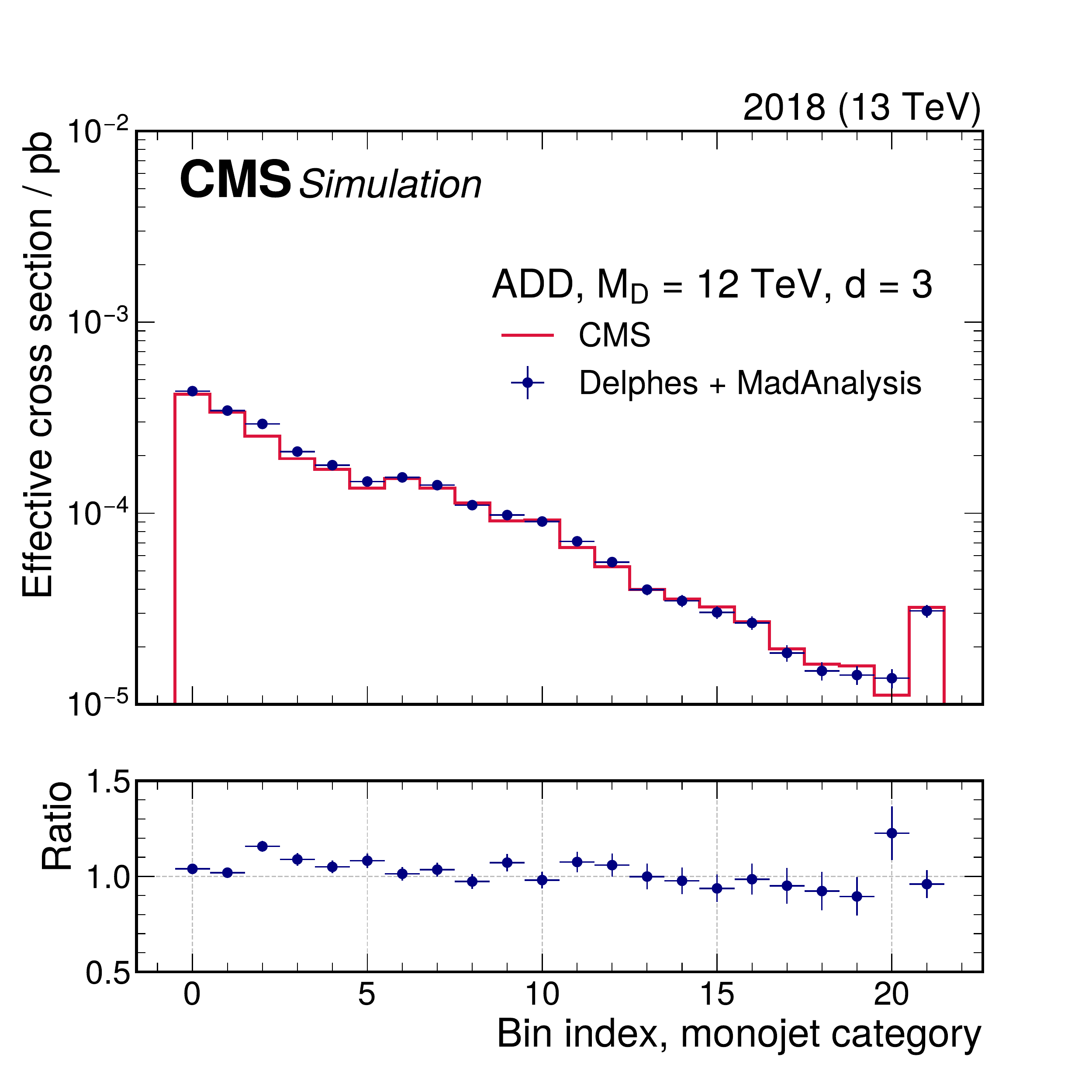}
        \caption{
            Comparison of the signal templates derived with \delphes and \madanalysis (dark blue points) and the CMS analysis work flow (red solid line).
            The panels show three example parameter points for the ADD interpretation, and showcase the selection procedure for different years (2016 in the upper left, 2017 in the upper right, and 2018 in the lower panels).
            The rightmost bin includes the overflow.
            In all cases, the average agreement is observed to be better than $10\%$, with maximum deviations up to $20\%$ in single bins.
        }
        \label{fig:reinterpretation_madanalysis_closure}

\end{figure*}

\subsection{Event display}

A graphical rendering of an observed high-\ptmiss collision event in the CMS detector is shown in Fig.~\ref{fig:supp_eventdisplay}.

\begin{figure*}[hbtp]
    \centering
        \includegraphics[width=1.0\textwidth]{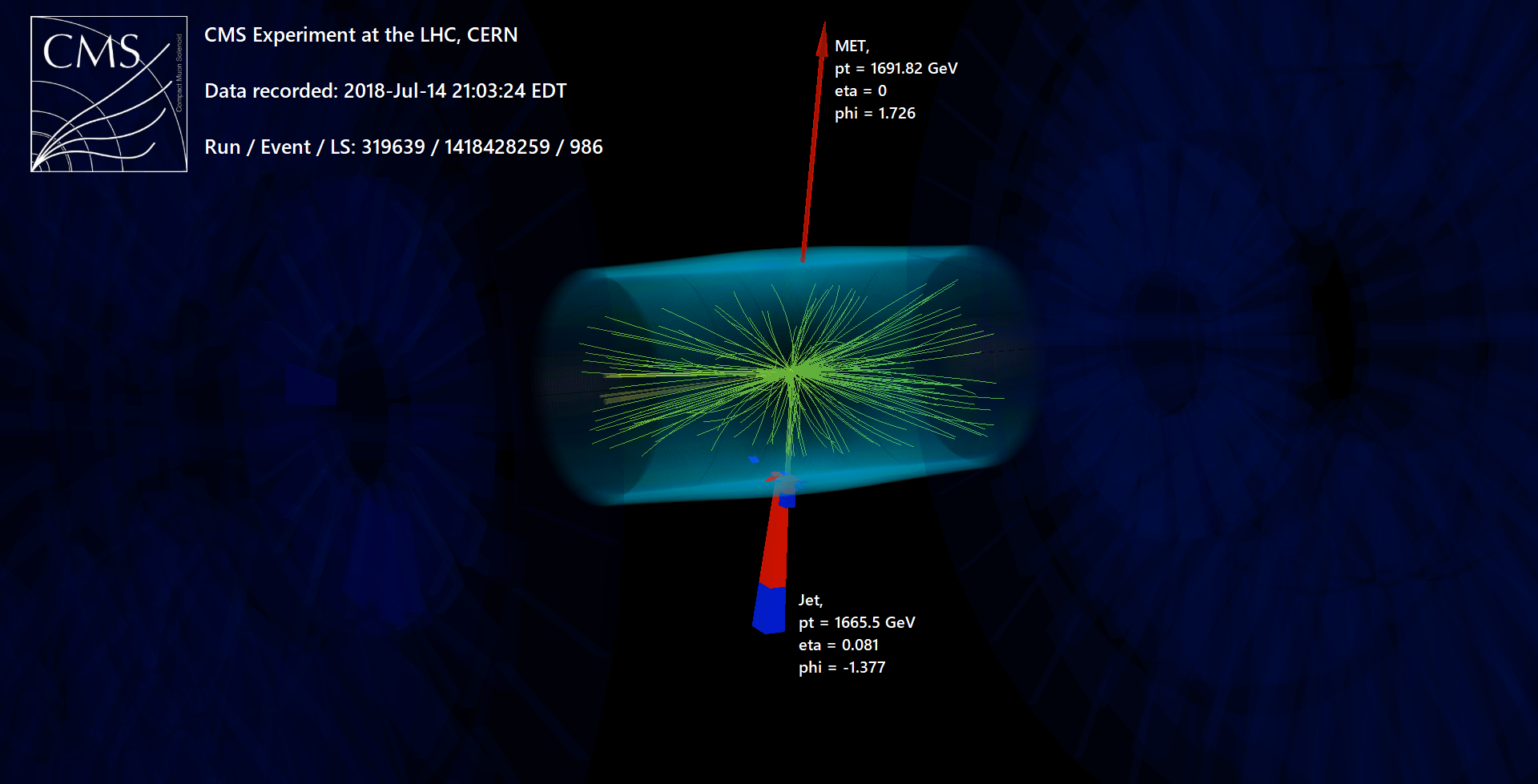}
        \caption{
            Display of a representative high-$\ptmiss$ event from the monojet category in the 2018 data set.
            In this event, a single high-\pt jet (calorimeter deposits indicated by the red and blue towers) recoils against large \ptmiss (indicated by the red arrow).
        }
        \label{fig:supp_eventdisplay}

\end{figure*}

}
\cleardoublepage \section{The CMS Collaboration \label{app:collab}}\begin{sloppypar}\hyphenpenalty=5000\widowpenalty=500\clubpenalty=5000\vskip\cmsinstskip
\textbf{Yerevan Physics Institute, Yerevan, Armenia}\\*[0pt]
A.~Tumasyan
\vskip\cmsinstskip
\textbf{Institut f\"{u}r Hochenergiephysik, Wien, Austria}\\*[0pt]
W.~Adam, J.W.~Andrejkovic, T.~Bergauer, S.~Chatterjee, M.~Dragicevic, A.~Escalante~Del~Valle, R.~Fr\"{u}hwirth\cmsAuthorMark{1}, M.~Jeitler\cmsAuthorMark{1}, N.~Krammer, L.~Lechner, D.~Liko, I.~Mikulec, P.~Paulitsch, F.M.~Pitters, J.~Schieck\cmsAuthorMark{1}, R.~Sch\"{o}fbeck, D.~Schwarz, S.~Templ, W.~Waltenberger, C.-E.~Wulz\cmsAuthorMark{1}
% ...
\vskip\cmsinstskip
% ...
1:  Also at TU Wien, Wien, Austria\\
% ...
\end{sloppypar}
\end{document}